

\input phyzzx
\Pubnum={\vbox{\hbox{CERN-TH.7174/94}\hbox{BONN-IR-94-02}
\hbox{hep-th/9404157}}}
\pubnum={\vbox{\hbox{CERN-TH.7174/94}\hbox{BONN-IR-94-02}
\hbox{hep-th/9404157}}}
\date={February, 1994}
\pubtype={}
\input epsf
\def\caption#1{\vskip 0.1in\centerline{\vbox{\hsize 4.8in\noindent
    \tenpoint\baselineskip=16pt\strut #1\strut}}}
\def\a{\alpha}
\def\b{\beta}
\def\g{\gamma}
\def\d{\delta}
\def\ep{\epsilon}
\def\ve{\varepsilon}
\def\z{\zeta}

\def\th{\theta}

\def\l{\lambda}
\def\m{\mu}
\def\n{\nu}
\def\x{\xi}

\def\s{\sigma}

\def\ph{\phi}

\def\G{\Gamma}
\def\D{\Delta}

\def\S{\Sigma}

\def\O{\Omega}
\def\o{\over}
\def\p{\partial}
\def\ri{\rightarrow}
\def\CE{\centerline}
\def\h{{1\over 2}}
\def\np{{\it Nucl.~Phys. }}
\def\pl{{\it Phys.~Lett. }}
\def\cmm{{\it Commun.~Math.~Phys. }}
\def\ijmp{{\it Int.~J.~Mod.~Phys. }}
\def\mpl{{\it Mod.~Phys.~Lett. }}
\def\prl{{\it Phys.~Rev.~Lett. }}
\def\p{\partial}

\def\b{\beta}
\def\g{\gamma}
\def\ap{\alpha_+}
\def\R{{\rm I\!R }}

\def\a{\alpha}
\def\G{\Gamma }
\def\D{\Delta}
\def\e{\varepsilon}
\def\no{\noindent}
\def\inbar{\,\vrule height1.5ex width.4pt depth0pt}
\def\IC{\relax\hbox{$\inbar\kern-.3em{\rm C}$}}
\def\IN{\relax{\rm I\kern-.18em N}}
\def\IR{\relax{\rm I\kern-.18em R}}
\font\cmss=cmss10 \font\cmsss=cmss10 at 7pt
\def\IZ{\relax\ifmmode\mathchoice
{\hbox{\cmss Z\kern-.4em Z}}{\hbox{\cmss Z\kern-.4em Z}}
{\lower.9pt\hbox{\cmsss Z\kern-.4em Z}}
{\lower1.2pt\hbox{\cmsss Z\kern-.4em Z}}\else{\cmss Z\kern-.4em
Z}\fi}
\def\sl{{SL(2,\IR)}}
\def\slU{{SL(2,\IR)/U(1)}}

\titlepage

\title{STRINGS, BLACK HOLES AND \break CONFORMAL FIELD \break
THEORY\foot{PhD Thesis at the University of Bonn.}}
\vskip 0.4cm
\author{Katrin Becker}
\vskip 0.4cm
\CE{\it Theory Division, CERN}
\CE{\it CH-1211, Geneva 23, Switzerland}
\vskip 0.4cm
\CE{\it and}
\vskip 0.4cm
\CE{\it Physikalisches Institut, Universit\"at Bonn,}
\CE{\it Nussallee 12, D-53115 Bonn, Germany.}
\vskip 1.0cm

\abstract{The {\slU} gauged Wess-Zumino-Witten model is an exact
conformal field theory describing a black hole in two-dimensional
space-time.
The free field approach of Bershadsky and Kutasov is a suitable
formulation of this CFT in order to compute physically interesting
quantities of this black hole. We find the space-time interpretation
of this model for $k=9/4$ and show that it reproduces the metric and
the dilaton found by Dijkgraaf, E.~Verlinde and H.~Verlinde in the
mini-superspace approximation.
We compute the two- and three-point functions of tachyons interacting
in the black hole background and analyse in detail the form of the
four-point tachyon scattering amplitude. We discuss the connection to
the $c=1$ matrix model and the deformed matrix model of Jevicki and
Yoneya.}

\vfill
\endpage
\def\chaone{INTRODUCTION}
\def\seconeone{String Theory and Black Holes}
\def\seconetwo{Outline of the Thesis}
\def\chatwo{STRINGS MOVING ON BLACK HOLES}
\def\sectwoone{Black Holes in General Relativity}
\def\sectwotwo{Strings in Curved Backgrounds}
\def\chathree{THE {\slU} GAUGED WZW MODEL }
\def\secthreeone{Lagrangian Formulation and Gauge Fixing}
\def\secthreeoneI{Current Algebra and the Representation Theory of
{\sl}}
\def\secthreeoneII{Physical States of the Coset Model and BRST
Quantization}
\def\secthreetwo{Witten's Semiclassical Interpretation as a 2D Black
Hole}
\def\secthreethree{The Exact Background of Dijkgraaf, Verlinde and
Verlinde}
\def\chafour{FREE FIELD APPROACH TO THE BLACK HOLE CFT}
\def\secfourone{The Operator Approach}
\def\secfouronetwo{The Gauss Decomposition}
\def\secfourtwo{TARGET SPACE GEOMETRY IN TERMS OF WAKIMOTO
COORDINATES}
\def\chafive{COHOMOLOGY OF THE EUCLIDEAN BLACK HOLE}
\def\secfiveone {The Classification of Distler and Nelson}
\def\secfivetwo {The Simplest Discrete States }
\def\secfivethree{The Complete Free Field Cohomology}
\def\chasix{CORRELATION FUNCTIONS IN THE BLACK HOLE BACKGROUND}
\def\secsixoneone{Scattering Amplitudes in String Theory}
\def\secsixonetwo{Conservation Laws and Zero-Mode Integrations}
\def\secsixonethree{Scattering in the Bulk}
\def\secsixtwo{Three-Point Function with One Highest Weight State}
\def\secsixthree{Three-Point Function of Arbitrary Tachyons}
\def\secsixfour{Illustrative Example}
\def\secsixfive{Two-Point Function}
\def\secsixsix{The Four-Point Function}

\endpage
\endpage
\footline{\hfil}
{\vbox{\vskip 20cm\hbox{\hskip 12cm Dedicated to my mother
and}\hbox{\hskip 12cm to the memory of my father}}}
\endpage
\footline={\hss\tenrm\folio\hss}
\pageno=-3
\CE{\bf STRINGS, BLACK HOLES AND }
\CE{\bf CONFORMAL FIELD}
\CE{\bf THEORY }
\vskip 1cm
\CE{\bf Katrin Becker}
\CE{\it Theory Division, CERN}
\CE{\it CH-1211, Geneva 23, Switzerland}
\CE{\it and}
\CE{\it Physikalisches Institut, Universit\"at Bonn,}
\CE{\it Nussallee 12, D-53115 Bonn, Germany.}
\vskip 3cm
\CE{\bf C{\tenpoint ONTENTS}}
\vskip 2cm
\settabs 5 \columns
\+ {\bf Chapter 1:} & {\bf I{\tenpoint NTRODUCTION}} &&&&& \hfill
{\bf 1} \cr
\+& 1.1. \seconeone &&&&& \hfill 1 \cr
\+& 1.2. \seconetwo &&&&& \hfill 6\cr
\vskip 0.75cm
\+ {\bf Chapter 2:} & {\bf S{\tenpoint TRINGS} M{\tenpoint OVING}
{\tenpoint ON} B{\tenpoint LACK} H{\tenpoint OLES}} &&&&& \hfill {\bf
10}\cr
\+& 2.1. \sectwoone\dotfill  &&&&& \hfill  10\cr
\+& 2.2. \sectwotwo \dotfill  &&&&& \hfill 13\cr
\vskip 0.75cm
\+ {\bf Chapter 3:} & {\bf T{\tenpoint HE} {\slU} G{\tenpoint AUGED}
WZW M{\tenpoint ODEL }}&&&&& \hfill {\bf 19} \cr
\+& 3.1. \secthreeone&&&&& \hfill  19 \cr
\+& 3.2. \secthreeoneI&&&&& \hfill 21 \cr
\+& 3.3. \secthreeoneII&&&&& \hfill 27 \cr
\+& 3.4. \secthreetwo&&&&& \hfill 31 \cr
\+& 3.5. \secthreethree&&&&& \hfill  34 \cr
\vskip 0.75cm
\+ {\bf Chapter 4:} & {\bf F{\tenpoint REE} F{\tenpoint IELD}
A{\tenpoint PPROACH} {\tenpoint TO} {\tenpoint THE} B{\tenpoint LACK}
H{\tenpoint OLE} CFT} &&&&& \hfill {\bf 41} \cr
\+& 4.1. \secfourone&&&&& \hfill 41 \cr
\+& 4.2. \secfouronetwo &&&&& \hfill 46\cr
\vskip 0.75cm
\+ {\bf Chapter 5:} &{\bf T{\tenpoint ARGET} S{\tenpoint PACE}
G{\tenpoint EOMETRY} {\tenpoint IN} T{\tenpoint ERMS} {\tenpoint OF}
W{\tenpoint AKIMOTO}} &&&&&  \cr
\+ & {\bf C{\tenpoint OORDINATES}}&&&&& \hfill  {\bf 47} \cr
\vskip 0.75cm
\+ {\bf Chapter 6:} & {\bf C{\tenpoint OHOMOLOGY} {\tenpoint OF}
{\tenpoint THE} E{\tenpoint UCLIDEAN} B{\tenpoint LACK} H{\tenpoint
OLE}} &&&&& \hfill {\bf 55} \cr
\+& 5.1. \secfiveone&&&&& \hfill 55  \cr
\+& 5.2. \secfivetwo&&&&& \hfill  57 \cr
\+& 5.3. \secfivethree&&&&& \hfill  61 \cr
\vskip 0.75cm
 \+ {\bf Chapter 7:} &{\bf C{\tenpoint ORRELATION} F{\tenpoint
UNCTIONS} {\tenpoint IN} {\tenpoint THE} B{\tenpoint LACK}
H{\tenpoint OLE}}\cr
\+ & {\bf B{\tenpoint ACKGROUND}}&&&&& \hfill  {\bf 64} \cr
\+& 6.1.  \secsixoneone&&&&& \hfill 64\cr
\+& 6.2.  \secsixonetwo&&&&& \hfill  65\cr
\+& 6.3.  \secsixonethree&&&&& \hfill  67\cr
\+& 6.4.  \secsixtwo&&&&& \hfill 72\cr
\+& 6.5.  \secsixthree&&&&& \hfill  74\cr
\+& 6.6.  \secsixfour&&&&& \hfill  80\cr
\+& 6.7.  \secsixfive&&&&& \hfill  83\cr
\+& 6.8.  \secsixsix&&&&& \hfill  87\cr
\vskip 0.75cm
\+ {\bf Chapter 8:} &{\bf C{\tenpoint ONCLUDING} R{\tenpoint
EMARKS}}&&&&& \hfill {\bf 96}\cr
\vskip 0.75cm
\+ {\bf Acknowledgements} &&&&& \hfill {\bf 99}\cr
\vskip 0.75cm
\+ {\bf A{\tenpoint PPENDIX} } &&&&& \hfill {\bf 100}\cr
\vskip 0.75cm
\+ {\bf R{\tenpoint EFERENCES}} &&&&& \hfill  {\bf 102}\cr
\endpage
\def\rhaw{S.~Hawking, ``Particle Creation by Black
Holes'', \cmm {\bf 43} (1975)~199.}

\def\rdvv{R.~Dijkgraaf, E.~Verlinde and H.~Verlinde,
``String Propagation in Black Hole Geometry'', \np {\bf B371} (1992)
269.}
\def\rbk{M.~Bershadsky and D.~Kutasov, ``Comment on
Gauged WZW Theory'', {\it Phys.~Lett.} {\bf B266} (1991) 345.}
\def\rms{E.~Martinec and S.~L.~Shatashvili, ``Black
Hole Physics and Liouville Theory'', \np {\bf B368} (1992) 338.}
\def\reky{T.~Eguchi, H.~Kanno and S.~Yang,
``$W_{\infty}$ Algebra in Two-Dimensional Black
Hole'', \pl {\bf B298} (1993) 73.}

\def\rgm{P.~Ginsparg and G.~Moore, ``Lectures in 2D
Gravity and 2D String Theory'', preprint
YCTP-P23-92, LA-UR-92-3479, hep-th/9304011.}
\def\rdn{J.~Distler and P.~Nelson, ``New Discrete
States of Strings Near a Black Hole'', \np {\bf B374} (1992) 123.}
\def\rw{M.~Wakimoto, ``Fock Representations of the
Affine Lie Algebra ${A_1}^{(1)}$'', {\it Commun.~Math. Phys.} {\bf
104} (1986) 605.}
\def\rd{V.~S.~Dotsenko, ``The Free Field
Representation of the SU(2) CFT'', \np {\bf B338}
(1990) 747; ``Solving the SU(2) CFT with the
Wakimoto Free Field Representation'', \np {\bf
B358} (1991) 547.}
\def\rdfk{P.~Di Francesco and D.~Kutasov, ``World
Sheet and Space Time Physics in Two Dimensional
(Super-)String Theory'', \np {\bf B375} (1992) 119;
``Correlation Functions in 2D String Theory'', \pl
{\bf B261} (1991) 385.}
\def\rdtwo{V.~S.~Dotsenko, ``Correlation Functions
of Local Operators in 2D Gravity Coupled to
Minimal Matter'', Summer School Carg\`ese 1991,
hep-th/9110030; ``Three-Point Correlation Functions
of the Minimal Conformal Field Theories Coupled to
2-D Gravity'', \mpl {\bf A6} (1991) 3601.}
\def\rgl{M.~Goulian and M.~Li, ``Correlation
Functions in Liouville Theory'',
\prl {\bf 66} (1991) 2051.}
\def\re{T.~Eguchi, ``$c=1$ Liouville Theory
Perturbed by the Black Hole Mass Operator'', {\it Phys.~Lett.} {\bf
B316} (1993) 74.}
\def\rjy{A.~Jevicki and T.~Yoneya, ``A Deformed
Matrix Model for the Black Hole Background in
Two-Dimensional String Theory'', {\it Nucl.~Phys.} {\bf B411} (1994)
64; K.~Demeterfi and J.~P.~Rodrigues, ``States and Quantum
Effects in the Collective Field Theory of a
Deformed Matrix Model'', {\it Nucl.~Phys.} {\bf B415} (1994) 3;
U.~Danielsson, ``A Matrix Model Black Hole'', {\it Nucl.~Phys.} {\bf
B395} (1993) 395; K.~Demeterfi, I.~Klebanov and J.~P.~Rodrigues,
``The exact $S$ Matrix of the Deformed
$c=1$ Matrix Model'', {\it Phys.~Rev.~Lett.} {\bf 71} (1993) 3409; U.
Danielsson, ``The Deformed Matrix Model at Finite Radius and a New
Duality Symmetry'', preprint CERN-TH-7021-93, hep-th/9309157.}
\def\rbih{W.~J.~Holman and L.~C.~Biedenharn, ``A
General Study of the Wigner Coefficients of
SU(1,1)'', {\it Ann.~Phys.} {\bf 47} (1968) 205;
V.~Bargman, ``Irreducible Unitary Representations
of the Lorentz Group'', {\it Ann. Math.} {\bf 48}
(1947) 568. }
\def\rdfa{V.~S.~Dotsenko and V.~A.~Fateev, ``Four-Point Correlation
Function and the Operator Algebra
in the 2D Conformal Invariant Theories with Central
Charge $c<1$'', \np {\bf B251} (1985) 691; V.~S.~Dotsenko and
V.~A.~Fateev, ``Conformal Algebra and
Multipoint Correlation Functions in 2D Statistical
Models'', \np {\bf B240} (1984) 312.}
\def\rfms{D.~Friedan, E.~Martinec and S.~Shenker,
``Conformal Invariance, Supersymmetry and String
Theory'', \np {\bf B271} (1986) 93.}

\def\rsei{N.~Seiberg, ``Notes on Quantum Liouville Theory and Quantum
Gravity'', {\it Prog. Theor.~Phys.~Suppl.} {\bf 102} (1990) 319.}
\def\rcl{S.~Chaudhuri and J.~Lykken, ``String
Theory, Black Holes and {\sl} Current Algebra'',
\np {\bf B396} (1993) 270.}
\def\remn{J.~Ellis, N.~E.~Mavromatos and D.~V.~Nanopoulos, ``Quantum
Coherence and Two-Dimensional
Black Holes'', \pl {\bf B267} (1991) 465.}
\def\rmo{N.~Marcus and Y.~Oz, ``The Spectrum of 2D
Black Hole or does the 2D Black Hole have
Tachyonic $W$ Hair?'', {\it Nucl.~Phys.} {\bf B407} (1993) 429.}
\def\rmoore{G.~Moore, ``Finite in All Directions'', preprint
YCTP-P12-93, hep-th/9305139.}
\def\rvm{S.~Mukhi and C.~Vafa, ``Two-Dimensional
Black Hole as a Topological Coset of $c=1$ String
Theory'', {\it Nucl.~Phys.} {\bf B407} (1993) 667.}

\def\rgs{M.~B.~Green and N.~Seiberg, ``Contact
Interactions in Superstring Theory'', {\np} {\bf B299} (1988) 559.}
\def\reder{A.~Erd\'elyi, ``Higher Transcendental
Functions'',  McGraw-Hill, New York, 1953, p. 188-189.}
\def\rvil{N.~Ja.~Vilenkin and A.~U.~Klimyk,
``Representation of Lie Groups and Special
Function'', (Kluwer Academic Publishers, Dordrecht, 1992).}
\def\rwadia{A.~Dhar, G.~Mandal and S.~Wadia, ``Wave
Propagation in Stringy Black Hole'', \mpl {\bf A8} (1993) 1701;
A.~Dhar, G.~Mandal and S.~Wadia,
``Stringy Quantum Effects in Two-dimensional Black
Hole'', \mpl {\bf A7} (1992) 3703.}
\def\rdas{S.~R.~Das, ``Matrix Models and Black
Holes'', \mpl {\bf A8} (1993) 69; S.~R.~Das,
``Matrix Models and Nonperturbative String
Propagation in Two-dimensional Black Hole
Backgrounds'', \mpl {\bf A8} (1993) 1331.}
\def\Ai{$$
S_{WZW}(g)={k \over 8\pi}\int_{\Sigma}d^2x\sqrt{h}
h^{ij}\,
{\rm tr}( g^{-1} \p_i g g^{-1} \p_j g)+ik\Gamma(g),
\eqn\Ai
$$}
\def\Aii{$$
\Gamma (g)={1\over 12 \pi}\int_B d^3 y
\varepsilon^{abc}\,
 {\rm tr}(g^{-1}\p_a g
g^{-1}\p_b g g^{-1} \p_c g),
\eqn\Aii
$$}
\def\Aiii{$$
S_{WZW}(g,A)=S_{WZW}(g)+{k\over 2\pi} \int
d^2z({\bar A}
{\rm tr}(Gg^{-1}\p g)+A {\rm tr} (G {\bar \p} g g^{-1})+A{\bar
A}
(-2+{\rm tr}( GgGg^{-1}))).
\eqn\Aiii
$$}
\def\Aiv{$$
S_{WZW}^{gf}=S_{WZW}(g)+S(X)+S(B,C).
\eqn\Aiv
$$}
\def\Av{$$
S(X)={k\over 4\pi}\int d^2z \p X {\bar \p}X,
\eqn\Av
$$}
\def\Avi{$$
S(B,C)=\int d^2 z (B{\bar \p} C +{\bar B}\p {\bar
C}).
\eqn\Avi
$$}
\def\Avii{$$
J=J^a t^a =-{k\over 2} \p g g^{-1}, \qquad  \qquad
{\bar J}={\bar J}^a t^a =-{k\over 2} g^{-1}
{\bar \p} g ,
\eqn\Avii
$$}
\def\Aviii{$$
J_{+}(z)J_{-}(w)={k\over (z-w)^2}-{2J_3(w)\over
(z-w)}
+\dots
$$
$$
J_3(z)J_{\pm}={\pm}{J_{\pm}(w)
\over (z-w)}+\dots
$$
$$
J_3(z)J_3(w)=-{{k\over 2}\over (z-w)^2}+\dots,
\eqn\Aviii
$$}
\def\Bi{$$
\langle \gamma(z) \beta(w) \rangle =-{1\over (z-w)}
,\qquad  \qquad
\langle \p \phi(z) \p \phi(w) \rangle =-{1\over
(z-w)^2},
\eqn\Bi
$$}
\def\Bii{
$$
\eqalign{
J_+(z)&=\b(z)\cr
&\cr
J_3(z)&=-\b(z)\g(z)-{\ap\over 2}\p \phi(z)\cr
&\cr
J_-(z)&=\b(z)\g^2(z)+\ap\g(z) \p\phi(z)+k\p\g(z).\cr}
\eqn\Bii
$$}
\def\Biii{$$
T_{{\rm SL(2,I\!R)}}=-{\Delta\over (k-2)}
,\qquad  \qquad
\Delta= -{1\over 2}: J_+J_-+J_-J_+:+:J_3J_3:.
\eqn\Biii
$$}
\def\Biv{$$
T_{{\rm SL(2,I\!R)}}=\b\p\g-{1\over 2}(\p \phi)^2-
{1\over \ap}\p^2\phi.
\eqn\Biv
$$}
\def\Bv{$$
c={3k\over k-2},
\eqn\Bv
$$}
\def\Bvi{$$
\Delta_0 |j,m\rangle =j(j+1)|j,m\rangle,
\qquad\qquad
J_0^3|j,m\rangle =m|j,m\rangle ,
\eqn\Bvi
$$}
\def\Bvii{$$
J_n^{\pm}|j,m\rangle=J_n^3|j,m\rangle =0\qquad
{\rm for}\quad n>0,
\eqn\Bvii
$$}
\def\Bviii{$$\eqalign{
J_0^+|j,m\rangle & =(-m+j)|j,m+1\rangle\cr
& \cr
J_0^{-}|j,m\rangle & =(-m-j)|j,m-1\rangle .\cr}
\eqn\Bviii
$$}

\def\Bix{$$
\eqalign{
&|{\rm HWS\; module }\rangle  =|j,m\rangle
\qquad m=j,j-1,\dots \cr
&|{\rm LWS\; module }\rangle  =|j,m\rangle
\qquad m=-j,-j+1,\dots \cr }
\eqn\Bix
$$}

\def\Bx{$$
(j_1, m_1|j_2\; m_2)=\delta^{j_1 j_2}\delta^{m_1
m_2}.
\eqn\Bx
$$}

\def\Bxi{$$
J_0^+ |j,m) =(j-m){N(j,m)\over N(j,m+1)}|j, m+1)
$$
$$
J_0^-|j,m) =-(j+m){N(j,m)\over N(j,m-1)}|j,m-1).
\eqn\Bxi
$$}

\def\Bxii{$$
(j,m+1|J_0^+ | j,m)^{\ast}=(j,m|J_0^-|j,m+1).
\eqn\Bxii
$$}

\def\Bxiii{$$
\left| (j-m){N(j,m)\over N(j,m+1)}\right|^2
=-(j+m+1)(j-m).
\eqn\Bxiii
$$}

\def\Bxiv{$$
|j,m) ={1 \over
\sqrt{\Gamma(j+m+1)\Gamma(j-m+1)}}|j,m\rangle.
\eqn\Bxiv
$$}

\def\Bxv{
$$
\eqalign{
&J_0^+|j,m)=\sqrt{(j-m)(j+m+1)}|j,m+1)\cr
&\cr
&J_0^-|j,m)=\sqrt{(j+m)(j-m+1)} |j,m-1)\cr
&\cr
&J_0^3|j,m)=m|j,m). \cr}
\eqn\Bxv
$$}

\def\Bxva{$$
|j,m\rangle=\sqrt{\G(j+m+1)\G(j-m+1)}|j,m).
\eqn\Bxva
$$}

\def\Bxvi{$$
T_{j\; m}(z)=:\g^{j-m}(z) e^{{2\over \ap}
j\phi(z)}:,
\eqn\Bxvi
$$}

\def\Bxvii{$$
\eqalign{
T(z)T_{j\;m}(w)&={h_{j,m}\over (z-w)^2}
T_{j\; m}(w) +{1\over (z-w)}\p
T_{j\;m}(w)+\dots\cr
&\cr
J^a(z)T_{j}(w)&={t^a_{(j,m)}\over (z-w)}
T_j(w)+\dots\cr}
\eqn\Bxvii
$$}

\def\Bxviii{$$
h_{j,m}=-{j(j+1)\over k-2}.
\eqn\Bxviii
$$}

\def\Bxix{$$
S={1\over 2\pi}\int{1\over 2} \p \phi{\bar \p}\phi -{2\over
\ap}
R^{(2)} \phi +\b{\bar \p}\g +{\bar \b}\p{\bar \g}.
\eqn\Bxix
$$}

\def\Bxx{$$
\eqalign{
&\langle T(z)T_1(z_1,{\bar z}_1)\dots
T_N(z_N,{\bar z}_N)\rangle =
\sum_{i=1}^N \left( {h_i\over (z-z_i)^2}+
{1\over (z-z_i)}{\p \over \p z_i}\right)
\langle T_1(z_1,{\bar z}_1)\dots
T_N(z_N,{\bar z}_N)\rangle\cr
&\cr
&\langle J^a(z) T_{j_1}(z_1,{\bar z}_1)
\dots T_{j_N}(z_N,{\bar z}_N)\rangle =
\sum_{i=1}^N {t^a_i\over (z-z_i)}
\langle T_{j_1}(z_1,{\bar z}_1)\dots
T_{j_N}(z_N,{\bar z}_N)\rangle .\cr}
\eqn\Bxx
$$}

\def\Bxxi{$$
{\cal Q}=\int d^2 z J(z,{\bar z})
,\qquad  \qquad
J(z,{\bar z})=\b(z){\bar \b}({\bar z})
e^{-{2\over \ap}\phi(z,{\bar z})} .
\eqn\Bxxi
$$}

\def\Bxxii{
$$
J_3(z)J(w,{\bar w})\sim {\rm reg.},
$$
$$
J_+(z)J(w,{\bar w})\sim {\rm reg.},
$$
$$
J_-(z)J(w,{\bar w})\sim {\p \over \p w}\left(
{e^{-{2\over \ap}
\phi(w,{\bar w})}\over (z-w)}\right).
\eqn\Bxxii
$$}

\def\Bxxiii{$$
S={1\over 2\pi}\int {1\over 2}\p \phi{\bar \p}\phi -{2\over
\ap}
 R^{(2)} \phi +\b{\bar \p}\g +{\bar \b}\p{\bar \g}
+2\pi M \b {\bar \b} e^{-{2\over \ap}\phi}.
\eqn\Bxxiii
$$}

\def\Bxxiv{$$
\b=-i\p v e^{iv-u} ,\qquad  \qquad \g=e^{u-iv},
\eqn\Bxxiv
$$}

\def\Bxxv{$$
\langle u(z) u(w)\rangle =\langle v(z) v(w)
\rangle=-\log(z-w).
\eqn\Bxxv
$$}

\def\Ci{$$
\langle \p X(z)\p X(w)\rangle =-{1\over (z-w)^2}
,\qquad  \qquad
\langle C(z)B(w)\rangle ={1\over (z-w)}.
\eqn\Ci
$$}

\def\Cii{$$
Q^{\rm U(1)}=\oint C(z)\Bigl(J^3-i\sqrt{{k\over 2}}
\p X\Bigr) dz ,
\eqn\Cii
$$}

\def\Ciii{$$
{\cal V}_{j \; m}=:\g^{j-m} e^{{2\over \ap} j\phi}
e^{im\sqrt{{2\over k}}X}:.
\eqn\Ciii
$$}

\def\Civ{$$
h_{j,m}=-{j(j+1)\over k-2}+{m^2\over k},
\eqn\Civ
$$}

\def\Cv{$$
T_{{\rm SL(2,I\!R)/U(1)}}=\b \p \g -{1\over 2}
(\p \phi)^2
-{1\over \ap}\p^2\phi -{1\over 2}(\p X)^2-B\p C.
\eqn\Cv
$$}

\def\Cvi{$$
c={3k\over k-2}-1
\eqn\Cvi
$$}

\def\Cvii{$$
T_{gh}=-2b\p c-\p b c.
\eqn\Cvii
$$}

\def\cv{$$
T_{{\rm SL(2,I\!\R)/U(1)}}=\b \p \g -{1\over 2}
(\p \phi)^2-{1\over \ap}\p^2\phi -{1\over 2}
(\p X)^2-B\p C-2b\p c-\p b c.
\eqn\cv
$$}

\def\Cviii{$$
Q^{Diff}=\oint c(z)\Bigl(T_{{\rm SL(2,\R)/U(1)}}+
{1\over 2}T_{gh}\Bigr).
\eqn\Cviii
$$}

\def\Cix{$$
(Q^{Diff})^2=0,\qquad \qquad \{ Q^{\rm U(1)},
 Q^{Diff}\}=0.
\eqn\Cix
$$}
\def\Cixa{$$
h_{j,m}=-{j(j+1)\over k-2}+{m^2\over k}=1;
\eqn\Cixa
$$}
\def\Cixb{$$
2j+1=\pm {2\over 3}m,
\eqn\Cixb
$$}
\def\Cx{$$\eqalign{
{\widetilde {\cal D}}^{\pm}&: \hskip 5mm p_X=\pm
{2s-4r-1\over 2\sqrt{2}},\qquad \qquad
p_\phi={2s+4r-5\over 2\sqrt{2}},\qquad {\cal N}=r(2s-1)\cr
&\cr
{\cal D}^{\mp}&: \hskip 5mm p_X=\pm {s-2r+1\over
\sqrt{2}}
,\qquad \hskip 2mm\qquad
p_\phi={s+2r-3\over \sqrt{2}},\qquad\;\; {\cal N}=s(2r-1)\cr
&\cr
{\cal C}&: \hskip 5mm p_X= {2(s-r)\over \sqrt{2}},
\hskip 9mm\qquad\qquad p_\phi={2(s+r-1)\over \sqrt{2}},\quad \;\;
\;{\cal N}=4sr, \cr}
\eqn\Cx
$$}
\def\Cxiii{$$
(L_0-{\bar L}_0)|j,m,{\bar m}\rangle
=0\qquad,\qquad
(L_0+{\bar L}_0-2)|j,m,{\bar m}\rangle =0,
\eqn\Cxiii
$$}
\def\Cxiv{$$
L_0|j,m,{\bar m}\rangle={\bar L}_0|j,m,{\bar m}
\rangle=|j,m,{\bar m}\rangle,
\eqn\Cxiv
$$}
\def\cxiv{$$
-{j(j+1)\over k-2}+{m^2\over k}=1,
\eqn\cxiv
$$}
\def\Cxv{$$
2j+1=\pm {2\over 3}m.
\eqn\Cxv
$$}
\def\Cxvi{$$
m={1\over 2}(n_1+n_2k),\qquad \qquad {\bar m}=
-{1\over 2}(n_1-n_2k),\qquad \qquad
n_1,n_2\in \IZ.
\eqn\Cxvi
$$}
\def\dia{$$
{\cal A}^{j_1.\,.\,.\,.\,j_N}_{m_1.\,.\,. m_N}
=\langle {\cal V}_{j_1\; m_1} \dots {\cal V}_{j_N\;
m_N}
\left(\int d^2z \b{\bar \b} e^{-{2\over \ap}\phi}
\right)^s\rangle_{M=0}
\eqn\dia
$$}
\def\di{$$
s=\sum_{i=1}^N j_i+(1-g).
\eqn\di
$$}
\def\dib{$$
\sum_{i=1}^N m_i=0.
\eqn\dib
$$}
\def\dii{$$
\# \b -\# \g =1.
\eqn\dii
$$}
\def\diii{$$
\Biggl\langle \prod_{i=1}^N {\cal V}_{{j_i}\;{m_i}}
\Biggr\rangle=M^s\G(-s)
\Biggl\langle \prod_{i=1}^N {\widetilde
{\cal V}_{{j_i}\;{m_i}}}\left( \int \b{\bar \b}
 e^{-{2\over \ap}
{\widetilde \phi}}d^2z \right)^s\Biggr\rangle_{M=0},
\eqn\diii
$$}
\def\div{$$
\eqalign{
\widetilde{\cal
A}^{j_1\;j_2\;\;\;j_3}_{j_1\;m_2\;m_3}=&
\int \prod_{i=1}^s d^2z_i
\bigl\langle e^{{2\over \ap} j_1
\widetilde{\phi}(0)}e^{{2\over \ap}
 j_2 \widetilde{\phi}(1)}
e^{{2\over \ap} j_3 \widetilde{\phi}(\infty)}
 e^{-{2\over \ap} \widetilde{\phi}
(z_i,{\bar z_i})}\bigr\rangle\cr
&\cr
&\langle \g^{j_2-m_2}(1)\g^{j_3-m_3}
(\infty)\b(z_i)\rangle
\langle {\bar \g}^{j_2-m_2}(1)
{\bar\g}^{j_3-m_3}(\infty){\bar \b} ({\bar
z}_i)\rangle.\cr}
\eqn\div
$$}
\def\dv{$$
{\cal P}=\prod_{i=1}^s(1-z_i)^{m_2-j_2}
\prod_{i<j}(z_i-z_j).
\eqn\dv
$$}
\def\dvi{$$
\widetilde{\cal
A}^{j_1\;j_2\;\;\;j_3}_{j_1\;m_2\;m_3}=
\int \prod_{i=1}^s d^2 z_i |z_i|^{-4\rho j_1}
|1-z_i|^{-4\rho j_2}\prod_{i<j}
|z_i-z_j|^{4\rho}
{\cal P}^{-1}{\p^s{\cal P}\over \p z_1\dots \p z_s}
{\bar {\cal P}}^{-1}
{\p^s{\bar {\cal P}}\over \p {\bar z_1}
\dots \p {\bar z_s}},
\eqn\dvi
$$}
\def\dvii{$$
{\cal P}^{-1}{\p^s{\cal P}\over \p z_1 \dots \p
z_s}
={\Gamma(-j_2+m_2+s)\over \Gamma(-j_2+m_2)}
\prod_{i=1}^s(1-z_i)^{-1}.
\eqn\dvii
$$}
\def\dviii{$$
\prod_{i=1}^s(z_i-z_j)=\sum_{\sigma(p(1),\dots,p(s))
}
 sign(p) z_{p(1)}^0\dots
z_{p(s)}^{s-1},
\eqn\dviii
$$}
\def\dix{$$
{\cal A}^{j_1\;j_2\;\;\;j_3}_{j_1\;m_2\;m_3}=
(-)^s\D(j_2-m_2+1) \D (j_3-m_3+1)
 {\cal I}(j_1,j_2,j_3,k).
\eqn\dix
$$}
\def\dx{$$
{\cal I}(j_1,j_2,j_3,k)=M^s
\G(-s)\int \prod_{i=1}^s d^2 z_i
|z_i|^{-4\rho j_1}|1-z_i|^{-4\rho j_2-2}
\prod_{i<j}|z_i-z_j|^{4\rho}
\eqn\dx
$$}

\def\dxi{$$
\left(\G(-j_2+m_2+s)\over \G(-j_2+m_2)\right)^2
=(-)^s\D(j_2-m_2+1)\D(j_3-m_3+1),
\eqn\dxi
$$}

\def\dxii{$$
e^{\a J^-_0}{\cal V}_{j\; m} e^{-\a J_0^-}=
\sum_{k=0}^\infty {\a^k\over k!}[J_0^-,{\cal
V}_{j\; m}]_k
\eqn\dxii
$$}

\def\dxiia{$$
e^{\a J^-_0}{\cal V}_{j\; m} e^{-\a J_0^-}=
\sum_{k=0}^\infty {\a^k\over k!}
[J_0^-,{\cal V}_{j\; m}]_k,
\eqn\dxiia
$$}

\def\dxiib{$$
[{J_0}^-,{\cal V}_{j\; m}]_{0}=
{\cal V}_{j\; m},\qquad \qquad [{J_0}^{-},{\cal
V}_{j\; m}]_{k}=[{J_0}^-,[{J_0}^-,{\cal V}_{j\;
m}]_{k-1}].
\eqn\dxiib
$$}

\def\dxiic{$$
[J^-_0,{\cal V}_{j\; m}]=-(j+m)
{\cal V}_{j\; m-1},\qquad\qquad
[J^-_0,{\cal V}_{j\; m}]_k=(-)^k
{\G(j+m+1)\over \G(j+m-k+1)}{\cal V}_{j\; m-k}.
\eqn\dxiic
$$}

\def\dxiii{$$
\eqalign{
\sum_{k_2,k_3=0}^{\infty}
{\a^{k_2+k_3}\over k_2! k_3!}
&{\Gamma(j_2+m_2+1) \over \Gamma(j_2+m_2+1-k_2)}
{\Gamma(j_3+m_3+1) \over \Gamma(j_3+m_3+1-k_3)}
{\cal
A}^{j_1\;j_2\;\;\;\;\;\;\;\;\;j_3}_{j_1\;m_2-k_2\;m_
3-k_3}\cr
&\cr
&=\sum_{k_1=0}^{\infty}{(-\a)^{k_1}\over k_1!}
{\Gamma(2j_1+1)\over \Gamma(2j_1+1-k_1)}
{\cal
A}^{j_1\;\;\;\;\;\;\;j_2\;\;\;j_3}_{j_1-k_1\;m_2\;m_
3}.\cr}
\eqn\dxiii
$$}

\def\dxiv{$$
{\cal
A}^{j_1\;\;\;\;\;\;\;j_2\;\;\;j_3}_{j_1-k_1\;m_2\;m_
3}=
\sum_{n=0}^{k_1}
\left({\Gamma(-2j_1)\over \Gamma(k_1-2j_1)}
{\Gamma (k_1+1)\over \G(n+1)\Gamma(k_1+1-n)}
{\Gamma(j_2+m_2+1)\over \Gamma(j_2+m_2+1-n)}
$$
$$
{\Gamma(j_3+m_3+1)\over \Gamma(j_3+m_3+1-k_1+n)}
{\G(j_2-m_2+n+s)\over \G(j_2-m_2+n)}\right)^2
I(j,k)
\eqn\dxiv
$$}

\def\dxvi{$$
{\cal A}^{j_1\;\;\;j_2\;\;\;j_3}_{m_1\;m_2\;m_3}
={\cal C}^2 {\cal I}(j_1,j_2,j_3,k),
\eqn\dxvi
$$}

\def\dxvib{$$
\eqalign{
{\cal
A}^{j_1\;\;\;\;\;\;\;j_2\;\;\;j_3}_{j_1-k_1\;m_2\;m_
3}=
\Biggl(&
{\G(-2j_1)\over \G(k_1-2j_1)}
\sum_{n=0}^{k_1} {1\over n!}
{\G(k_1+1)\over \G(k_1+1-n)}
{\G(j_2+m_2+1)\over\G(j_2+m_2+1-n)}\cr
&\cr
&{\G(j_3+m_3+1)\over \G(j_3+m_3+1-k_1+n)}
{\G(-j_2+m_2-n+s)\over \G(-j_2+m_2-n)}
\Biggl)^2 {\cal I}(j_1,j_2,j_3,k).\cr}
\eqn\dxvib
$$}

\def\dxvic{$$
{\cal C}=
{\G(j_1+m_1+1)\over \G(2j_1+1)}
{\G(j_3+m_3+1)\over \G(j_1-m_1)\G(-m_2-j_2)}
$$
$$
\sum_{n=0}^{k_1}
{\G(n+j_2-m_2+1)\G(j_1-m_1+n)\G(-j_2-m_2+n)
\over \G(n+1-j_1+j_3-m_2) \G(n-j_1-j_3-m_2)\G(n+1)}
\eqn\dxvic
$$}

\def\dxvii{$$
{\cal C}={\G(-2j_1)\over \G(-j_1-m_1)}
{\G(j_3+m_3+1)\over
\G(-j_2+m_2)} {\G(j_1+j_3+m_2+1)\over
\G(-j_1+j_3-m_2+1)}
$$
$$
\lim_{x\rightarrow 1}
{_3F_2}(j_2-m_2+1,m_1-j_1,-j_2-m_2;-j_1-j_3-m_2,-j_1
+j_3-m_2+1|x).
\eqn\dxvii
$$}
\def\dxix{$$\eqalign{
{\cal I}(j_1,j_2,j_3,k)&=
M^s \G(-s)\int \prod_{i=1}^s d^2 z_i |z_i|^{-4\rho
j_1}
|1-z_i|^{-4\rho
j_2-2}\prod_{i<j}|z_i-z_j|^{4\rho}\cr
&=\rho^{2s}(-\pi \D(1-\rho)M)^s\left( {\G(-2j_1+s)
\over \G(-2j_1)}\right)^2 {\cal Y}_{13}{\cal
Y}_{2}\;\;,\cr}
\eqn\dxix
$$}

\def\dxx{$$
{\cal Y}_{13}=\prod_{i=0}^{s-1}\D(-2j_1\rho+i\rho)
\D(-2j_3\rho +i\rho),
\quad  \quad {\cal Y}_{2}=
\G(-s)\G(s+1)\prod_{i=0}^{s-1}\D((i+1)\rho)
\D(-2j_2\rho+i\rho).
\eqn\dxx
$$}

\def\dxxi{$$
2j_2+1=-{2\over 3}m_2 ,\qquad \qquad
2j_i+1={2\over 3}m_i\quad {\rm for} \quad i=1,3.
\eqn\dxxi
$$}

\def\dxxia{$$
j_2={s\over 2}-{1\over 4},\qquad  \qquad m_2=
-{3\over 2}s-{3\over 4}.
\eqn\dxxia
$$}

\def\dxxib{$$
{\cal C}=-{1\over 2}{\G(-2j_1)\over \G(-2j_1+s)}
{\G(2j_1+{5\over 4})\over \G(2j_1+{5\over 4}-s)}
{\G({1\over 2})\G({1\over 4})\over
\G(-2s-{1\over 2})\G(s+{5\over 4})}.
\eqn\dxxib
$$}

\def\dxxic{$$
{\cal C}^2=\rho^{2s}\left({\G(-2j_1)\over
\G(-2j_1+s)}\right)^2{\G({1\over 4})\over
\G({3\over 4})}\prod_{i=1}^3\D\Bigl(2j_i+{5\over
4}\Bigr).
\eqn\dxxic
$$}

\def\dxxii{$$
2j+1=\pm {2\over 3}m+\varepsilon r(j,\g),
\eqn\dxxii
$$}

\def\dxxiii{$$
\prod_{i=0}^{n-1}(i+x)={\G(n+x)\over \G(x)}
\qquad ,\qquad
\G(-n+\varepsilon)={(-)^n\over \varepsilon
\G(n+1)}+
{\cal O}(1)\quad{\rm for}\quad
n\in {\rm I\!N}
\eqn\dxxiii
$$}

\def\dxxiv{$$
{\cal Y}_{13}=\rho^{2s-2}
\prod_{i=1,3}
\D(-8j_i-2)\D\Bigl(2j_i+{3\over 4}\Bigr).
\eqn\dxxiv
$$}

\def\dxxv{$$
s=2j_2+{1\over 2}+{\e\over 2}\widetilde{r}
\eqn\dxxv
$$}

\def\dxxvi{$$
{\cal Y}_2 =
(-)^{s+1}\rho^{2s+1}{\G({3\over 4})\over \G({1\over
4})}\D(-8j_2-2)\D\Bigl(2j_2+{3\over 4}\Bigr).
\eqn\dxxvi
$$}

\def\dxxvia{$$
{\cal R}={\G(s+{\widetilde{r}+4\over 8})\over
\G({\widetilde{r}+4\over 8})\G(s+1)},
\eqn\dxxvia
$$}

\def\dxxvii{$$
{\cal I}(j_1,j_2,j_3,k)=-\rho^{6s-1}M^s
(\pi \D(-\rho))^s\left({\G(-2j_1+s)\over
\G(-2j_1)}\right)^2\prod_{i=1}^3\D(-8j_i-2)
\D(2j_i+{3\over 4})
\eqn\dxxvii
$$}

\def\dxxviia{$$
{\cal A}(j_1,j_2,j_3)=
-\rho^{6s-1} (-\pi M\D(-\rho))^s\prod_{i=1}^3
\D(-8j_i-2)\D\Bigl(2j_i+{3\over 4}\Bigr)\D
\Bigl(2j_i+{5\over 4}\Bigr),
\eqn\dxxviia
$$}

\def\dxxviii{$$
{\cal A}(j_1,j_2,j_3)=
\widetilde{M}^s\prod_{i=1}^3\D(-4j_i-1).
\eqn\dxxviii
$$}

\def\dxxviiia{$$
{\cal V}_{p}^{\pm}=e^{(-\sqrt{2}\pm |p|)\phi +ipX}.
\eqn\dxxviiia
$$}

\def\dxxviiib{$$
{\cal A}(p_1,p_2,p_3)=
\widetilde{M}^{s}\prod_{i=1}^3\D(1-\sqrt{2}|p_i|)
,\qquad  \qquad
s=1+{1\over 2}\sum_{i=1}^3\left({|p_i|\over
\sqrt{2}}-1\right).
\eqn\dxxviiib
$$}

\def\dxxviiic{$$
A(p_1,p_2,p_3)
=[\mu \D(-\rho)]^s\prod_{i=1}^3 (-\pi)
\D(1-\sqrt{2}|p_i|),\quad\quad
{s\over 2}=1+{1\over 2}\sum_{i=1}^3
\left({|p_i|\over \sqrt{2}}-1\right).
\eqn\dxxviiic
$$}

\def\gi{$$
{\langle {\cal V}_{j_2\; m_2}
{\cal V}_{j_3 \;m_3+1} \rangle \over
\langle {\cal V}_{j_2\; m_2+1}
{\cal V}_{j_3 \;m_3} \rangle} =
-{j_2-m_2\over j_3-m_3} =
-{j_3+m_3+1 \over j_2+m_2+1}.
\eqn\gi
$$}

\def\gii{$$
{\cal A}^{j_2\;\;\;j_3}_{m_2\;m_3}=
(-)^s\D(1+j_2-m_2)\D(1+j_3-m_3)I(j_2,j_3,\rho),
\eqn\gii
$$}

\def\giii{$$
I(j_2,j_3,k)=M^s \G(-s)\lim_{\varepsilon\rightarrow
0}
 \int \prod_{n=1}^s d^2 z_n
|z_n|^{-4\rho \varepsilon i}|1-z_n|^{-4\rho j_2-2}
\prod_{n<m}|z_n-z_m|^{4\rho}
$$
$$
=-(\pi M\D(-\rho))^s  \D(1-s)\D(\rho s)
\lim_{\varepsilon\rightarrow 0}
\D(1-2\rho\varepsilon i)\D((\varepsilon
i-j_2+j_3)\rho))
\D((\varepsilon i+j_2-j_3)\rho).
\eqn\giii
$$}

\def\giv{$$
\delta(j_2-j_3)=\lim_{\varepsilon\rightarrow 0}
{1\over \pi}{\varepsilon \over
\varepsilon^2+(j_2-j_3)^2}.
\eqn\giv
$$}

\def\gv{$$
{\cal I}(j_2,j_3,k)=2\pi i\rho\delta(j_2-j_3)
(\pi M\D(-\rho))^s  \D(1-s)\D(\rho s)
\eqn\gv
$$}

\def\gvi{$$
\left( {\G(-j_2+m_2+s)\over \G(-j_2+m_2)}\right)^2=
\D(1+j_2-m_2)\D (1+j_3-m_3)
\eqn\gvi
$$}

\def\gvii{$$
{\cal A}^{j_2\;\;\;j_3}_{m_2\;m_3}=
2\pi i\rho \delta(j_2-j_3)(-\pi M \D(-\rho))^s
 \D(1-s)\D(\rho s)
\D(1+j_2-m_2)\D(1+j_3-m_3),
\eqn\gvii
$$}

\def\gviiia{$$
\langle {\cal V}_{j_1\; m_1}(0)
{\cal V}_{j_2\; m_2}(1)\rangle=
\lim_{z_s\rightarrow \infty}\langle
{\cal V}_{j_1\; m_1}(0){\cal V}_{j_2\; m_2}(1)
\,\,\b(z_s){\bar \b}({\bar z}_s)e^{-{2\over \ap}
 \phi(z_s,{\bar z_s})}\,\,{\cal Q} ^{s-1}\,\rangle.
\eqn\gviiia
$$}

\def\gviiib{$$
s=j_1+j_2+1,\qquad\qquad m_1+m_2=0.
\eqn\gviiib
$$}

\def\gviii{$$
{\cal P}^{-1}{\p^s{\cal P}\over \p z_1
\dots \p z_s}
{\bar {\cal P}}^{-1}{\p^s{\bar {\cal P}}\over
 \p {\bar z}_1 \dots \p {\bar z}_s}=
(-)^s\,\D(1+j_1-m_1)\D(1+j_2-m_2)\prod_{i=1}^s
|z_i|^{-2}\,|1-z_i|^{-2},
\eqn\gviii
$$}

\def\gix{$$
{\cal
P}=\prod_{i=1}^sz_i^{m_1-j_1}(1-z_i)^{m_2-j_2}
\prod_{i<j}^s(z_i-z_j).
\eqn\gix
$$}

\def\gx{$$
\eqalign{
\langle {\cal V}_{j_1\; m_1}(0){\cal V}_{j_2\;
m_2}(1)\rangle=
&(-)^s M^s\G(-s)\D(1+j_1-m_1)\D(1+j_2-m_2)\cr
&\cr
&\int \prod_{i=1}^{s-1} d^2z_i|z_i|^{-4\rho j_1-2}
|1-z_i|^{-4\rho j_2-2}\prod_{i<j}^{s-1}
|z_i-z_j|^{4\rho}.\cr}
\eqn\gx
$$}

\def\gxa{$$
\langle {\cal V}_{j_1\; m_1}(0){\cal V}_{j_2\; m_2}
(1)\rangle=(-)^s
M^s\G(-s)\D(1+j_1-m_1)\D(1+j_2-m_2)\G(s)
$$
$$
(\pi\D(1-\rho))^{s-1}
\prod_{i=1}^{s-1}\D(i\rho)\prod_{i=0}^{s-2}
\D(-2j_1\rho+i\rho)
\D(-2j_2\rho+i\rho)\D(1+\rho(s-i)).
\eqn\gxa
$$}

\def\gxi{$$
\langle {\cal V}_{j\; m}(0){\cal V}_{j\;
-m}(1)\rangle=
(-\pi
M\D(-\rho))^s\D(1+j-m)\D(1+j+m)s\D(1-s)\D(\rho s).
\eqn\gxi
$$}

\def\gxii{$$
\langle {\cal V}_{j\; m}(0){\cal V}_{j\;
-m}(1)\rangle=
{\widetilde{M}^{2j+1}\over 2j+1}(\D(-4j-1))^2,
\eqn\gxii
$$}

\def\gxiii{$$
\langle {\cal V}_p^+(0){\cal V}_{-p}^+(1)\rangle=
{\widetilde{M}^{|p|\over \sqrt{2}}\over
\sqrt{2}|p|}
(\D(1-\sqrt{2}|p|))^2,
\eqn\gxiii
$$}

\def\gxiv{$$
\langle V_p^+(0)V_{-p}^+(1)\rangle=
{\mu^{\sqrt{2}|p|}\over \sqrt{2}|p|}
(\D(1-\sqrt{2}|p|))^2.
\eqn\gxiv
$$}

\def\gxv{$$
\langle V_p^+(0)V_{-p}^+(1)\rangle
\sim {M^{|p|\over \sqrt{2}}\over \sqrt{2}|p|}\D
\left(1-{|p|\over \sqrt{2}}\right)^2.
\eqn\gxv
$$}

\def\gxvi{$$
{\cal V}_p^+\rightarrow {{\cal V}_p^+ \over \D
\bigl({1\over 2}-{|p|\over \sqrt{2}}\bigr)}
\eqn\gxvi
$$}

\def\hi{$$
m_1+m_2+m_3=j_1+j_2+j_3=0
\eqn\xxi
$$}

\def\hii{$$
{\cal A}^{j_1\;\;\;j_2\;\;\;j_3}_{m_1\;m_2\;m_3}=
\int d^2 z\;\langle \,{\cal V}_{j_1\; m_1}(0)
{\cal V}_{j_2\; m_2}(1)
{\cal V}_{j_3\; m_3}(\infty)\,\,\b(z){\bar \b}
({\bar z})\, e^{-{2\over \ap}\,\phi(z,{\bar z}) }\rangle_{M=0}
$$
$$
=M\G(-s)\int d^2 z\;\left( {j_1-m_1\over z}-
{j_2-m_2 \over 1-z}\right)\left({j_1-m_1\over
{\bar z}}-{j_2-m_2\over 1-{\bar z}}\right)
|z|^{{-4\rho }j_1}\, |1-z|^{{-4\rho}j_2}.
\eqn\xxii
$$}

\def\hiii{$$
{\cal A}^{j_1\;\;\;j_2\;\;\;j_3}_{m_1\;m_2\;m_3}
=M\G(-s)\left( {-m_1j_2+m_2 j_1\over
j_1}\right)^2\int d^2 z
 |z|^{{-4\rho}j_1} |1-z|^{{-4\rho} j_2-2}+{\cal
B}(j_1,j_2)
+{\cal B}^\ast (j_1,j_2)
\eqn\xxiii
$$}

\def\hiiia{$$
{\cal B}(j_1,j_2)\sim \int d^2z {\p \over \p{\bar
z}} \Bigl(
|z|^{-4 \rho j_1}|1-z|^{-4\rho j_2-2}(1-{\bar
z})\Bigr).
\eqn\xxiv
$$}

\def\hiiib{$$
\int_{\Sigma}d^2z{\p\over \p{\bar z}}(f(z,{\bar
z}))=
-{i\over 2}\int_{\p \Sigma} dz f .
\eqn\xxv
$$}

\def\hiiic{$$
{\cal B}(j_1,j_2)\sim-{i\over 2}\,\
\lim_{\e\rightarrow 0} \oint_\e dz\,
|z|^{-4\rho j_1}\,|1-z|^{-4\rho j_2 -2}\,
(1-{\bar z})=\pi\lim_{\e\rightarrow 0}
\e^{-8\,\rho j_2},
\eqn\xxx
$$}

\def\dxviii{$$
\eqalign{
{\cal C}=&(-)^s{\G(-2j_1)\G(s+1)\over \G(-2j_1+s)}
\sum_{i=0}^s{\G(m_1-j_1+i)\over \G(m_1-j_1)}
{\G(-m_2-j_2+i)\over\G(-m_2-j_2)}\cr
&\cr
&
{\G(-m_1-j_1+s-i)\over \G(-m_1-j_1)}
{\G(m_2-j_2+s-i)\over \G(m_2-j_2)}
{(-)^i\over (s-i)!i!},\cr}
\eqn\xxxiii
$$}

\def\hva{$$
{\cal C}_{s=1}={j_1m_2-m_1j_2\over j_1}.
\eqn\xxxiv
$$}

\def\hvi{$$
{\cal A}^{j_1\;\;\;j_2\;\;\;j_3}_{m_1\;m_2\;m_3}
=\lim_{\varepsilon\rightarrow 0}{\cal
A}^{j_1\;\;\;j_2\;\;\;j_3}_{m_1\;m_2\;m_3}=
-{9\over
4}\widetilde{M}\D(-1-4j_1)\D(-1-4j_3)\G(-s)
\lim_{\varepsilon\rightarrow 0}\D(-2\rho
j_2)\D(\rho)
\eqn\hvi
$$}

\def\hvii{$$
\lim_{\e\rightarrow 0}\D(-2\rho
j_2)\D(\rho)\G(-s)=\G(-2j_2-1/2)
\lim_{\e\rightarrow 0}\D(2+(-8+2A)\e)\D(-4+16\e)
\eqn\hvii
$$}

\def\ki{$$
\Gamma(1+z-n)=(-1)^n{\Gamma(1+z)\Gamma(-z)
\over \Gamma(n-z)}\qquad {\rm for} \qquad n\in {\rm
I\!N},
\eqno{\rm (A.1)}
$$}

\def\kia{$$
\prod_{i=0}^{n-1}(i+x)={\G(n+x)\over \G(x)},
\;\;\;\;
\D(x)\D(-x)=-{1\over x^2},\;\;\;\;
\D(x)\D(1-x)=1,\;\;\;\;\D(1+x)=-x^2\D(x)
\eqno{\rm (A.2)}
$$}

\def\kiii{$$
\G(2x)={2^{2x-1}\over \sqrt{\pi}}\G(x)\G
\Bigl(x+{1\over 2}\Bigr),\qquad\qquad
\D(2x)=2^{4x-1}\D(x)\D\Bigl(x+{1\over 2}\Bigr).
\eqno{\rm (A.3)}
$$}

\def\kii{$$
\lim _{\e\rightarrow 0}\G(-n+\e)={(-)^n\over
\e\G(n+1)}+{\cal O}(1)\quad {\rm for}\quad n\in
{\rm I\!N}.
\eqno{\rm (A.4)}
$$}

\def\dxvia{$$
{_3F_2}(\a,\b,\gamma;\rho,\sigma|x)=
{\G(\rho)\G(\sigma)\over \G(\a)\G(\b)\G(\g)}
\sum_{\nu=0}^\infty{\G(\a+\nu)\G(\b+\nu)
\G(\g+\nu)\over
\G(\rho+\nu)\G(\sigma+\nu)}
{x^{\nu}\over \nu!}.
\eqno{\rm (A.5)}
$$}

\def\kv{$$
_3F_2(a,b,c;1+a-b,1+a-c|1)=
{\G(1+{a\over 2})\G(1+a-b)\G(1+a-c)
\G(1-b-c+{a\over 2})
\over
\G(1+a)\G(1-b+{a\over 2})\G(1-c+{a\over 2})
\G(1+a-b-c)}.
\eqno{\rm (A.6)}
$$}

\def\kvi{$$
{1\over m!}\int \prod_{i=1}^m\Bigl({1\over 2}
i dz_i d{\bar z_i}\Bigr)
\prod_{i=1}^m|z_i|^{2\a}|1-z_i|^{2\b}
\prod_{i<j}^m|z_i-z_j|^{4\rho}=
\pi^m\bigl(\D(1-\rho)\bigr)^m
$$
$$
\prod_{i=1}^m\D(i\rho)\prod_{i=0}^{m-1}
\D(1+\a+i\rho)\D(1+\b+i\rho)
\D(-1-\a-\b-(m-1+i)\rho).
\eqno{\rm (A.7)}
$$}

\pageno=1

\chapter{\chaone}
\sec{\seconeone}

\no The classical theory of general relativity predicts the existence
of black holes as solutions to Einstein's equations. These are
regions of space-time from which, classically, it is not possible to
escape to infinity; they are separated from the exterior by a null
hypersurface called the event horizon.
General relativity provides an adequate description of black holes
that are much bigger than the Planck mass. However, much smaller
black holes could have been formed in the early universe. For these
black holes a description in terms of general relativity breaks down
and it has to be replaced by a quantum theory of gravity.

It is undoubtedly true that such a  formulation in terms of quantum
gravity is important to solve many puzzles connected with the
``information loss'' that takes place in the presence of a black
hole. The classical ``no hair'' theorem states that a black hole can
be characterized by just three conserved charges that are: mass,
angular momentum and electric charge. In principle all the
information on the matter that formed the black hole is therefore
hidden behind the horizon. This is not a worry in the purely
classical theory, since  the information is still present behind the
horizon, even if we cannot get at it. The situation is different in
the quantum theory. In 1974 Hawking \REF\haw{\rhaw}[\haw] discovered
that  due to quantum mechanical pair production at the horizon a
black hole can radiate and loose mass. The outgoing radiation is
thermal, as if the black hole were a black body with a temperature
proportional to the surface gravity. Eventually the black hole
evaporates completely and carries all the information of the
collapsing matter with it. This implies that pure states can evolve
into mixed states and the laws of quantum mechanics would be
violated. Again, we expect that quantum gravity will help us to solve
this paradox. When the black hole is of the order of the Planck mass
we expect corrections from quantum gravitational effects, so that
Hawking's semiclassical calculation is no longer valid. There are
several possibilities for what could happen next in the evaporation
process
\REF\Strominger{J.~A. Harvey and A.~Strominger, ``Quantum Aspects of
Black Holes'', Lectures given at the ``{\it Trieste Spring School on
String Theory and Quantum Gravity}'', Trieste (1992) and at TASI 92:
``{\it From Black Holes and Strings to Particles}'', Boulder (1992),
preprint EFI-92-41, hep-th/9209055; G.~T.~Horowitz, ``The Dark Side
of String Theory: Black Holes and Black Strings'',  Lectures given at
the ``{\it Trieste Spring School on String Theory and Quantum
Gravity}'', Trieste (1992), preprint UCSBTH-92-32, hep-th/9210119; S.
Hawking, ``Hawking on the Big Bang and Black Holes'', Advanced Series
in Astrophysics and Cosmology-Vol.8, World Scientific, 1993.}
[\Strominger]:

\item{\triangleright} Hawking's proposal
\REF\Hawkingtwo{S.~W.~Hawking, ``Breakdown of Predictability in
Gravitational Collapse'', {\it Phys.~Rev.} {\bf D14} (1976) 2460.}
[\Hawkingtwo] is radical and it states that the black hole evaporates
completely and carries all the information on the in-falling matter
with it. Quantum mechanics is therefore not deterministic and our
basic physical laws have to be reformulated.

\item{\triangleright}The black hole does not evaporate completely but
leaves a stable remnant that could contain the information.
The final state (remnant plus radiation) is pure. This proposal
raises some conceptual difficulties because, for example,  it clearly
violates CPT since the black hole can form but never disappear
completely.

\item{\triangleright}The black hole disappears completely but the
outgoing radiation is correlated with the in-falling matter and
radiation in such a way that the final state, consisting of pure
radiation, is pure. Quantum coherence could be restored by the
radiation emitted in the final stages of the evaporation where
unknown laws of quantum gravity become relevant.

\item{\triangleright}None of the above.

Quantum gravity plays an important near the singularity of the black
hole. As long as the singularity is hidden inside the event horizon
it does not affect the exterior world, because the two regions are
causally disconnected.
The situation is different if naked singularities do appear since in
this case a description in terms of general relativity is no longer
valid. This is clearly unsatisfactory and several solutions to this
problem have been proposed. One of them is the Cosmic Censorship
Hypothesis of Penrose, which states that naked singularities do not
appear in nature. However, although we have no counter-examples to
this conjecture a general proof is still lacking, so that we have to
look for a different way to solve this breakdown of general
relativity. One of the answers could be that these singularities
would not be present, if instead of considering a classical theory of
gravitation, a description in terms of quantum gravity were made.

{}From this discussion it becomes clear that quantum gravity plays an
essential role in every theory of extremely strong gravitational
fields such as a black hole. This brings us to the question of
whether we know how to reconcile general relativity with quantum
mechanics. As far as four-dimensional gravity is concerned, we
encounter many difficulties when we try to quantize the theory. All
of them are related to the fact that quantum gravity is a
non-renormalizable theory in four dimensions. At present our most
promising candidate to be a consistent theory of quantum gravity is
string theory. According to this it appears natural to analyse the
connection between string theory and black hole physics.

In this context two interesting 2D black hole toy models have been
extensively studied during the last three years. Here the technical
complexities of four-dimensional quantum black holes are simplified
drastically, while the conceptual difficulties of the problem are
still kept.
One of these models is the $(1+1)$-dimensional dilaton gravity model
of Callan, Giddings, Harvey and Strominger
\REF\CGHS{C.~Callan, S.~Giddings, J.~Harvey and A.~Strominger,
``Evanescent Black Holes'', {\it Phys.~Rev.} {\bf D45} (1992) 1005.}
[\CGHS],
in which the gravitational collapse can be studied using a
two-dimensional quantum field theory. This model incorporates Hawking
radiation, and it solves the $\beta$-function equations of the string
to first order in the string coupling constant. It has been
successfully solved in the semiclassical limit while at the quantum
level many puzzles are still unresolved.

Analysing the classical solutions of string theory
\REF\Wit{E.~Witten, ``String Theory and Black Holes'',
{\it Phys.~Rev.} {\bf D44} (1991) 314. }
\REF\MSW {G.~Mandal, A.~Sengupta and S.~Wadia, ``Classical Solutions
of Two-Dimensional String Theory'', {\it Mod.~Phys.~Lett.} {\bf A6}
(1991) 1685.}
[\Wit,\MSW]
it has been found that the graviton-dilaton equations admit a
Schwarzschild-like solution and this was the starting point to find
the exact solution of the $\beta$-function equations.
If we are considering the gravitational applications of string theory
it is important to be able to go beyond the leading orders in the
expansion in $\alpha'$ (the string coupling constant). The leading
order may not correctly describe the strong curvature regions near
singularities.
Witten [\Wit] proposed that the exact conformal field theory (to all
orders in $\alpha'$)
that describes the above black hole solution can be formulated in
terms of an {\sl} gauged Wess-Zumino-Witten (WZW) model. Depending on
whether the subgroup that is gauged is compact or not we get the
Euclidean version of the black hole or its Lorentzian continuation.
In the semiclassical limit $k\rightarrow \infty$ (where $k$ is the
level of the Kac-Moody algebra) the Minkowski version of this model
has a maximally extended space-time analogue to the Schwarzschild
black hole of four-dimensional general relativity.

The exact background metric of Witten's black hole
has been determined by Dijkgraaf, E. Verlinde and
H. Verlinde
\REF\dvv{\rdvv}
[\dvv]. This has been done in the ``$L_0$-approach'' in which a
mini-superspace description of the problem was made. We are going to
explain this quantum mechanical approach in some detail in
section~3.4. It has been checked by Tseytlin
\REF\Tseytlin{A.~A.~Tseytlin, ``On the Form of the Black Hole
Solution in $D=2$ Theory'', {\it Phys.~Lett.} {\bf B268} (1991) 175.}
[\Tseytlin]
and by Jack et al.
\REF\Jack{I.~Jack, D.~R.~T.~Jones and J.~Panvel, ``Exact Bosonic and
Supersymmetric String Black Hole Solutions'', {\it Nucl.~Phys.} {\bf
B393} (1993) 95.}
[\Jack]
that this metric indeed solves the $\beta$-function equations of the
string perturbatively up to three and four loops. The maximally
extended space-time of this geometry has been considered by Perry and
Teo
\REF\teo{M. Perry and E. Teo, ``Nonsingularity of the Exact
Two-Dimensional String Black Hole'', {\it Phys.~Rev.~Lett.} {\bf 70}
(1993) 2669.}
[\teo] and by Yi
\REF\yi{P. Yi, ``Nonsingular 2D Black Holes and Classical String
Backgrounds'', {\it Phys.~Rev.} {\bf D48} (1993) 2777.}
[\yi].
It consists of an infinite number of universes connected by
wormholes. There are no singularities.

It has been claimed by Witten
\REF\Wittwo{E.~Witten, ``Two-Dimensional String Theory and Black
Holes'',
Lecture given at the ``{\it Conference on Topics in Quantum
Gravity}'', Cincinnati (1992), preprint IASSNS-HEP-92-24,
hep-th/9206069.}
[\Wit,\Wittwo]
that the black hole looses mass due to Hawking radiation. The
end-point of this radiation process is described by the standard
$c=1$ matrix model that can be regarded as an analogue of the extreme
Reissner-Nordstr\" om black hole. However, it has been argued by
Seiberg and Shenker
\REF\Seiberg{N.~Seiberg and S.~Shenker, ``A Note on Background
(In-)Dependence'', {\it Phys.~Rev.} {\bf D45} (1992) 4581.}
[\Seiberg]
that the black hole mass operator has a non-normalizable wave
function, implying the stability of the black hole. The precise
relation between the black hole conformal field theory and the
standard non-critical string theory is an open problem.

{}From the matrix model point of view there have been two different
main approaches in the literature to describe the black hole. The
first one uses  the  formulation of the $c=1$ conventional matrix
model
\REF\das{\rdas}
\REF\wadia{\rwadia}
[\das,\wadia]
in terms of non-relativistic fermions. In this approach the
black-hole singularity is identified with the Fermi surface in the
phase space of the fermion and it is a consequence of the
semi-classical approximation.  They concluded that stringy quantum
effects wash out the classical singularity.
In the approach of Jevicki et al.
\REF\jy{\rjy}
[\jy]
it is conjectured that Witten's black hole is described in terms of a
deformation of the usual $c=1$ matrix model. The working hypothesis
of these authors consists of two key ingredients, namely a non-local
redefinition of the tachyon field and a deformation of the $c=1$
matrix model at $\m=0$.
The space-time interpretation of this model has not been worked out
so far, so that the relation to the {\slU} gauged WZW is unclear.
However, the ``deformed'' matrix model is interesting in its own
right because it is a different model than the conventional $c=1$
model, that can be solved non-perturbatively using a free fermion
picture.

The formulation in terms of matrix models is important since these
models allow us to take into account higher genus effects
\REF\doubles{E.~Br\'ezin and V.~A.~Kazakov,
``Exactly Solvable Field Theories of Closed Strings'',
{\it Phys.~Lett.} {\bf 236B} (1990) 144; M.~R.~Douglas and
S.~H.~Shenker,
``Strings in Less than One Dimension'',
{\it Nucl.~Phys.} {\bf B335} (1990) 635; D.~J.~Gross and
A.~A.~Migdal,
``Non-perturbative Two-Dimensional Quantum Gravity'', {\it
Phys.~Rev.~Lett.} {\bf 64} (1990) 127;
D.~J.~Gross and A.~A.~Migdal, ``A Non-perturbative Treatment of
Two-Dimensional Quantum Gravity'', {\it Nucl.~Phys.} {\bf B340}
(1990) 333.}
[\doubles], even in the (physically more interesting) supersymmetric
theories
\REF\three{L.~Alvarez-Gaum\'e, K.~Becker and M.~Becker, ``Superloop
Equations in the Double Scaling Limit'', preprint CERN-TH.6575/92
(July 1992).}
\REF\five{L.~Alvarez-Gaum\'e, K.~Becker, M.~Becker, R.~Empar\'an and
J.~Ma\~nes, ``Double Scaling Limit of the Super-Virasoro
Constraints'',
{\it Int.~J.~Mod.~Phys.} {\bf A8} (1993) 2297.}
\REF\salama{L.~Alvarez-Gaum\'e, K.~Becker and M.~Becker,
``Super-Virasoro Constraints and Two-Dimensional Supergravity'', talk
given at the XIX International Colloquium on Group Theoretical
Methods in Physics, Salamanca, Spain, 1992, preprint
CERN-TH.6720/92.}
\REF\two{K.~Becker and M.~Becker, ``Non-perturbative Solution of the
Super-Virasoro Constraints'', \mpl {\bf A8} (1993) 1205.}
[\three,\five,\salama,\two].

Given these powerful non-perturbative formulations, it is certainly
quite important to understand the precise relation between
non-critical string theory and the black hole CFT in the continuum
approach, because this will help us to understand better the discrete
approach. This is one of the purposes of this thesis.

The connection between the BRST cohomologies of both theories has
been studied extensively in the literature.
Distler and Nelson
\REF\dn{\rdn}
[\dn]
used the representation theory of {\sl} to show that the black hole
has the same physical states as the $c=1$ theory plus some new
discrete states that do not have any counterpart in $c=1$. However,
they also stated that it is possible that the real spectrum of the
black hole is actually only a truncation of what is allowed by
representation theory so that both cohomologies could in principle
agree. The structure of discrete states for the black hole is as rich
as the one of $c=1$. So for example, extending standard techniques of
the Kac-Moody current algebras to the non-compact case, Chaudhuri and
Lykken
\REF\cl{\rcl}
[\cl]
constructed the elements of the ground ring and showed that the
discrete states
of the black hole form a $W_\infty$-type algebra.
Eguchi et al.
\REF\eky{\reky}
[\eky]
have shown that the free-field BRST cohomologies of $c=1$ coupled to
Liouville theory and the black hole coincide. The important point in
their proof is the existence of an isomorphism between the states of
$c=1$ and those of the black hole. It can be derived from the fact
that both energy-momentum tensors agree up to BRST commutators. The
appearance of the $W_\infty$ algebra and the ground ring is then as
natural as it is for $c=1$.
The role that the $W_\infty$ algebra plays in the black hole context
has been the subject of many controversies. Ellis et al.
\REF\em{\remn}
[\em]
have argued that due to the infinite number of conserved quantities
quantum coherence would be maintained during the black hole
evaporation process.
A different point of view has been presented in ref.
\REF\mo{\rmo}
[\mo].

In spite of this extensive analysis of the cohomologies, the problems
of calculating the $\cal S$-matrix in the full quantum field theory
of tachyons in the black hole background and its relation to the
scattering amplitudes of $c=1$ have not been considered so far. This
is one of the problems that we are going to address here.

We will consider Witten's 2D black hole using an explicit
representation of the fields in terms of Wakimoto coordinates. This
representation of Witten's black hole has been introduced by
Bershadsky and Kutasov
\REF\bk{\rbk}
[\bk]
and it provides us with a suitable prescription of how to evaluate
scattering amplitudes using a free-field approach. The relation to
the gauged WZW model is very clear in this formulation, specially
through the Gauss decomposition
\REF\rgmom{A.~Gerasimov, A.~Morozov, M.~Olshanetsky
and A.~Marshakov, ``Wess-Zumino-Witten Model as a
Theory of Free-Fields'', \ijmp {\bf A5} (1990) 2495.}
\REF\vv{C.~Vafa and E.~Verlinde, unpublished.}[\rgmom, \vv]
and the space-time interpretation for finite $k$
\REF\becker{K.~Becker and M.~Becker, ``More Comments on Gauged WZW
Theory'', preprint CERN-TH.7213/94, in preparation.}
[\becker], for which we will later show that we get agreement with
the metric found be Dijkgraaf et al. [\dvv].

The evaluation of scattering amplitudes can therefore be done with
familiar techniques used in the Liouville approach to 2D quantum
gravity. After the zero mode integration of the fields
\REF\gl{\rgl}
[\gl],
the path integral is reduced to the one of a free theory, where the
screening charges are just new insertions. The amplitudes that can be
calculated in this way are those where the number of screening
charges is an integer. More general amplitudes are determined by
analytic continuation from the integers. The {\sl} screening charge
has been identified with the operator that creates the mass of the
black hole [\eky,\bk]. In ref. [\bk] it has been shown that the bulk
amplitudes of tachyons, i.e. those amplitudes that do not need
screening charges to satisfy the charge conservation, agree with
tachyon correlation functions of $c=1$ matter coupled to Liouville
theory. This correspondence is easy to understand if one takes into
account that the representation of Witten's black hole in terms of
Wakimoto coordinates can be formulated in the $c=1$ language. The
tachyon vertex operators have identical form in $c=1$ and the black
hole, the only difference is the perturbation considered in the
action. While for ordinary Liouville theory the perturbation is a
tachyon operator (the cosmological constant), the black hole mass
operator corresponds to the discrete state $W_{1,0}^-$$\bar
W_{1,0}^-$ of $c=1$. This is an operator on the wrong branch whose
wave function is not normalizable. However, we will later see that
correlation functions of tachyon vertex operators in both theories do
indeed have very similar features, also if one takes these two
different perturbations into account \REF\wir{K.~Becker and
M.~Becker, ``Interactions in the {\slU} Black Hole Background'', {\it
Nucl.~Phys.} {\bf B418} (1994) 206.}
[\wir,\becker].
We will calculate explicitly the two-, three-, and four-point
functions, where the remarkable analogy between the scattering
amplitudes will become clear. Our methods can be applied to $N$-point
functions.

Finally, we would like to mention that a connection between $c=1$ and
the black hole has been considered by Martinec and Shatashvili
\REF\ms{\rms}[\ms] from a different point of view. They analysed the
Hamiltonian path integral quantization of the gauged WZW model and
showed that there appears a relation to the Liouville theory coupled
to a free scalar field. This connection could in principle be used to
determine the scattering amplitudes of the black hole in terms of the
$c=1$ correlators. It would be nice to see a direct connection
between our results and those of ref. [\das,\wadia,\ms].

\sec{\seconetwo}

This thesis is organized as follows:

\item{{\rm II.}}In chapter 2 we are going to introduce some basic
notions that we need in order to describe the propagation of strings
in a black hole background. First of all we are going to see in
section~1.1 how black holes appear as a solution of Einstein's
equations in the theory of classical general relativity. The
Schwarzschild solution of the gravitational field equations in empty
space is explained in some detail as well as its generalizations
characterized by the mass, charge and angular momentum. In
section~1.2 we will see how the Polyakov action that describes string
propagation in flat space-time has to be generalized in order to
consider strings moving on more general manifolds ${\cal M}$. The
resulting action is a non-linear $\s$-model. The values of the
background fields of this action are fixed by demanding conformal
invariance of the quantum theory. This condition is satisfied if the
$\beta$-function equations of the non-linear $\s$-model vanish. There
exist an infinite number of solutions that satisfy this condition.
The WZW models will be specially interesting in this context, since
these are exactly solvable theories.

\item{{\rm III.}} In chapter 3 we explain the {\slU} coset model,
following closely [\Wit,\dvv]. This model can be interpreted as
describing the propagation of strings in the black hole background.
In section~3.1 we present the {\slU} gauged WZW model. We will see
that after choosing the Lorentz gauge the quantization procedure of
the theory gets simple. In section~3.2 we explain this quantization
and review some basic facts about the representation theory of {\sl},
which is relevant to classify the physical states of the black hole
CFT. The form of the physical states of the coset model can be
determined with the BRST quantization procedure. This is explained in
section~3.3. In section~3.4 we present Witten's semiclassical
interpretation of the target space geometry as a 2D black hole, which
has a similar structure as the Schwarzschild solution of Einstein's
equations. In section~3.5 we will see how these results get
corrections in $1/k$ as shown in the mini-superspace description of
Dijkgraaf et al.

\item{{\rm IV.}} In chapter 4 we are going to introduce the
free-field representation of the black hole CFT. In section~4.1 we
will see how the {\sl} operator algebra of Kac-Moody currents can be
realized through the Wakimoto representation of \sl, thus making it
clear that the {\sl} symmetry is manifest in the free-field approach.
Using the Sugawara prescription we can construct the energy-momentum
tensor and therefore the action in terms of these coordinates. We
introduce the {\sl} screening charge, that is an operator that
guarantees the charge conservation of the correlation functions and
added to the action is considered as the interaction of the model.
Then it is simple to obtain the form of the gauged-fixed action and
the form of the Kac-Moody primaries in terms of these coordinates. In
section~4.2 we will see how this model can be obtained from the
Lagrangian of the {\sl} gauged WZW model choosing a concrete
parametrization of the \sl-valued field $g$ in terms of the Gauss
decomposition.

\item{{\rm V.}}
In chapter~5 we are going to make the space-time interpretation of
the {\slU} gauged WZW model in terms of Wakimoto coordinates
[\becker]. We are able to find the connection to the space-time
interpretation of ref. [\dvv]. This connection for $k=9/4$ is
important, since it is only in this case that we are able to evaluate
the scattering amplitudes in the free-field approach, as we will see
in the next chapters [\wir,\becker].

\item{{\rm VI.}}
To see which are the characteristics that the $c=1$ model coupled to
Liouville theory shares with the black hole CFT we present, as a
starting point, the comparison between the cohomologies of the two
models. In section~6.1 we write down the classification, of Distler
and Nelson, of all the physical states that are allowed from the
representation theory [\dn]. In section~6.2 we show how some of the
discrete states of the previous classification look like in terms of
Wakimoto coordinates. The simplest, new, discrete state that has no
analogue in $c=1$ coupled to Liouville turns out to be BRST-trivial
[\bk]. The black hole mass operator is the discrete state $W_{1,0}^-$
$\bar W_{1,0}^-$ of $c=1$ [\bk,\eky]. A systematic analysis of the
cohomology of the free-field model can be made through the
isomorphism between the energy-momentum tensors of both theories as
shown by Eguchi et al. [\eky]. This will be presented in section~6.3.

\item{{\rm VII.}}In chapter 7 we consider the $\cal S$-matrix of
tachyons interacting in the black hole background [\wir,\becker].
Using the free-field approach we are able to perform a direct
computation of the correlation functions. In section~7.1 we formulate
the problem and in section~7.2 we show how, after performing the zero
mode integration of the fields, we are left with correlators of a
free theory [\bk]. In section~7.3 we consider the correlation
functions in the bulk; they are special since they satisfy the energy
conservation [\bk]. These correlators have an obvious relation to the
tachyon correlation functions in standard non-critical string theory.
We then consider correlation functions where the number of screening
charges is different from zero. In section~7.4 we begin with the
correlation function containing one highest-weight state and can
obtain
the general three-point function using the {\sl} Ward identities. As
a result of our computation we observe that these correlators share a
remarkable analogy with tachyon correlators of standard non-critical
string theory. They factorize in leg factors that have poles in
intermediate channels where all the discrete states belonging to the
BRST cohomology of $c=1$ are present. The new discrete states of
Distler and Nelson do not appear. They are therefore BRST-trivial and
decouple from the correlation functions. This means that the BRST
cohomology of the 2D black hole is the one of $c=1$, which is in
agreement with the result of ref. [\eky]. The parameter $M$ (that is
related to the mass of the black hole) has to be renormalized in
order to get non-vanishing correlation functions; this is a fact
known from $c=1$ matter coupled to Liouville theory perturbed by the
cosmological constant, where the parameter $\mu$ is infinitely
renormalized. The integrals that have to be evaluated are singular
for $k=9/4$, so that a careful treatment of the regularisation and
renormalization is in order. In section~7.5 we calculate the
three-point function with one screening as an illustrative example
and analyse contact term interactions that arise in our computations
in detail. To see whether the characteristics that appeared are
generic for all the amplitudes, we compute  the two-point function of
(not necessarily) on-shell
tachyons. Here a similar result is found, which indicates that the
$N$-point functions might also have these characteristics. We
explicitly compute the four-point tachyon amplitude with the
chirality configuration $(+,+,+,-)$. We apply similar techniques as
those used in ref.
\REF\dfk{\rdfk}
[\dfk]; i.e. we compute the pole structure, the asymptotic behavior
and the symmetries of the amplitude to determine its final form. The
three-point function of on-shell tachyons is one of the basic
ingredients.

\item{{\rm VIII.}}Chapter 8 contains our conclusions and outlook.

\item{}We review the most
important formulas for our calculations in the Appendix.

\endpage

\chapter{\chatwo}

\no In this chapter we will introduce some basic notions in order to
describe the propagation of strings in the black hole background.
We first review shortly the classical solution to Einstein's
equations. It is generally accepted that these equations have to be
modified in a proper description in terms of quantum gravity. We will
see how string theory modifies general relativity at a scale
determined by the string coupling constant $\a'$, which is believed
to be of the order of the Planck scale.

This short introduction is not supposed to be a complete review on
the subject because this can be found in textbooks about gravitation
and black holes
\REF\misner{See for example: C.~W.~Misner, K.~S.~Thorne and
J.~A.~Wheeler, ``Gravitation'', Freeman, San Francisco (1973); or
S.~L.~Shapiro and S.~A.~Teukolsky, ``Black Holes, White Dwarfs and
Neutron Stars'', John Wiley \& Sons, (1983).} [\misner]
and about string theory
\REF\gsw{See for example: M.~B.~Green, J.~H.~Schwarz and E.~Witten,
``Superstring Theory'', Cambridge (1987).} [\gsw]. Furthermore there
exist some excellent review articles on the subject
\REF\marti{E. Martinec, ``An Introduction to 2-d gravity and Solvable
String Models'', Lectures given at the ``{\it Trieste Spring School
on String Theory and Quantum Gravity}'', Trieste (1991), preprint
RU-91-51, hep-th/9112019.}
[\Strominger,\marti].

\sec{\sectwoone}

Nearly all the work done in general relativity before 1960 was
concerned with solving the Einstein equation in a particular
coordinate system. In the early 1970's there appeared the ``no hair''
theorems that state that a stationary black hole is uniquely
determined by it's mass $M$, charge $Q$ and its intrinsic angular
momentum $J$. We are first going to see how the simplest black hole
solution of Einstein's equations look like. It has no charge and no
angular momentum. The gravitational field equations take in empty
space the following form
$$
{\cal R}_{ik}=0.
\eqn\zi
$$

\no A solution in four-dimensional space-time is the Schwarzschild
metric represented by the line element

$$
ds^2=-\left( 1-{2M\o r}\right) dt^2+  \left( 1-{2M\o
r}\right)^{-1}dr^2+r^2d\O^2,
\eqn\zii
$$

\no where $d\O^2=d\th^2+\sin^2\th d\ph^2$ and $(r,\th,\ph)$ are the
three-dimensional spherical coordinates. This is the unique
spherically symmetric vacuum solution of Einstein's equation and is
often used to represent the empty space region surrounding a
spherical star or collapsing body of mass $M$. This metric appears to
have a singularity at $r=2M$. But, by looking at the curvature
invariant

$$
I={\cal R}_{ijkl}{\cal R}^{ijkl}=48{M^2 \o r^6},
\eqn\zzii
$$

\no this point turns out to be just a coordinate singularity. It can
be removed by a simple change of coordinates. There are many
different coordinate transformations that can be done to show that
$r=2M$ is not a physical singularity. One of them is the so-called
Kruskal coordinates:

$$
\eqalign{
\bar u &=-4M\exp\left({r^*-t\o 4M}\right)\cr
& \cr
\bar v &=4M \exp\left({r^*+t\o 4M}\right).
\cr}
\eqn\ziii
$$

\no Here $r^*$ represents the tortoise coordinate

$$
r^*=r+2M\ln \left({r\o 2M}-1\right),
\eqn\zziii
$$

\no which satisfies $dr=(1-2M/r) dr^*$. The line element can be
written in terms of these coordinates in the following form

$$
ds^2=-{2M\o r} e^ {-r/ 2M} d\bar u d \bar v +r^2 d\O^2.
\eqn\ziv
$$

\no Clearly this metric is non-singular at $r=2M$. However, $r=0$ is
still a singularity, as can be seen by comparing with the form of the
scalar curvature {\zzii}. This is a point with an infinite
gravitational field strengths.

The original Schwarzschild coordinate system covers only part of the
space-time manifold. The region $r\geq 2M$ corresponds to $-\infty <
\bar u <0$ and $0< \bar v < \infty$. In Kruskal coordinates we can
analytically continue this solution to the whole region $-\infty <
\bar u$, $\bar v < \infty$. The resulting Kruskal diagram is an
extension of the Schwarzschild black hole. It consists of several
different regions. Region I represents the region outside the horizon
and is asymptotically flat. The horizons of the black hole are given
by the two lines ${\bar u}=0$ and ${\bar v}=0$. The physical
singularity is located at ${\bar u} {\bar v}=1$, so that it has the
form of an hyperbola. Region III is the region between the black hole
singularity and the horizon.
Once an observer has entered this region he can (classically) never
escape from it. Region IV has the ``time-reversed'' properties of
region III and is called a white hole. Region II is asymptotically
flat, with $r\geq 2M$. Region V and VI are the regions of negative
mass above and below the singularities.

\midinsert
\epsfysize=3.5in
\centerline{\hskip 0.5cm \epsffile{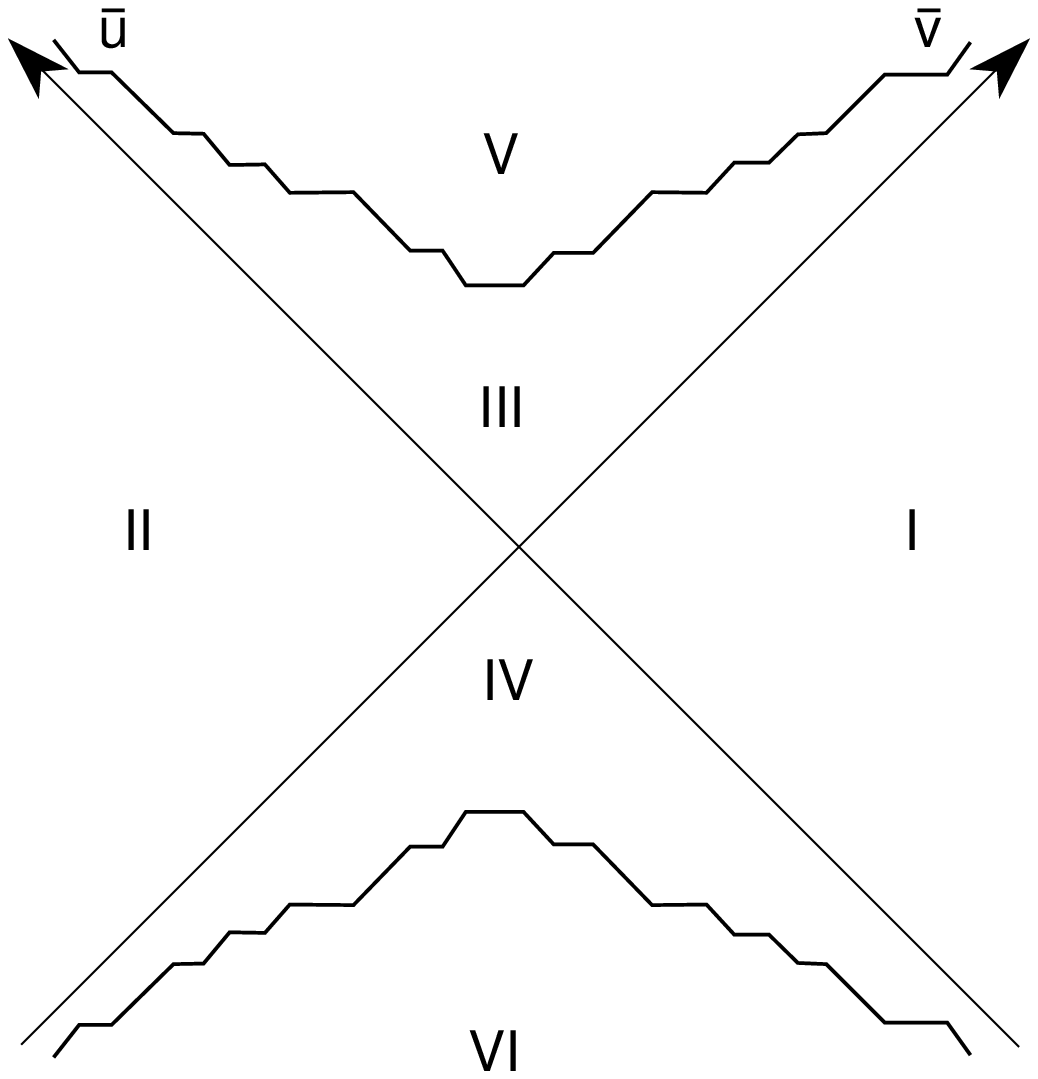}}
\caption{\hskip 2.0cm{\bf Fig. 1}: Kruskal diagram of the
Schwarzschild metric.}
\endinsert

It is important to realize that the full analytically continued
Schwarzschild metric is merely a mathematical solution of Einstein's
equations. For a black hole formed by gravitational collapse, part of
the space-time must contain the collapsing matter. The Schwarzschild
metric is a solution of the gravitational field equations in empty
space. The outside of a collapsing star is still described by the
Schwarzschild metric. Thus the world-line of a point on the surface
of the star will be the boundary of the physically meaningful part of
the Kruskal diagram. The ``white hole'' and the regions II, IV, V and
VI are not present in real black holes.

A more general black hole solution to Einstein's equations that is
characterized by $(M,J,Q)$ is called the Kerr-Newman black hole. A
special case of this solution is the Reissner-Nordstr\" om black
hole, which has no angular momentum $J=0$. It is characterized by the
line element

$$
ds^2=-\left( 1-{2M\o r}+{Q^2\o r^2}\right) dt^2+  \left( 1-{2M\o
r}+{Q^2\o r^2}\right)^{-1}dr^2+r^2d\O^2.
\eqn\zzzz
$$

\no This space-time has a curvature singularity at $r=0$ as for
Schwarzschild but it has in addition two horizons, where $ 1-{2M/
r}+{Q^2/ r^2}$ vanishes:

$$
r_{\pm}=M\pm (M^2-Q^2)^{1/2}.
\eqn\zzzzi
$$

\no The extremal Reissner-Nordstr\" om black hole corresponds to the
case $M=Q$ and plays a special role in connection with Witten's 2D
black hole solution  and the $c=1$ matrix model [\Wit,\teo,\yi].

\sec{\sectwotwo}
To address the connection between string theory, singularities and
strong gravitational fields, it is important to study strings in
curved backgrounds. We are going to see that WZW models naturally
appear in this context.

The propagation of a string in flat Minkowski space-time is described
by the action:

$$
{\cal S}_0=-{1\o 4\pi\a'}\int d^2\s \sqrt{h}h^{\a\b} \p_\a X^\mu
\p_\b X^\nu \eta_{\mu \nu},
\eqn\zvi
$$

\no where $h_{\a\b}$ is the world-sheet metric which, is regarded, in
a string theory as a dynamical variable and $\eta_{\mu \nu}$ is the
Minkowski metric. The parameter $\a'$ is the string coupling constant
that is a free parameter of dimension $(length)^2$ that makes the
expressions dimensionless. It plays the role of Planck's constant and
the classical limit corresponds to small $\a'$. Quantum mechanical
perturbation theory is therefore an expansion in $\a'$. The variables
$X^\mu(\sigma)$, where $\mu=1,\dots,d$, are scalar fields. The above
action is known as the Polyakov action. It is classically invariant
under the reparametrizations of the string world-sheet $\s
\rightarrow \s'$ and it has a local Weyl symmetry or conformal
symmetry that is manifest through the vanishing of the classical
energy-momentum tensor:

$$
T_{\a\b}=-{2\pi \o \sqrt{h}} {\delta {\cal S}_0\o \delta h^{\a
\b}}=0.
\eqn\zzvii
$$

\no This equation is known as the Virasoro condition.
If we would like to consider string propagation on an arbitrary
manifold we would have to consider the following, more general,
action called a non-linear $\s$-model:

$$
{\cal S}=-{1\o 4\pi\a'}\int d^2\s \sqrt{h}h^{\a\b} \p_\a X^\mu \p_\b
X^\nu {\cal G}_{\mu \nu}(X).
\eqn\zvii
$$

\no As in the flat space-time, this action is invariant under
reparametrizations of the world-sheet coordinates $\s$ and it has a
classical local Weyl invariance.
In a more systematic way one can include all of the massless states
of the closed string (and not only the graviton) as part of the
background [\gsw]. The action takes then the following form

$$
\eqalign{
{\cal S}=&-{1\o 4\pi\a'}\int d^2\s \sqrt h ( h^{\a \b } \p_\a X^\mu
\p_\b X^\nu {\cal G}_{\mu\nu}(X)+T(X))\cr
&\cr
&-{1\o 4\pi\a'}\int d^2\s \epsilon^{ab} \p_a X^\mu \p_b X^\nu {\cal
B}_{\mu\nu}(X)
+{1\o 4\pi} \int d^2 \s \sqrt{h}{R} \Phi(X).\cr}
\eqn\zviii
$$

\no The above action is a functional of the world-sheet metric
$h_{\a\b}$ and the $d$ space-time coordinates $X^{\mu}$. The previous
ansatz for the action is the most general one, involving only
renormalizable interactions. The couplings ${\cal G}_{\mu \nu}(X)$,
$\Phi(X)$, $T(X)$ and ${\cal B}_{\m\n}(X)$ can be identified with the
graviton, dilaton, tachyon and the antisymmetric tensor respectively.
Their values are restricted by demanding conformal invariance or
local scale invariance of the theory. The best way to impose this
condition is to consider the theory in $2+\epsilon$ dimensions and to
calculate those terms of the action that break the symmetry at the
quantum level in the limit $\epsilon \rightarrow 0$. Demanding these
terms to vanish restricts the values of the couplings of the theory.
These constraints can be formulated in terms of the so-called
$\b$-functions. There is one $\b$-function for each of the fields
${\cal O}_i$, and the trace of the energy-momentum tensor is
formulated in terms of these functions

$$
\langle T_{z {\bar z}}\rangle  =\int e^{-{\cal S}} \b_i {\cal O}^i.
\eqn\zzviii
$$

\no Therefore the statement of conformal invariance translates into
the requirement that the $\b$-functions associated with the
background fields vanish.
These $\b$-functions can be calculated using background field
perturbation theory, i.e. as an expansion in $\a'$, but no closed
expression is known that holds to all orders in the string coupling
constant. To first order in $\a'$ these equations have the following
form
\REF\cfmp{C.~G.~Callan, E.~J.~Martinec, M.~J.~Perry and D.~Friedan,
``Strings in Background Fields'', {\it Nucl.~Phys.} {\bf B262} (1985)
593.}
[\cfmp]:

$$\eqalign{
&\b_{\mu \nu} ^{\cal G}={\cal R}_{\mu\nu}-{1\o 4} H_\mu^{\l \s}
H_{\nu \l \s} +2\nabla_\mu \nabla_\nu \Phi -\nabla_\mu T\nabla_\nu
T+O(\a')=0 \cr
&\cr
&\b^\Phi={d-26\o 48 \pi^2 }-{\a'\o 16 \pi^2} \left(4(\nabla
\Phi)^2-4\nabla^2\Phi   -{\cal R} +{1\o 12} H^2+(\nabla
T)^2+V(T)\right)+O(\a'^2)=0\cr
&\cr
& \b^T=-2\nabla^2T +4\nabla \Phi \nabla T+V'(T)+O(\a') =0\cr
&\cr
&\b_{\mu \nu}^{\cal B} =\nabla_\l H_{\mu \nu}^\l -2(\nabla_\l \Phi)
H_{\mu \nu}^\l +O(\a')=0.\cr}
\eqn\zix
$$

\no Here ${\cal R}$ is the curvature of ${\cal G}_{\mu \nu}$,
$V(T)=-2T^2+O(T^3)$ is the tachyon potential and $H_{\mu \nu
\l}=\nabla_\mu {\cal B}_{\nu \l} +\nabla_\l {\cal B}_{\mu \nu}
+\nabla_\nu {\cal B}_{\l\mu}$ is the antisymmetric tensor field
strength.

Combining the $\b$-functions for $T$ and $\Phi$ it is possible to
write
\REF\friedanphd{D.~H.~Friedan, ``Nonlinear Models in $2+\epsilon$
Dimensions'',
{\it Ann.~Phys.} {\bf 163} (1985) 318.}
[\cfmp,\friedanphd]:

$$
\b_{\mu \nu}^{\cal G} +8 \pi^2 {\cal G}_{\mu \nu} {\b^\Phi \o \a'}
=\left({\cal R}_{\mu \nu} -\h {\cal G}_{\mu \nu} {\cal R} \right)
-T_{\mu \nu}^{ matter}=0
\eqn\zzix
$$

\no where

$$
T_{\mu \nu}^{ matter}={1\o 4} \left[ H_{\mu \nu}^2 -{1\o 6} {\cal
G}_{\mu \nu} H^2\right] -2\nabla_\mu \nabla _\nu \Phi +2{\cal G}_{\mu
\nu} \nabla^2 \Phi -2{\cal G}_{\mu \nu} (\nabla \Phi)^2.
\eqn\zzx
$$

\no This is the Einstein equation for the background metric ${\cal
G}_{\mu\nu}$. However, if we include two-loop corrections we have to
add a term of the form

$$
{\a' \o 2} {\cal R}_{\mu \l \s\tau}{\cal R}_{\nu}^{\l \s \tau},
\eqn\zzx
$$

\no to equation {\zzix}, which gives us corrections to the Einstein
equation coming from string theory.

Conformal invariance imposes strong constraints on the quantum
theory, but it does not fix the theory uniquely. Two solutions of the
$\b$-function equations that are believed to be exact i.e. to all
orders in $\a'$ are Witten's eternal black hole [\Wit] and the
Liouville solution in flat space found by David
\REF\david{F.~David, ``Conformal Field Theories Coupled to 2-d
Gravity in the Conformal Gauge'', {\it Mod.~Phys.~Lett.} {\bf A3}
(1988) 1651.}
[\david]
and Distler and Kawai (DDK)
\REF\diska{J.~Distler and H.~Kawai, ``Conformal Field Theory and 2D
Quantum Gravity or Who's Afraid of Joseph Liouville?, {\it
Nucl.~Phys.} {\bf B321} (1989) 509.}
[\diska]. The latter has the following form

$$
\eqalign{
&T={\mu \o 2\g^2} e^{\g \ph} \qquad {\rm with} \qquad Q={2\o \g} +\g
=\sqrt{26-d\o 3} \cr
&\cr
&\Phi={Q \o 2} \ph \cr
&\cr
&{\cal G}_{\mu \nu}=\delta_{\mu \nu}\cr}
\eqn\zzxi
$$

\no where $\phi = X_0$. These expressions solve the lowest order in
$\a'$ of the $\b$-function equations if the lowest order in $T$ is
taken into account but they are not a solution if higher powers of
$T$ are considered. One might hope that higher orders in $\a'$ of the
$\b$-functions correct this problem and that probably after a
redefinition of the fields the above solution is exact
\REF\banks{T.~Banks, ``The Tachyon Potential in String Theory'', {\it
Nucl.~Phys.} {\bf B361} (1991) 166.}
[\banks].

Inserting these expressions into the $\s$-model action {\zviii}, we
get the familiar expression for the Liouville action

$$
{\cal S}={1\o 8\pi} \int d^2 \s \sqrt{\widehat h}\left( {\widehat
h}^{ab} \p_a\ph \p_b\ph
+Q  {\widehat R}\ph +{\mu \o \g^2} e^{\g \ph}\right),
\eqn\zzx
$$

\no where we have considered the analytic continuation to the
Euclidean theory with $\a'=2$ and we have used the conformal gauge
$h={\widehat h}e^{\g\phi}$. Equations {\zix} can be derived as
Euler-Lagrange equations of the following effective action

$$
{\cal S}_{eff}=\int d^2\s \sqrt{{\cal G}} e^{-2\Phi}((\nabla
T)^2-2T^2-4(\nabla \Phi)^2-{1\o 12}H^2 -{\cal R}+(d-26)/ 3+\dots).
\eqn\zx
$$

\no If we consider weak field tachyons for which we can surely
neglect higher powers of $T$, the equation of motion of this field
can be written in the form of a Klein-Gordon equation

$$
(L_0-1) T=0,
\eqn\zxii
$$

\no where $L_0$ is given by

$$
L_0=-{1\o 2 e^\Phi \sqrt{\cal G}}\p_i e^\Phi \sqrt{\cal G} {\cal
G}^{ij} \p_j.
\eqn\zxiii
$$

\no This equation will be used later to identify the background
metric ${\cal G}_{\mu \nu}$ and the dilaton $\Phi$  of the {\slU} WZW
model [\dvv].

The low-energy effective action without the tachyon contribution
describes a $(1+1)$-dimensional black hole toy model known as dilaton
gravity [\CGHS]. Without coupling this model to additional matter it
is equivalent to the lowest order in $\a$' of the black hole that was
first studied by Witten [\Wit]. However, when we couple this model to
$N$ scalar fields one can study the formation of a 2D black hole and
the process of Hawking radiation.

Corrections to the low-energy effective action can be computed by
calculating higher-order corrections to the $\beta$-functions and can
involve, for example, higher derivative terms or higher powers of the
tachyon. We will mainly be concerned with the exact solution of the
$\b$-function equation that describes Witten's black hole. It has
been found in ref. [\Wit] to be described in terms of a gauged WZW
model.

Naively, the connection between the previous $\s$-model and the WZW
model can be motivated in the following way.
Solving the model described by the action {\zvii} is in general very
difficult, so that one has to make simplifications for the background
metric. One of these simplifications is to assume that the string
propagates on a group manifold of a semisimple Lie group $G$. If $g$
is an element of $G$, then in analogy to the $\s$-model the first
guess for the action in terms of $g$ would be
\REF\Wittwo{E.Witten, ``Non-Abelian Bosonization in Two Dimensions'',
{\it Commun.~Math.~Phys.} {\bf 92} (1984) 455.}
\REF\Poly{A.~M.~Polyakov and P.~B.~Wiegmann, ``Theory of Non-Abelian
Goldstone Bosons'', {\it Phys.~Lett.} {\bf 131B} (1983) 121.}
\REF\kz{V.~G.~Knizhnik and A.~B.~Zamolodchikov, ``Current Algebra and
Wess-Zumino Model in Two Dimensions'', {\it Nucl.~Phys.} {\bf B247}
(1984) 83.}
\REF\kaku{M. Kaku, ``Strings, Conformal Fields, and Topology: an
Introduction'', Springer (1991).}
[\Wittwo,\Poly,\kz,\kaku]

$$
{\cal S}=\int_{\S} {\rm tr}( \p_a g^{-1} \p^a g ) d^2\x.
\eqn\zzxi
$$

\no The field $g$ is some function of the string field $X_\mu$ in
terms of which we can express the metric $\cal G_{\m\n}$. This naive
choice of the action needs corrections because it is not
conformal-invariant. We can modify the action

$$
{\cal S}={1\o 4\l^2}\int {\rm tr} (\p_a g^{-1} \p^a g) d^2\x
+ik\G(g),
\eqn\zzxii
$$

\no by adding the so-called Wess-Zumino term

$$
\G(g)={1\o 24\pi} \int d^3 X  \epsilon^{\a\b\g}{\rm tr} [(g^{-1}\p_\a
g)(g^{-1}\p_\b g)(g^{-1}\p_\g g)].
\eqn\zzxiii
$$
\no In the above formula we integrate over a three-dimensional
manifold with boundary equal to $\S$. The characteristics of the
resulting action are very different from those of the action with
$k=0$. It represents a conformal-invariant $\s$-model for special
values of $\l$:
$$
\l^2 ={4\pi \o k}.
\eqn\zzxiv
$$

\no This action is called the Wess-Zumino-Witten model.

\endpage
\chapter{\chathree}
\no In this chapter we are going to see how an exact conformal field
theory, the {\slU} gauged WZW model, can be regarded as a model
describing the propagation of strings in the black hole background.
We review the Lagrangian formulation of the {\slU}conformal field
theory and present the semiclassical approach of Witten [\Wit], which
relates the background of the WZW model to a Schwarzschild-like
space-time. We will explain the mini-superspace approach of
Dijkgraaf, H. Verlinde and E. Verlinde [\dvv]. Here it will be clear
how the semiclassical metric found by Witten gets corrections of
order $1/k$.

 \sec{\secthreeone}
The conformal field theory that describes a black
hole in two-dimensional target space-time has a
Lagrangian formulation in terms of a gauged
WZW model based on the non-compact group {\sl} [\Wit]. The ungauged
{\sl} WZW model is described by following action:

\Ai

\noindent where $\Sigma$ is a Riemann surface with
metric tensor $h^{ij}$, $g:\Sigma \rightarrow${\sl}
is an {\sl}-valued field on $\Sigma$ and $k$ is a
real and positive number. The Wess-Zumino term,
which guarantees conformal invariance, is represented by:

\Aii

\noindent where $B$ is a three-dimensional manifold
with boundary $\Sigma$.

The action {\Ai} possesses a global {\sl$\times$\sl}
symmetry, since it is invariant under $g\rightarrow
a g b^{-1}$ with $a,b\in $\sl. The reason is that the products that
appear in the action are of the form

$$
g^{-1}\p g\rightarrow b g^{-1} \p g b^{-1}
\eqn\AAii
$$

\no and the trace is clearly invariant under this change of basis. To
get the interpretation as a 2D black hole we are interested in the
gauging of a subgroup of this symmetry group. Depending on which
subgroup we gauge, we get the Euclidean version of the black hole or
its Lorentzian counterpart. These solutions can also be obtained as
analytic continuation from one to the other. To get the Euclidean
version of the black hole we gauge an Abelian subgroup $h$ of {\sl}
that is compact, while for the Minkowski version this subgroup is
non-compact. Therefore, in the Euclidean theory we set

$$
a=b^{-1}=h=\left(\matrix{\cos\epsilon &\sin\epsilon \cr
-\sin\epsilon &\cos\epsilon \cr}\right).
\eqn\AAiii
$$

\no For $\epsilon$ small we can represent $h=1+\epsilon G$, where

$$
G=\left(\matrix{0&1\cr
-1&0\cr}\right).
\eqn\AAiv
$$

\no Under the transformation $g\rightarrow hgh$ the field $g$ is not
invariant but transforms as

$$
\delta g=\ep (Gg+gG).
\eqn\AAv
$$

\no To show that the action {\Ai} is invariant under the above
transformation, we have to assume that the parameter $\epsilon$ is
not dependent on the coordinates, i.e. that we have a global
symmetry. If we would like to make this symmetry local we must
introduce a gauge field $A$ that satisfies
$\delta A_i=-\p_i\epsilon$. The gauge-invariant generalization of the
WZW action then takes the following form:

\Aiii

\no Now we can choose a gauge. If we
impose the Lorentz gauge condition
$\p_{\alpha}A^{\alpha}=0$, the gauge slice can be
parametrized as $A^{\alpha}=\varepsilon^{\alpha
\beta}\p_{\beta}X$ or

$$
A=\p X\qquad {\rm and} \qquad {\bar A}=-{\bar \p} X.
\eqn\AAiii
$$

\no The complete gauge-fixed
action of the Euclidean theory is then given by [\dvv]:

\Aiv

\noindent Here $X$ is a free scalar field:

\Av

\no which for the Euclidean theory is compact, with compactification
radius $R=\sqrt k$ in units of the self-dual radius.
\no The fields $(B,C)$ are a spin $(1,0)$ system of fermionic
ghosts that come from the Jacobian of the redefinition {\AAiii}:

\Avi

\no This form for the action can be proved either using an explicit
representation in terms of Euler angles [\dvv], as we will see later
or more generally as done by Gawedzki and Kupiainen
\REF\GK {K.~Gawedzki and A.~Kupiainen, ``Coset Construction From
Functional Integrals'', {\it Nucl.~Phys.} {\bf B320} (1989) 625.}
[\GK]
without using an explicit representation for $g$.
{}From {\Aiv} we see that the gauged WZW model can be
expressed through the ungauged {\sl} WZW model and
the action of the free-fields $X$, $B$ and $C$. As
already remarked in ref. [\dvv] this makes the
quantization straightforward.

\sec{\secthreeoneI}

To quantize the ungauged theory [\dvv] one notices that the {\sl}
symmetry gives rise to the conserved currents $\p {\bar J}=0$ and
${\bar \p} J=0$:

\Avii

\noindent where $t^1=i\sigma_1/2$, $t^2=\sigma_2/2$
and $t^3=i\sigma_3/2$ are the generators of {\sl} and $\sigma_i$ are
the Pauli
matrices. The above form of the currents follows as equations of
motion of the action {\Ai}, considering the variation $\d S_{WZW}$
under the transformation $g \rightarrow g+\d g$. The modes of these
currents satisfy the
{\sl} current algebra of level $k$, which is
equivalent to the following operator product
expansion (OPE):

\Aviii

\no which, after expanding in modes, can be written as

$$
\eqalign{
&[J_n^3,J_m^3]=-\h kn \delta _{n+m,0}\cr
&\cr
&[J_n^3,J_m^\pm]=\pm J_{n+m}^\pm \cr
&\cr
&[J_n^+,J_m^-]=kn \delta_{n+m,0}-2J_{n+m}^3.\cr}
\eqn\AAviii
$$

\no Here we have introduced the notation $J_{+}=J_{1}+iJ_{2}$ and
$J_{-}=J_{1}-iJ_{2}$.

The representation theory of Kac-Moody algebras shares many features
with the Virasoro algebra. In the following we are going to explain
some basic notions about the representation theory
of {\sl}
\REF\bars{I.~Bars, ``String Propagation on Black Holes'', preprint
USC-91-HEP-B3, May 1991.}
\REF\vil{\rvil}
[\bars,\vil]
that we will need later. The basic fields
from which we can build all the other states are
the Kac-Moody primaries
that satisfy:

\Bvii
and are characterized by the zero-mode Casimir
eigenvalue $j$ and by the eigenvalue of $J^3_0$:

\Bvi

\noindent where $\Delta_0= {1\over
2}(J_0^+J_0^-+J_0^-J_0^+)-(J_0^3)^2$. Here we have introduced a
holomorphic notation, but the same holds for the antiholomorphic
part. We can
construct the representation by acting with raising
and lowering operators $J_0^+$ and $J_0^-$, as we know for the
ordinary harmonic oscillator. A
solution of {\Bvi} is

\Bviii

\noindent If $j\in \IR$ it is standard in the
representation theory of {\sl} [\vil] to introduce new states
$|j,m)$ that satisfy

\Bxv

\noindent These recursion relations are satisfied
up to a function depending only on $j$, if we
normalize the states as:

\Bxva

\vskip 0.5cm

\noindent We will later work with the states that satisfy
{\Bviii} unless otherwise stated. These Kac-Moody
primaries define an irreducible representation of
{\sl}, on which we can impose two types of constraints
[\dn]:

\item{\it 1.} {\it Hermiticity constraints.} We
demand that $\Delta_0$ and $J^3_0$ should have real
eigenvalues. This means $m\in\R$ and $j=-{1\over
2}+i\lambda $ or $j\in {\rm \IR}$.
The types of Hermitian representations of {\sl} can
be classified as follows \REF\bih{\rbih}[\bih]:

\item{\bullet} Principal discrete series:
Highest-weight or lowest-weight representation.
Contain a state annihilated by ${J_0}^+$ and
${J_0}^-$, respectively. They are one-sided and
infinite-dimensional. From {\Bviii} we see that
these modules satisfy either $(j+m)$ or $(j-m)$ is
an integer and:
\Bix
\no If in addition $2j$ is an integer, the
representation is double-sided.

\item{\bullet} Principal continuous series.
Satisfies $j=-{1\over 2}+i\lambda$ with $\lambda$,
$m\in \R$.

\item{\bullet} Supplementary series. In this case
$j\in \R$, but neither $j+m$ nor $j-m$ is an
integer.

\item{\it 2.}{\it Unitarity constraints.} The
states are constrained to have positive norm, i.e.
$J^+_0J^-_0$ and $J^-_0J^+_0$ should have positive
eigenvalues. This imposes restrictions on the
allowed values of $j$. We will not impose any
constraints of unitarity on our states.

\no We will later see that on-shell states of the Euclidean black
hole belong to the discrete and supplementary series, while the
principal continuous series is off-shell. In the Minkowski theory the
on-shell states are those corresponding to the principal continuous
series.

We can create the above states by acting with a vertex operator
$T_{j\; m}$ on the SL(2)-invariant vacuum

$$
|j,m\rangle =\lim _{z\rightarrow 0} T_{j\; m}(z) \; |0\rangle.
\eqn\BBxvi
$$

\no These vertex operators have
the following OPE with the energy-momentum tensor
and the currents:

\Bxvii

\vskip 0.5cm

\no Where by $T_j$ we mean the multiplet of states with fixed $j$.
The correlation functions of these vertex operators satisfy the {\sl}
Ward identities that have the following form:

\Bxx

\no They can be derived by pushing a contour integral through the
correlator, where the contour encloses each of the points $z_i$.
Deforming the contour to a sum over small contours (each of them
enclosing one of the $z_i$'s), we get the above result.

Given a Kac-Moody algebra we can always construct a corresponding
enveloping Virasoro algebra.
The stress tensor follows from the Sugawara
construction and is given by the following
expression:

\Biii

\no The above prescription is a generalization of the U(1) case,
where the current is $j=\p \phi(z)$ and the energy-momentum tensor is
given by $T(z)=-{1 \o 2}:j(z)j(z):$. The normalization constant
$1/(k-2)$  in the expression {\Biii} is fixed by requiring that the
currents are indeed $(1,0)$ primary fields. The modes of the stress
tensor can be expressed through the currents

$$
L_n=-{1\o k-2}\sum_{m=-\infty}^\infty :J_{m+n}^a J_{-m}^a:.
$$

\no These currents have the following OPE with the energy-momentum
tensor

$$
T(z) J^a(w)={J^a(w)\o (z-w)^2}+{\p J^a(w)\o (z-w)}+\dots,
\eqn\BBiii
$$

\no that is equivalent to the commutation relation

$$
[L_m,J_n^a]=-n J_{m+n}^a.
\eqn\BBiv
$$

\noindent The stress tensor has a central
charge as a function of the level $k$ of the
Kac-Moody algebra:

\Bv

\no that can be obtained from the OPE

$$
T(z) T(w)\sim {c/2 \o (z-w)^4}+\dots.
\eqn\BBv
$$

\no Using the previously introduced OPEs, it is also easy to deduce
that the conformal weight $h_{j,m}$ of a primary field field $T_{j\;
m}$ is given by

$$
h_{j,m}=-{j(j+1)\over k-2}.
\eqn\zzzvi
$$

\no In the case of the {\slU} coset model the energy-momentum tensor
that follows from the gauged fixed action is

$$
T_{\rm SL(2,\R)/U(1)} =-{\D\o (k-2)} -\h (\p X)^2 -B\p C,
\eqn\zzi
$$

\no which has a central charge

$$
c={3k \over{k-2}}-1.
\eqn\zii
$$

\no There appear two interesting regions. The limit $k\rightarrow
\infty$ corresponds to the semiclassical limit in which the
$\sigma$-model is weakly coupled. From {\Aiii} it is clear that $k$
plays the role of $\hbar^{-1}$ in the quantum theory. If we choose
$k=9/4$ the central charge will be $c=26$ and the theory describes a
critical bosonic string in a curved background.

The total central charge of the theory is zero if we take into
account the reparametrization ghosts. This is a fermionic system of
ghosts described by the action

$$
S(b,c)=\int d^2 z (b{\bar \p} c +{\bar b}\p {\bar
c}).
\eqn\zziii
$$

\no The field $b$ has conformal dimensions $(h,\bar h)=(2,0)$, $c$
has dimension $(h,\bar h)=(-1,0)$, and the two-point function of
these fields is

$$
\langle b(z) c(w)\rangle =\langle c(z) b(w)\rangle={1\o (z-w)}.
\eqn\zziv
$$

\no The energy-momentum tensor of this ghost system has a central
charge of $c=-26$ and is given by

$$
T^{b,c}(z)=-2b(z) \p c(z)-\p b(z) c(z).
\eqn\zzv
$$

\no Gauged WZW models are the Lagrangian formulation of coset CFTs.
Such a coset construction can be associated to any current algebra.
Given a symmetry group $G$ and a subgroup $H$ of $G$, the stress
tensor of the coset CFT is then the difference of the two stress
tensors

$$
T_{G/H}=T_G-T_H.
\eqn\z
$$

\no Each of these operators can be constructed using the
corresponding currents and the Sugawara prescription. The central
charge of the coset theory is the difference between the two central
charges:
$$
c_{G/H}=c_G-c_H.
\eqn\zi
$$

\no The connection between this coset construction and the gauged WZW
models has been explained in
\REF\kara{D.~Karabali,  Q-H.~Park,  H.~Schnitzer, Z.~Yang, ``A GKO
Construction Based on a Path Integral Formulation of Gauged
Wess-Zumino-Witten Actions'', {\it Phys.~Lett.} {\bf B216} (1989)
307; D.~Karabali and H.~Schnitzer, ``BRST Quantization of the Gauged
W-Z-W Action and Coset Conformal Field Theories'', {\it Nucl.~Phys.}
{\bf B329} (1990) 649.}
[\GK,\kara].

\sec{\secthreeoneII}

In general, to obtain the form of the physical states of the quantum
theory we can use the BRST quantization procedure. The BRST symmetry
is a symmetry of the gauged-fixed action. Associated with this
symmetry we have a nilpotent charge $Q_{\rm BRST}$ that commutes with
the energy-momentum tensor

$$
\d T=[Q_{\rm BRST},T]=0.
\eqn\zzvi
$$

\no A state is said to be BRST-invariant if it is annihilated by the
BRST charge. This is a necessary condition for gauge invariance and
should therefore be satisfied by physical states. A state $|\phi
\rangle$ that satisfies

$$
| \ph \rangle =Q_{\rm BRST}| \ph ' \rangle
$$

\no is said to be BRST-trivial. It is annihilated by $Q_{\rm BRST}$
because of the nilpotency of this operator, and it decouples in $\cal
S$-matrix elements. Therefore, physical states have to be
BRST-invariant but not trivial. This BRST invariance is equivalent to
the condition

$$
[Q_{\rm BRST}, \phi (z)]= {\rm total \;\;derivative}.
\eqn\zzziii
$$

\no In the case of the black hole CFT, we have two BRST charges
corresponding to the U(1) symmetry and the diffeomorphism invariance
that are called $Q^{\rm U(1)}$ and $Q^{Diff}$ respectively. Their
expressions are given by

\Cii

\no and

\Cviii

 If we would like to consider the anti-holomorphic BRST-constraints,
we have to take into account that there are two possible choices of
sign for the current

$$
{\bar J}_3^{total} ={\bar J}_3 \pm i\sqrt{{k\over 2}}.
\eqn\mnm
$$

\no Both choices are related by a duality transformation.

It is easy to verify that $T_{\rm SL(2,\R)/U(1)}$ commutes with each
of the two BRST charges. These operators satisfy

\Cix

\no If we consider states of ghost number zero, the condition
{\zzziii} is equivalent up to a total derivative to the physical
on-shell condition, which demands the physical state to have
dimension $(1,1)$. More explicitly we see that the expression

$$
\eqalign{
\left[Q^{Diff},\ph(z)\right]&=\oint {dw \o 2\pi i} c(w) T(w)\ph(z)\cr
&\cr
& =\oint {dw \o 2\pi i} c(w) \left( {h\ph(z)\o (w-z)^2}+{\p \ph(z)\o
(w-z)}+\dots \right)\cr
&\cr
& = h(\p c) \ph(z)+c \p \ph(z)\cr}
\eqn\zzziv
$$

\no is a total derivative if $h=1$. The physical-state condition can
be equivalently formulated as the following two equations:

$$
\eqalign{
(L_0+{\bar L}_0-2)|\phi\rangle&=0\cr
&\cr
(L_0-{\bar L}_0)|\phi\rangle&=0.
\cr}
\eqn\wxx
$$

The invariance under the two BRST charges implies that physical
states of the coset theory are of the form

$$
{\cal V}_{j\; m \;\bar m}=T_{j\; m \;\bar m} \exp
\left(i\sqrt{2\o k}\left(m X(z)+{\bar m} {\bar X}({\bar
z})\right)\right),
\eqn\Cixai
$$

\no where $T_{j\; m \;\bar m}$ are the fields of the ungauged theory.
The quantities $m$ and $\bar m$ denote the eigenvalues of the zero
modes of the currents $J_3(z)$ and $\bar J_3(\bar z)$ and we have
used the notation $X(z, \bar z)=X(z)+\bar X(\bar z)$. It is easy to
verify that the above field is invariant under $Q^{\rm U(1)}$. The
invariance under $Q^{Diff}$ up to a total derivative implies the
on-shell condition

\Cixa

\no for $k=9/4$ this condition takes the form:

\Cixb

\no and the same for the antiholomorphic component.

In the Minkowski black hole the scalar field $X(z, \bar z)$ is
uncompactified, so that we have the restriction $m=\bar m$.
For the Euclidean theory this field is compactified. This implies
that the eigenvalues $m$ and $\bar m$ have to be on the lattice
[\gsw]:

\Cxvi

\no States with $n_1=0$ are called winding
modes, while states with $n_2=0$ are called momentum modes.
The spectrum of the Euclidean black hole CFT is therefore given by

$$
h_{n_1n_2}^j=-{j(j+1) \o k-2}+{(n_1+n_2k)^2 \o 4k},
\eqn\Cxviyz
$$

$$
\bar h_{n_1n_2}^j=-{j(j+1) \o k-2}+{(n_1-n_2k)^2 \o 4k}.
\eqn\Cxviyzy
$$

\no This coincides with the spectrum of a Liouville field with
momenta $\a=\sqrt {2\o {k-2}}j$ and a background charge $Q={2\o
{k-2}}$ coupled to a scalar field with compactification radius

$$
R= \sqrt k R_0,
\eqn\Cxviyzyz
$$

\no where $R_0=1/ {\sqrt 2}$ is the self dual radius.

We will discuss the cohomology of the 2D black hole CFT in more
detail in chapter~5. To continue we will choose a concrete
representation for the {\sl}-valued field $g$ or equivalently for the
currents. We will now see why this model describes strings in a black
hole background.

\endpage

\sec{\secthreetwo}

In  Witten's semiclassical description of the background, the gauge
field $A$ of the gauge-invariant action {\Aiii} is integrated out,
using its equations of motion. This procedure is of course only valid
in the semiclassical limit and we expect to get corrections in $1 /
k$.

We will first consider the Euclidean theory. In this case gauge
invariance is fixed by setting

$$
g=\cosh r +\sinh r \left(\matrix{\cos \th & \sin \th \cr\sin \th &
-\cos \th\cr}\right).
\eqn\wi
$$

\no The resulting action then takes the form:

$$\eqalign{
S(r,\th)&={k\o 2\pi} \int d^2 z (\p r{\bar \p} r +\tanh ^2r \p \th
{\bar \p} \th)\cr
&\cr
&={k\o 4\pi} \int d^2 x\sqrt{h} h^{ij}  (\p_i r \p_j r +\tanh ^2r
\p_i \th \p_j \th).\cr}
\eqn\wii
$$

\no The Wess-Zumino term is a total derivative in this gauge and it
has been dropped. By comparing {\wii} with {\zvii} the target space
metric ${\cal G}_{\mu \nu}$ can be computed. It is given by

$$
{\cal G}_{\mu \nu }=\left(\matrix{{\cal G}_{rr}& {\cal G}_{r\th } \cr
{\cal G}_{\th r} &  {\cal G}_{\th \th} \cr}\right)=
\left(\matrix{1& 0\cr 0  & \tanh ^2r \cr}\right).
\eqn\wiii
$$

\no We therefore obtain the following line element

$$
ds^2={\cal G}_{\mu \nu } dX^\mu  dX^\nu =dr^2+\tanh ^2 r d\th^2.
\eqn\wix
$$

\no This metric corresponds to a surface of a semi-infinite cigar
(see Fig. 2) that, in the asymptotic region $r\rightarrow \infty$,
gets $\R \times S^1$.

\topinsert
\vskip -3cm
\epsfysize=3.0in
\centerline{\hskip 0.5cm \epsffile{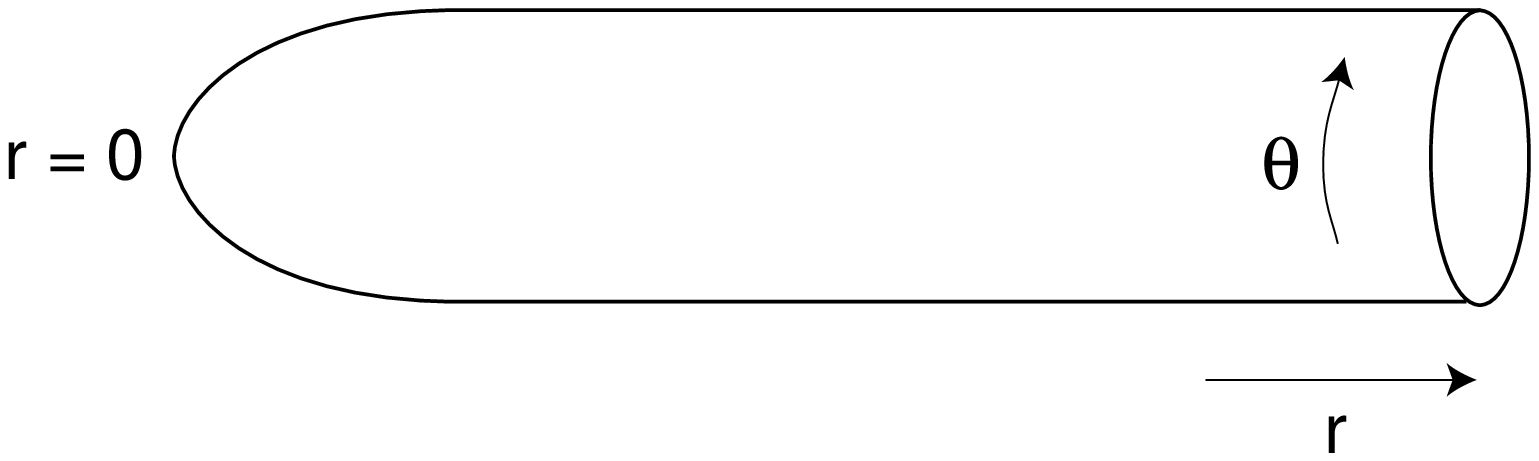}}
\caption{{\bf Fig. 2}: Target space geometry of the Euclidean black
hole in the semiclassical limit (analogue to regions I and II of the
Kruskal diagram of the Minkowskian Schwarzschild metric).}
\endinsert

{}From the measure in the integration over $A$ there appears a finite
correction that gives rise to the target space dilaton, so that the
classical action has the form

$$\eqalign{
S(r,\th)&={k\o 2\pi} \int d^2 z (\p r{\bar \p} r +\tanh ^2r \p \th
{\bar \p} \th)\cr
&\cr
&={k\o 4\pi} \int d^2 x\sqrt{h} h^{ij}  (\p_i r \p_j r +\tanh ^2r
\p_i \th \p_j \th)-{1\o 8\pi} \int d^2 x \sqrt{h} \Phi (r,\th) R
.\cr}
\eqn\wwwix
$$

The expression for the dilaton can be obtained demanding that the
$\b$-function equation to one loop order\foot{Note that this equation
has a different normalization w.r.t. {\zix} and that the tachyon is
supposed to be small, so that contributions of $O(T^2)$ are
neglected.}

$$
R_{ab}=D_a D_b \Phi
\eqn\wwix
$$

\no should be satisfied. The result is

$$
\Phi=2 \log \cosh r+{\rm const.}
\eqn\wwx
$$

\no The antisymmetric tensor field $B_{\m\n}$ can be gauged away in a
$(1+1)$-dimensional target space so that it has not to be taken into
account. The condition that one is considering small tachyons gives
an equation of motion for this field that allows us to determine its
form as a function the coordinates [\Wit]. We will do this in section
3.5.

If we compare the obtained action with Liouville theory {\zzx}
coupled to $c=1$ matter, we see that for large $r$ this field can be
identified with the Liouville field $\phi$. The precise relation
between the two theories will be one important point that we will
later explore in more detail.

To obtain the space-time interpretation as a 2D black hole we will
make the analytic continuation to Lorentz signature. Naively, the
Minkowski black hole can be obtained with the redefinition $\th =i
t$, so that the line element has the form

$$
ds^2=dr^2-\tanh ^2 rdt^2.
\eqn\wx
$$

\no The above metric has a singularity at $r=0$ that turns out to be
only a coordinate singularity, since the scalar curvature

$$
{\cal R}={4\over {\cosh^2r}}
$$

\no is regular at this point. In order to get a parametrization of
the complete space-time, including the regions past the singularity,
we will make the coordinate transformations

$$
2v=e^{r'+t},\qquad 2u=-e^{r'-t}, \qquad {\rm where} \qquad r'=r+\ln
(1-e^{-2r}).
\eqn\wxi
$$

\no The line element then has the form

$$
ds^2=-{du dv \o 1-uv}.
\eqn\wxii
$$

\no This metric exhibits all the space-time regions from the ordinary
Schwarzschild solution {\ziv}, with the horizons at $uv=0$ as well as
a curvature singularity at $uv=1$.

Instead of considering the formal analytic continuation $\th=it$, the
Minkowski version of the black hole can also be obtained gauging a
different subgroup of \sl. In this case we consider the noncompact
one parameter group generated by

$$
\d g= \e \left[\left(\matrix{1&0\cr0&-1\cr}\right)g +g
\left(\matrix{1&0\cr0&-1\cr}\right)\right].
\eqn\wwxiii
$$

\no We can parametrize the {\sl} group element as

$$
g=\left(\matrix{a&u\cr
-v&b\cr}\right), \qquad {\rm with} \qquad ab+uv=1.
\eqn\wwxii
$$

\no This is a global  parametrization.
The {\sl} coset is a double cover of the $(u,v)$ plane, depending on
the sign of $a$ and $b$. These variables can be eliminated with a
particular choice of gauge. In the region $1-uv>0$, one can fix the
gauge $a=b$ because either $a,b>0$ or $a,b<0$ holds. In the region
$1-uv<0$, we can choose the gauge $a=-b$. In both cases we obtain
with {\Aiii} and {\wwxii} the action

$$
{\cal S}=-{k\o 4\pi} \int d^2 x \sqrt{h} {h^{ij} \p_i u\p_j v \o
1-uv}.
\eqn\wwxiv
$$

\no The corresponding line element is represented by {\wxii}. This
action describes the regions I-VI of the Lorentzian black hole.

\sec{\secthreethree}

The target space geometry of the quantum theory for finite $k$ was
found by Dijkgraaf, H. Verlinde and E. Verlinde [\dvv]. In their
approach the gauge field $A$ is not integrated out, so that
corrections in $1/k$ can be taken into account. The form of the exact
metric follows by comparing the form of the Klein-Gordon operator
{\zxiii} with the $L_0$-operator that follows from the group theory
of {\sl}, as we will now see.
The action of the {\sl}-gauged WZW model can be written by
parametrizing $g$ in terms of Euler angles:

$$
g=e^{i \th_L \s_2 /2} e^{r \s_1 /2} e^{i \th_R \s_2 /2}.
\eqn\wxvi
$$

\no Here $\sigma_i$ are Pauli matrices and the ranges of the fields
are $0\leq r\leq \infty$, $0\leq \th_L < 2\pi$ and $-2\pi \leq \th_R
< 2\pi$. This is a suitable parametrization for the Euclidean black
hole, while for the Minkowski case we have to use

$$
g=e^{i t_L \s_3 /2} e^{r \s_1 /2} e^{i t_R \s_3 /2},
\eqn\wxviy
$$

\no where $0\leq r\leq \infty$, $-\infty \leq t_{R,L} < \infty$.
After introducing the form of $g$, the action {\Aiv} for the
Euclidean theory can be written in the form:

$$
S_{WZW}^{gf}={k \o 4\pi} \int d^2 z ({\bar \p} r \p r -{\bar \p} \th
_L \p \th_L-{\bar \p} \th_R \p \th_R -2 \cosh r {\bar \p} \th_L \p
\th_R)+S(X)+S(B,C).
\eqn\wxvii
$$

The Lorentz gauge condition for $A$ has been imposed, so that the
fields $X$ and $(B,C)$ appear as previously explained. We can
calculate the form of the conserved  currents {\Avii} in terms of
Euler angles. The expressions are given by

$$
\eqalign{
J^3(z)&=k(\p\th_L+\cosh r\p \th_R)\cr
&\cr
J^\pm(z)&=ke^{\pm i\th_L}(\p r\pm i\sinh r\p \th_R).
\cr}
\eqn\wwxvii
$$

The primary fields of the ungauged theory can be represented using
the matrix elements of the different representations of {\sl} [\vil]:

$$
T(r,\theta_R, \theta_L)=\langle j,\omega_L|
g(r,\th_L,\th_R)|j,\omega_R\rangle.
\eqn\wwxviii
$$

\no The quantum numbers $\omega_L$ and $\omega_R$ are the eigenvalues
of the currents $J_0^3$ and $\bar J_0^3$ respectively. The form of
the primary fields of the coset theory is determined by {\Cixai}. The
invariance under $Q^{\rm U(1)}$ implies the relations

$$
\omega_L+ m=0\qquad {\rm and} \qquad \omega_R-{\bar m}=0.
\eqn\wwxix
$$

\no We can shift the fields $\theta _L \rightarrow \theta _L-X$ and
$\theta _R \rightarrow \theta _R+\bar X$ so that the dependence on
the field $X$ disappears from the primary fields. For the Euclidean
theory primary fields can be represented through the Jacobi functions

$$
T(r,\th_L,\th_R)={\cal P}_{\omega_L\; \omega_R}^j(\cosh
r)e^{i\omega_L\th_L+i\omega_R \th_R} .
\eqn\wwxvi
$$

We will now restrict ourselves to the mini-superspace description of
the problem. This means that we keep only the zero-mode algebra:

$$
\eqalign{
&[{\cal J}_3,{\cal J}_\pm ]=\pm {\cal J}_\pm \cr
&\cr
&[{\cal J}_+, {\cal J}_-]=-2{\cal J}_3.}
\eqn\wxvi
$$

\no These zero modes are represented as differential operators in the
following way:

$$
\eqalign{
{\cal J}_3&=-i{\p \o \p \th_L},\cr
&\cr
{\cal J}_\pm&=e^{\pm i\th_L}\left( {\p \o \p r}\mp {i\o \sinh
r}\left( {\p \o \p \th_R}-\cosh r {\p \o \p \th_L}\right)\right).\cr}
\eqn\wxvii
$$

\no In this representation the zero-mode Casimir of {\sl} that
follows from the Sugawara prescription takes the form:

$$
\D_0={\p^2\o \p r^2}+\coth r{\p \o \p r}+{1\o \sinh^2 r}\left({\p^2
\o \p \th^2_L}-2\cosh r {\p^2\o \p \th_L \p \th_R}+{\p^2 \o \p
\th_R^2}\right).
\eqn\wxviii
$$

\no The complete $L_0$ and ${\bar L}_0$ operators follow, taking into
account the free boson $X$. Their expressions are

$$
\eqalign{
L_0&=-{\D_0\o k-2}-{1\o k} {\p^2 \o \p \th_L^2},\cr
&\cr
{\bar L}_0&=-{\D_0\o k-2}-{1\o k} {\p^2 \o \p \th_R^2}.
\cr}
\eqn\wxix
$$

The on-shell conditions {\wxx} imply that a physical state is
annihilated by the operator:

$$
L_0-{\bar L}_0={1\o k}\left( {\p^2\o \p \th_R^2}-{\p^2\o \p
\th_L^2}\right).
\eqn\wxxi
$$

\no Therefore, these states can be split in the following way:

$$
T(r,\th_L,\th_R)=T(r,\th)+\widetilde T(r,{\widetilde \th}),
\eqn\wxxii
$$

\no where

$$
\th=\h (\th_L+\th_R) \qquad \qquad
\widetilde \th=\h (\th_L-\th_R).
\eqn\wxxiii
$$

\no States that depend only on $\th$ are the momentum modes, while
the winding modes depend only on $\widetilde \th$.
The $L_0$ operator has a different form depending on whether it acts
on $T(r,\th)$ or $\widetilde T(r,\widetilde \th) $. Therefore both
operators will lead to different target spaces that correspond to
dual manifolds. When the operator $L_0$ acts on $T(r,\th)$ it takes
the form

$$
L_0=-{1\o k-2}\left( {\p^2 \o \p r^2}+\coth r {\p \o \p r}+\left(
\coth^2{r\o 2} -{2\o k} \right) {\p^2 \o \p \th ^2}\right).
\eqn\wxxiv
$$

The expressions for the metric and the dilaton for finite $k$ follow
from the comparison with {\zxiii} and take the form:

$$
ds^2={k-2\o 2} \left(dr^2 +\b^2(r) d\th^2\right)\qquad {\rm and}
\qquad \Phi=\log \left( {\sinh r \o \b(r) } \right),
\eqn\wxxvi
$$

\no where $\b(r)$ is given by

$$
\b(r)=2\left(\coth^2 {r\o 2}-{2\o k}\right)^{-\h}.
\eqn\wwxxvi
$$

\no To leading order in $1/ k$ this coincides with the semiclassical
metric, found by Witten {\wix} after a simple rescaling of the
coordinate r. The space- time diagram in this limit is the
semi-infinite cigar represented in Fig. 2. It has been verified that
these expressions are the perturbative solution of the
$\beta$-function equations up to three [\Tseytlin] and four loops
[\Jack]. When it acts on  $\widetilde T(r,\widetilde \th)$ the
operator has the form:

$$
L_0=-{1\o k-2}\left( {\p^2 \o \p r^2}+\coth r {\p \o \p r}+\left(
\tanh^2{r\o 2} -{2\o k} \right) {\p^2 \o \p \widetilde \th
^2}\right).
\eqn\wxxv
$$

\no It leads to a background metric and dilaton {\wxxvi} with

$$
\b(r)=2\left( \tanh^2 {r\o 2}-{2\o k}\right)^{-\h}.
\eqn\wwxxvii
$$

\no This metric and the corresponding dilaton have a singularity at
$r={\rm arc}\; {\rm tanh} \; \sqrt{2/k}$. In the semiclassical limit
$k\rightarrow \infty$ the line element takes the form

$$
ds^2=dr^2+4\coth ^2{r\o 2} d{\widetilde \th}^2,
\eqn\wwxxviii
$$

\no so that the winding modes propagate on a manifold that looks like
a ``trumpet'' (Fig. 3). We have a real singularity at $r=0$.

\topinsert
\epsfysize=1.5in
\centerline{\hskip 0.0cm \epsffile{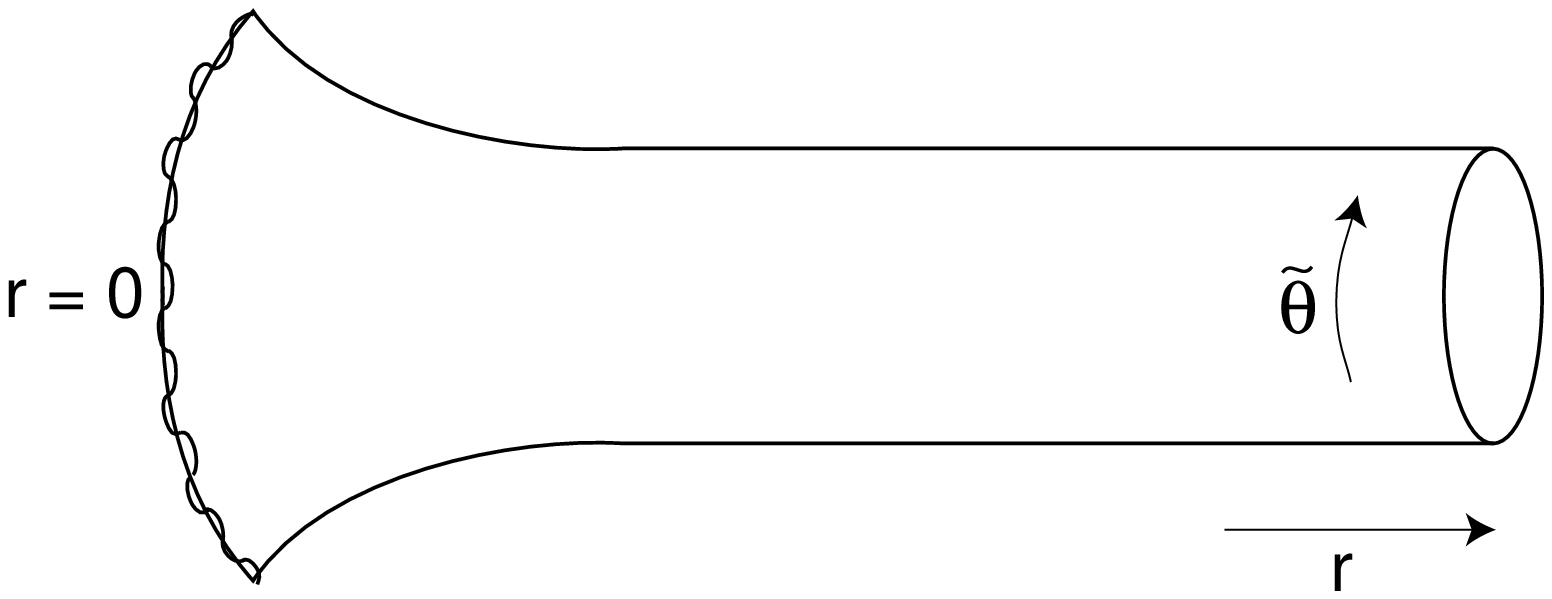}}
\vskip 1.0cm
\caption{{\bf Fig. 3}: Target space geometry of the Euclidean black
hole on which winding modes propagate in the limit $k\rightarrow
\infty$.}
\endinsert

The metric that describes the propagation of winding modes
{\wwxxviii} can be obtained from the one describing the propagation
of momentum modes {\wxxvi} to all orders in $1/k$ by the
transformation

$$
r\rightarrow r+{i \pi} \qquad {\rm  and} \qquad  \th\rightarrow
{\widetilde \th},
\eqn\wwxxix
$$

\no or equivalently $r\rightarrow r+{i \pi /2}$ in the notation
{\wix}.

In the last section we have seen that if we choose a suitable
parametrization of the \sl-valued field $g$ in terms of the
coordinates $(u,v)$ and gauging a noncompact subgroup of this
symmetry group, we are able to obtain all the regions of the
maximally extended Minkowskian version of the Schwarzschild black
hole. Instead of formulating the maximal extension of the Minkowski
black hole in $(u,v)$ coordinates, we can use the $(r,t)$ coordinates
by choosing a suitable range of the fields. By introducing the
variables

$$
r=\log \left(\sqrt{1-uv}+\sqrt{-uv}\right)\qquad {\rm and} \qquad
t={1\o 2}\log{\left(-{u \o v}\right)},
\eqn\wwxxxii
$$

\no where $r$ is real or imaginary depending on which region of the
Schwarzschild black hole we want to describe, this metric can also be
written in the form
$$
ds^2=dr^2-\tanh^2  r \, d t^2.
\eqn\wwxxxi
$$
\no $r$ takes the values:
$$
r\in \cases{ [0,\infty ] & in region I where $uv\leq 0$\cr
i[0,\pi/2]& in region III where $0\leq uv < 1$\cr
[0,\infty ]+i \pi/2& in region V where $1\geq uv$\cr}
\eqn\wwxxxiii
$$

\no The metric {\wwxxxi} can then be written the following form
[\dvv]

$$
ds^2=\cases{ dr^2-\tanh^2r dt^2 & in region I with $r\in
\{0,\infty\}$ \cr
-dr^2+\tan^2r dt^2 & in region III with $r\in \{ 0,\pi /2\}$\cr
dr^2-\coth^2r dt^2  & in region V with $r\in \{ 0,\infty \}$\cr}
\eqn\wwxxxv
$$

\no where we have redefined $r\rightarrow r/i$ in region III and
$r\rightarrow r-i\pi/2$ in region V. We can now make a correspondence
with the different regions of the Euclidean black hole. We observe
that regions I and V of the Minkowski black hole correspond to the
cigar and the trumpet in the Euclidean theory. The metric of region
III can be seen as the analytic continuation  of a metric of the
form:

$$
ds^2=dr^2+\tan^2 r d\th^2
\eqn\wwwxxxv
$$

The manifold that corresponds to this metric has the shape of a
``hat''\foot{I thank H.~Verlinde for pointing this out to me.} as
shown in Fig.~4. As a CFT it can be regarded as the SU(2)/U(1) coset,
that is a parafermionic model.
\REF\giveon{A. Giveon, ``Target Space Duality and Stringy Black
Holes'', {\it Mod.~Phys.~Lett.} {\bf A6} (1991) 2843.}
[\dvv,\giveon]. This model can be shown to be self-dual. The form of
the Euclidean metric in regions I, III and V is represented in
Fig.~4.

\midinsert
\epsfysize=3.3in
\centerline{\hskip -3cm \epsffile{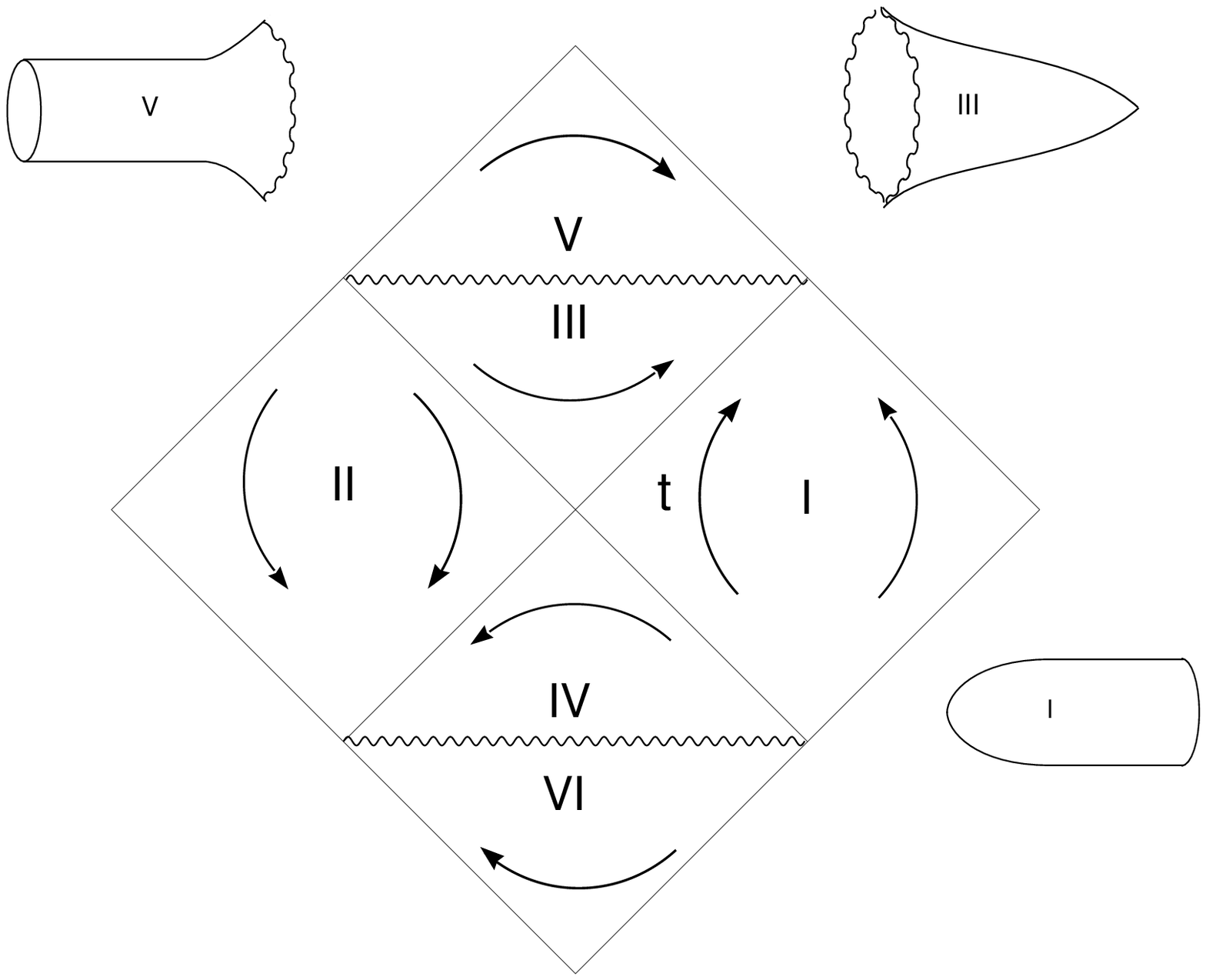}}
\vskip 1cm
\caption{{\bf Fig. 4}: Penrose diagram of the maximally extended
Minkowski black hole and the corresponding regions in the Euclidean
black hole. The arrows indicate the time flow $\p_t=u\p_u+v\p_v$.}
\endinsert

\endpage

\chapter{\chafour}

\no In the mini-superspace description of the black hole CFT, we have
considered field configurations that are independent of the space
coordinate and we kept only the zero mode of the fields. If we would
like to study the full string theory, we have to find a suitable
formulation of the problem that allows us to actually make
computations and, at the same time, to keep all the modes of the
fields. Such a formulation is the representation of the CFT in terms
of free-fields. This representation has already played an important
role in the computation of correlation functions of minimal models or
the amplitudes of gravitationally dressed fields in non-critical
string theory.

The WZW model can be written in terms of free fields using the
Gauss-decomposition [\rgmom,\vv]. Equivalently, one can write the
{\sl} current algebra in terms of the Wakimoto representation
\REF\w{\rw}
[\w].
For the SU(2) CFT, Dotsenko
\REF\d{\rd}
[\d]
has used this representation to compute certain correlation functions
of primary fields for spherical topologies. The obtained correlators
agree with those computed by  Fateev and A. B. Zamolodchikov
\REF\fz{V.~A.~Fateev and A.~B.~Zamolodchikov, ``Operator Algebra and
Correlation Functions in the Two-Dimensional SU(2)$\times$ SU(2)
Chiral Wess-Zumino Model'', {\it Sov. J. Nucl. Phys.} {\bf 43} (4)
(1986) 637.}
[\fz]
using the SU(2) Ward identities. In the case of the black hole CFT,
Bershadsky and Kutasov [\bk] have proposed to use the Wakimoto
representation of {\sl} to compute interesting quantities for the
black hole, as for example the $\cal S$-matrix of tachyons
interacting in the black hole background
[\wir]. We will now present the free-field approach to the {\slU}
coset model.

\section{\secfourone}

\no The {\sl} Wakimoto representation $\foot{In the following we will
use the conventions of ref. [\eky]}$ is formulated in terms of three
free-fields $\beta$, $\gamma$ and $\phi$ . The chiral bosonic
superconformal ghosts $\beta$-$\gamma$ have spin 1 and 0,
respectively, and they are described by the action

$$
S={1\o 2\pi} \int d^2 z (\b {\bar \p} \g +{\bar \b} \p {\bar \g}).
\eqn\BBii
$$

\no The two-point function of these fields is

$$
\langle \gamma(z) \beta(w) \rangle =-{1\over (z-w)},
$$

\no and the same for the antiholomorphic fields ${\bar \g}$ and
${\bar \b}$. The other OPEs are regular and we have no contractions
between holomorphic and antiholomorphic ghosts.
To perform concrete calculations it will be
useful to bosonize the $\b$-$\g$ system as follows
\REF\fms{\rfms}[\fms]:

\Bxxiv
\noindent where $u$ and $v$ are ordinary bosons

\Bxxv

\no From this bosonization prescription we see that we are able to
define non-integer powers of the operator $\gamma$, while we can only
define positive integer powers of $\beta$ since this operator
contains derivatives. The field $\phi$ is an ordinary non-compact
free boson with a background charge

$$
S={1\over 2\pi}\int{1\over 2} \p \phi{\bar \p}\phi -{2\over
\ap}
R^{(2)} \phi
\eqn\BBiii
$$

\no and the propagator

$$
\langle \p \phi(z) \p \phi(w) \rangle =-{1\over
(z-w)^2}.
$$

\no We have introduced the notation $\ap^2=2k-4$. The
currents that satisfy the OPE {\Aviii} have the
following form:

\Bii

\no This is easy to check, using the previous two-point functions.
The energy-momentum tensor follows from the Sugawara prescription.
After inserting the above form of the currents, we obtain

\Biv

\noindent It has a central
charge as a function of the level $k$ of the
Kac-Moody algebra:

$$
c={3k\over {k-2}}.
\eqn\zzzv
$$

\no The complete action associated with the energy-momentum
tensor {\Biv} is therefore:

\Bxix

\no As we mentioned in section~3.2, the basic fields are the
Kac-Moody primaries $|j,m\rangle$.
In the Wakimoto representation they are created  by the action of
the following (not normalized) ``tachyon'' vertex
operator on the SL(2) vacuum:

\Bxvi

\noindent and the same expression for the
antiholomorphic part.
\noindent Using the free-field representation of
the currents {\Bii}, it is easy to check that
$T_{j\; m}$ satisfies the definition of a
Kac-Moody primary {\Bvii}, as well as {\Bvi} and
{\Bviii}. The conformal dimension of $T_{j\;m}$ is given by {\zzzvi}.

If we would like to calculate correlation functions
of the vertex operators $T_{j\; m}$ we would
need a screening charge, in order to guarantee the charge
conservation arising from the zero mode
integration. This screening operator can be determined from the
observation that the ungauged model has the {\sl} symmetry, so that
the  correlators have to satisfy the Ward Identities.
Since these identities should be
satisfied in the free-field representation, the
screening charge must have a regular OPE with the
stress tensor and the currents. We must also take
into account, the fact that only positive integer powers of
$\b$ are well defined through bosonization.
The screening that satisfies these conditions can
be represented as the following surface integral:

\Bxxi

\noindent The operator $J(z,\bar z)$ is no longer a Kac-Moody
primary. It is one of the simplest operators at higher mass level, as
we will discuss in more detail in the next chapter. One can easily
check the identities [\d]

\Bxxii

\vskip 0.5cm

\noindent The total derivative appearing in the
last OPE requires a careful treatment of contact
terms, as we will see when we calculate correlation functions in
chapter 6. This comes from the
fact that we are working with a surface integral
and not with a contour integral, where this
contribution generically vanishes. As is known to
be correct for a Coulomb gas model
\REF\dfa{\rdfa}
or Liouville theory
\REF\gl{\rgl}
\REF\dtwo{\rdtwo}
[\dfa,\gl,\dfk,\dtwo],
the screening charge has to be added to the action and considered as
the
interaction of the model [\bk]. We will see in section 4.3 that this
prescription indeed reproduces the correct values for the metric and
the dilaton for finite $k$ found in ref. [\dvv]. The complete action
of the ungauged model is therefore
\Bxxiii

\noindent In the above expression there appears a free parameter $M$
that  is related to the black hole mass
[\bk,\eky]. This becomes clear from the space-time interpretation of
this model already at the semiclassical level.

As we previously saw, in order to construct the conformal field
theory of the
Euclidean black hole we are interested in the coset
\slU. This gauging must be done as explained in section~3.2. To gauge
the U(1) subgroup, we introduce a
gauge boson $X$ and a pair of fermionic ghosts $B,
C$ of spin $1, 0$ respectively. The complete energy-momentum tensor
of the U(1) gauged theory is therefore

\Cv

\no Taking into account the fermionic diffeomorphism ghosts $(b,c)$
the total central charge of the theory is zero.
This means that the complete gauge-fixed action (without the
fermionic ghosts) is$\foot{The fermionic ghost contribution can be
generically factorized out of the correlation functions on the
sphere.}$

$$
S={1\over 2\pi}\int {1\over 2}\p \phi{\bar \p}\phi+{1\over 2}\p
X{\bar \p}X -{2\over
\ap}
 R^{(2)} \phi +\b{\bar \p}\g +{\bar \b}\p{\bar \g}
+2\pi M \b {\bar \b} e^{-{2\over \ap}\phi}.
\eqn\Cvvvv
$$

\no  For the Euclidean theory the gauge boson $X$ is compact.

The two BRST charges are given by {\Cii} and {\Cviii} with the
previous representation for the currents and the energy-momentum
tensor.
The tachyon states of the gauged theory
are the dressed ghost number zero primary fields,
which are invariant under $Q^{U(1)}$\foot{Since $j$
and $m$ are arbitrary at this moment, these
operators will be called tachyons.}

\Ciii

\no The characteristics of these fields have been  mentioned in
section~3.2.

\section{\secfouronetwo}
\no Briefly, we would like to mention that the model introduced by
Bershadsky and Kutasov can be obtained directly from the Lagrangian
formulation of the WZW model, using the Gauss decomposition
\REF\vafa{C. Vafa, ``Strings and Singularities'', preprint
HUTP-93-A028, hep-th/9310069.} [\rgmom,\vv,\ms,\vafa]\foot{I thank E.
Verlinde for providing me a copy of his notes.}. The Wakimoto
representation of the current algebra is equivalent to the choice of
a particular representation for the {\sl} valued field $g$. As we
previously explained this representation is suitable to evaluate the
full string theory. The {\sl}-valued field $g$ is parametrized with
the Gauss decomposition as follows [\vv,\vafa]

$$
g=g(\g, \ph, {\bar \g})=\left(\matrix{1& 0\cr \g  & 1 \cr}\right)
\left(\matrix{e^{\ph/2} & 0\cr 0  & e^{-\ph/2} \cr}\right)
\left(\matrix{1& {\bar \g} \cr 0  & 1 \cr}\right).
\eqn\rxvii
$$

\no Here all the fields are real and the boson $\phi$ is non-compact
with a range $-\infty< \phi <\infty$. We can insert this expression
for $g$ into the ungauged WZW action {\Bxxiii}.
Using the Polyakov-Wiegmann formula [\kz]

$$
S_{WZW}(GH^{-1})=S_{WZW}(G)+S_{WZW}(H)+{1\o 16\pi} \int {\rm tr}
\left(G^{-1} {\bar \p} G H^{-1} \p H \right)d^2 \x
\eqn\rxviii
$$

\no the ungauged model can be written as

$$
S_{WZW}={1\o 16\pi} \int \left(e^{-\ph} {\bar \p} \g \p {\bar \g} +\p
\ph {\bar \p}\ph\right).
\eqn\rxix
$$

\no If we introduce two auxiliary fields $\b$ and ${\bar \b}$ the
action becomes

$$
S_{WZW}={1\o 16\pi}\int \left(\p \ph {\bar \p}\ph +\b {\bar \p} \g
+{\bar \b} \p {\bar \g}-\b {\bar \b} e^{-\ph}\right).
\eqn\rrxix
$$

\no This is precisely the action of Bershadsky and Kutasov after a
trivial rescaling of the fields and up to quantum corrections. Of
course, we have to be careful with the insertion of equations of
motion if we pass to the quantum Lagrangian. Then we have to keep
Jacobians into account after making a change of variables [\rgmom].
This is, for example, the origin of the scalar curvature term in the
action {\Bxxiii}. The free parameter corresponding to the black hole
mass $M$ can be obtained from a shift in $\phi$.

\endpage

\chapter{\secfourtwo}

\no In this section we are going to show that the {\slU} gauged WZW
model, formulated in terms of Wakimoto coordinates, has the same
space-time interpretation for finite $k$ as the one found by
Dijkgraaf et al. [\becker]. We will derive the expressions for the
metric and the dilaton for finite $k$ of the 2D black hole in terms
of free-field coordinates identifying the $L_0$ operator with the
target space Laplacian of the $\s$-model$\foot{We would like to thank
E.~Verlinde for helping us to understand this space-time
interpretation.}$ .

The mini-superspace approximation of the currents {\Bii} takes the
following form:

$$
\eqalign{
{\cal J}_+ & ={\p \o \p \g}\cr
& \cr
{\cal J}_3&=-\g {\p \o \p \g}+\h {\p \o \p \ph}\cr
&\cr
{\cal J}_-&=\g^2{\p \o \p \g}-\g {\p \o \p \ph}-e^{-2\ph}{\p \o \p
{\bar \g}}.\cr}
\eqn\ri
$$

\no We would like to remark that the zero mode algebra is also
satisfied if we parametrize $J_-$ as

$$
{\cal J}_-=\g^2{\p \o \p \g}-\g {\p \o \p \ph}+{\epsilon}
e^{-2\ph}{\p \o \p {\bar \g}},
\eqn\rimnm
$$

\no where $\epsilon$ is an arbitrary free parameter, related to the
black hole mass. We will discuss this possibility at the end and we
will use {\ri} for the moment being.

The expressions {\ri} can be computed following the steps that lead
to the quantum-mechanical description of Liouville theory
\REF\sei{\rsei}
\REF\gm{\rgm} [\sei,\gm].
First, we have to  use the canonical quantization procedure and map
the complex plane $(z, \bar z)$ to the cylinder $(t,\s)$

$$
z=e^{t+i\s},\qquad \qquad {\bar z}=e^{t-i\s}.
\eqn\rii
$$

\no Then we expand the fields using a Fourier decomposition. For
$\ph(\s,t)$ and its canonical conjugate momentum $\Pi(\s,t)$, this
expansion takes the form

$$
\eqalign{
\phi(\s, t)&=\phi_0(t)+\sum_{n\neq 0} {i\o n} (a_n(t) e^{-i n
\s}+b_n(t) e^{in\s})\cr
&\cr
\Pi(\s,t)&=p_0(t)+\sum_{n\neq 0} {1\o 4\pi} (a_n(t) e^{-i n
\s}+b_n(t)e^{in\s}).\cr}
\eqn\riii
$$

\no Since from this point of view $\phi(\s, t)$ is not a free-field
the time dependence of the components might be complicated.

 In order to quantize the theory we impose the equal-time commutator

$$
[\phi(\s,t),\Pi(\s',t)]= \delta(\s-\s').
\eqn\riiv
$$

In the mini-superspace description we consider field configurations
that are not dependent on $\s$ and we keep only the zero modes of the
fields $\phi_0(t)$ and $p_0(t)$. We replace the canonical-conjugate
momenta by derivatives with respect to the fields\foot{We drop the
subindex $0$ of the zero modes.}:

$$
\eqalign{
\Pi_\ph&={\p \ph \o \p t}=-{\p \o \p \ph}\cr
&\cr
\Pi_{\g}&=\b={\p \o \p \g} .\cr}
\eqn\rv
$$

\no To obtain the complete expression for ${\cal J}_-$ we have to
replace $\p \g$ by an appropriate expression. Here we use the
equation of motion for $\b$ and find:

$$
{\p \g \o \p t}=-{\bar \b} e^{-{2\o \a_+}\ph}=-e^{-{2\o \a_+}\ph} {\p
\o \p {\bar \g}}.
\eqn\rvi
$$

\no With these substitutions, we obtain the expression {\ri} for the
currents, after rescaling $\ph$ by a factor of $\a_+$,
$\ph\rightarrow \a_+ \ph$. For the antiholomorphic currents we follow
exactly the same procedure and obtain

$$
\eqalign{
{\bar {\cal J}}_+ & ={\p \o \p {\bar \g}}\cr
& \cr
{\bar {\cal J}}_3&=-{\bar \g} {\p \o \p {\bar \g}}+\h {\p \o \p
\ph}\cr
&\cr
{\bar {\cal J}}_-&={\bar \g}^2{\p \o \p {\bar \g}}-{\bar \g} {\p \o
\p \ph}-e^{-2\ph}{\p \o \p  \g}.\cr}
\eqn\rvii
$$

\no We note that although ${\cal J}_-$ contains a ${\bar \g}$
dependence all the antiholomorphic currents $\bar {\cal J}_i$ commute
with the holomorphic currents ${\cal J}_i$. Here we have chosen a
particular normal-ordering prescription for the $\b$-$\g$ system;
other normal-ordering prescriptions are equivalent to this one after
a redefinition of the wave-functions.

Independently, we can construct these currents using some elements of
group theory. The currents are the generators of the three
one-parameter subgroups of {\sl} represented by the following
matrices [\vil]:

$$
\eqalign{
\omega_1(t) & =g(t,0,0)=\left(\matrix{1& t\cr 0  & 1 \cr}\right)\cr
& \cr
\omega_2(t) & =g(0,t,0)=\left(\matrix{e^t & 0\cr 0  & e^{-t}
\cr}\right)\cr
& \cr
\omega_3(t) & =g(0,0,t)=\left(\matrix{1 & 0\cr t  & 1
\cr}\right).\cr}
\eqn\rxx
$$

\no The infinitesimal generators ${\widehat A}_i$ corresponding to
these transformations are defined by

$$
{\widehat A}_i=\left({\p\g(t) \o \p t}{\p \o \p \g}+{\p \ph(t)\o \p
t}{\p \o \p \ph}+{\p{\bar \g}(t) \o \p t}{\p \o \p {\bar
\g}}\right)_{t=0}.
\eqn\rxxi
$$

\no Here $\g(t)$, $\ph(t)$ and ${\bar \g}(t)$ are identified from the
definition:

$$
g(\g(t),\ph(t), {\bar \g}(t))=\omega_i(t)\cdot g(\g,\ph,{\bar \g}).
\eqn\rxxii
$$

\no The expressions that we obtain for the infinitesimal generators
therefore are:

$$
\eqalign{
{\widehat A}_1&=e^{-2\ph}{\p \o \p {\bar \g}} +\g {\p \o \p
\ph}-\g^2{\p \o \p \g}\cr
&\cr
{\widehat A}_2&={\p \o \p \ph}-2\g{\p \o \p  \g}\cr
&\cr
{\widehat A}_3&={\p \o \p  \g}.\cr}
\eqn\rxxiii
$$

\no This means that, after identifying

$$
{\widehat A}_1=-{\cal J}_-,\qquad \qquad
{\widehat A}_2=2{\cal J}_3, \qquad \qquad
{\widehat A}_3={\cal J}_+,
\eqn\rxxiv
$$

\no we obtain the same result as we had from the mini-superspace
approximation of the currents. The antiholomorphic generators are
obtained in the same way. In this case we have to set

$$
g(\g(t),\ph(t), {\bar \g}(t))= g(\g,\ph,{\bar \g}) \cdot
\omega_i(t),
\eqn\rxxii
$$

\no which gives us the antiholomorphic currents {\rvii}.

With the mini-superspace approximations for the currents, we now
obtain the expressions for the zero-mode Casimir using Sugawara's
prescription {\Biii}

$$
\D_0={\p ^2 \o \p \ph^2}+{\p \o \p \ph}+e^{-\ph}{\p^2\o \p \g \p
{\bar \g}}.
\eqn\rviii
$$

\no Here we have rescaled $\ph$ by a factor of 2, $\ph \rightarrow
\ph/2$.
This means that the complete  Klein-Gordon operators $L_0$ and ${\bar
L_0}$ are given by

$$
\eqalign{
L_0=-{\D_0\o k-2}-{1\o k}{\p^2\o \p X^2}\cr
&\cr
{\bar L}_0=-{\D_0\o k-2}-{1\o k}{\p^2\o \p {\bar X}^2}.\cr}
\eqn\rix
$$

\no Therefore we are left with five fields: $\ph, \g ,{\bar \g}, X,
{\bar X}$. If we take the BRST constraints into account we see that
not all of these variables are independent. We will eliminate $\g$
and ${\bar \g}$ using the BRST conditions and keep only the variables
$\ph, X, \bar X$ to make the space-time interpretation. The
diffeomorphism invariance means that the physical states are
dimension $(1,1)$ fields, so that they satisfy

$$
(L_0-{\bar L}_0)\Psi =\left({\p^2 \o \p X^2}-{\p^2 \o \p {\bar
X}^2}\right)\Psi =4{\p ^2 \Psi \o \p X^+ \p X^-} =0,
\eqn\rxi
$$

\no where we have introduced coordinates $X^\pm = X\pm {\bar X}$ and
$\Psi$ are  the wave functions associated to the states. This means
that we have two types of wave functions, one that depends only on
$X^+$ and one that depends only on $X^-$:

$$
\Psi^+=\Psi^+(\g,{\bar \g}, \ph, X^+) \qquad {\rm and }\qquad
\Psi^-=\Psi^- (\g,{\bar \g}, \ph , X^-).
\eqn\rrxi
$$

\no The total wave function is then the sum $\Psi= \Psi^++\Psi^-$.
 One of these wave functions will represent the winding modes and the
other one the momentum modes.
The U(1) constraint eliminates the $\g$, ${\bar \g}$ dependence of
the $L_0$ and ${\bar L_0}$ operators, as we will now see. First we
notice that for $\Psi$ this constraint implies\foot{Here we have two
choices for ${\bar {\cal J}}_3^{total} ={\bar {\cal J}}_3 \pm i \p /
\p X$. We have taken the ``-'' sign. The two solutions are again
related by a duality transformation. }

$$
\eqalign{
{\cal J}_3\Psi &=\left( {\p \o \p \ph}-\g {\p \o \p \g}-i{\p \o \p
X}\right)\Psi=0 \cr
&\cr
{\bar {\cal J}}_3\Psi &=\left( {\p \o \p \ph}-{\bar \g} {\p \o \p
{\bar \g}}-i{\p \o \p {\bar X}}\right)\Psi=0.\cr}
\eqn\rx
$$

\no When acting on $\Psi^+$ or $\Psi^-$ we therefore obtain the
conditions:

$$
\eqalign{
{\cal J}_3\Psi^+ &=\left( {\p \o \p \ph}-\g {\p \o \p \g}-i{\p \o \p
X^+}\right)\Psi^+=0,
\qquad {\cal J}_3\Psi^- =\left( {\p \o \p \ph}-\g {\p \o \p \g}-i{\p
\o \p X^-}\right)\Psi^- =0 ,\cr
&\cr
{\bar {\cal J}}_3\Psi^+ &=\left( {\p \o \p \ph}-{\bar \g} {\p \o \p
{\bar \g}}-i{\p \o \p X^+}\right)\Psi^+=0,
\qquad {\bar {\cal J}}_3\Psi^- =\left( {\p \o \p \ph}-{\bar \g} {\p
\o \p {\bar \g}}+i{\p \o \p X^-}\right)\Psi^- =0 .\cr}
\eqn\rxi
$$

\no This means that the $\g$ and ${\bar \g}$ dependence of the wave
functions $\Psi^+$ and $\Psi^-$ can be eliminated and their
dependence on the Wakimoto coordinates is:

$$
\Psi^+=\Psi^+ \left(\g {\bar \g}e^{\h (\ph +iX^+)},  e^{\h(\ph
-iX^+)}\right)=
\Psi^+\left(e^{\h ({\widetilde \ph} +i{\widetilde X^+})},e^{\h
({\widetilde \ph} -i{\widetilde X}^+)}\right)
\eqn\rxii
$$

\no and

$$
\Psi^-=\Psi^- \left(\g e^{\h (\ph -iX^-)}, {\bar \g} e^{\h (\ph
+iX^-)}\right)=
\Psi^-\left(e^{\h (\widetilde{\ph} -i{\widetilde X}^-)},e^{\h
({\widetilde \ph}  +i{\widetilde X}^-)}\right),
\eqn\rxii
$$

\no where we have defined new variables ${\widetilde \ph}$ and
${\widetilde X^\pm}$. In terms of these coordinates the $\g$ and
${\bar \g}$ dependence of the of the $L_0$ and $\bar L_0$ operators
has been eliminated and the {\sl} zero-mode Casimir operators take
the following form:

$$
\D_0^+={\p^2 \o \p {\widetilde \ph}^2}
+{\p \o \p {\widetilde \ph}} +e^{-{\widetilde \ph}} \left( {\p \o \p
{\widetilde \ph}} -i{\p \o \p {\widetilde X}^+}\right)^2,
\eqn\rxiii
$$

\no and

$$
\D_0^-={\p^2 \o \p {\widetilde \ph}^2}
+{\p \o \p {\widetilde \ph}} +e^{-{\widetilde \ph}} \left( {\p^2 \o
\p {\widetilde \ph}^2}+{\p^2 \o \p {\widetilde X^{-^2}} }\right).
\eqn\rxiii
$$

\no The wave functions $\Psi^-$ and $\Psi^+$ play a role analogous to
that of the wave functions $T(r,\th)$ and ${\widetilde
T}(r,{\widetilde \th})$, which describe the propagation of momentum
and of winding modes respectively. This means that each of them will
generate a different background corresponding to the cigar and the
trumpet respectively. We introduce the variable $r$ as

$$
r=2\;{\rm ar coth} \sqrt{1+e^{-{\widetilde \ph}}}\qquad {\rm or}
\qquad {\widetilde \ph}=2\log \sinh {r\o 2},
\eqn\rxiv
$$

\no and we observe that with the above redefinition the condition
$-\infty < \phi < \infty$ holds since $0<r<\infty$ is satisfied. We
obtain, after the redefinition

$$
{\widetilde \theta} = {\widetilde X}^+ -i \log (e^{-\widetilde \ph}
+1),
\eqn\rxv
$$

\no the following form of the $L_0$ operator when acting on the
momentum modes  $\Psi^+$:

$$
L_0\Psi ^+=-{1\o k-2} \left( {\p^2 \o \p r^2}+\coth r {\p \o \p
r}+\left(
{\rm tanh}^2 {r\o 2}-{2\o k}\right) {\p^2\o \p {\widetilde \theta
}^{^2}}\right)\Psi ^+.
\eqn\rxvi
$$

\no If this operator acts on the winding modes described by $\Psi^-$,
we obtain

$$
L_0\Psi ^- =-{1\o k-2} \left( {\p^2 \o \p r^2}+\coth r{\p \o \p
r}+\left(
{\rm coth}^2 {r\o 2}-{2\o k}\right){\p^2\o \p {\theta}^{^2}}
\right)\Psi^-.
\eqn\rxvi
$$

\no This is precisely the form of the $L_0$ operators of ref. [\dvv].
We have introduced the notation ${\widetilde X}^-\equiv \th$. This
implies  that we will have the same background interpretation, i.e.
the same metric and dilaton {\wxxvi} for finite $k$ as was found in
ref. [\dvv]. The dilaton written in terms of $\widetilde \ph$ takes
the form:

$$
\Phi^+={\widetilde \ph} +\log \sqrt{ 1-{2\o k}
\left(1+e^{-{\widetilde\phi} }\right)}
\eqn\Rxvi
$$

\no and

$$
\Phi^-={\widetilde \ph}+\log\sqrt{\left(e^{-{\widetilde
\ph}}+1\right)\left(e^{-{\widetilde \ph}}+1-2/k\right)}.
\eqn\rrxvi
$$

\no From the above expression we observe that only in the limit $k
\rightarrow \infty$  (and $\widetilde \ph \rightarrow \infty$ for the
second expression) does the field $\widetilde \ph$ coincide with the
dilaton. The first expression for the dilaton coincides with the one
computed in [\bk] in the semiclassical limit.

We could now consider the possibility of changing the sign of the
black hole mass, by choosing $\epsilon=-1$ in eq.{\rimnm}.
We then get the relation

$$
{\widetilde \ph}=2\log \cosh {r\o 2}
\eqn\rrxvimnm
$$
\no In this case the role of $\widetilde X^+$ and $\widetilde X^-$ is
interchange.

In this calculation the screening charge proposed in ref. [\bk] was
an important ingredient {\rvi}.
We conclude that this operator generates the correct background, even
for finite $k$. This agreement with the background of ref. [\dvv] is
important since we are going to evaluate the correlation functions of
this model for $k=9/4$, so that a space-time interpretation for
finite $k$ is the one of relevance.
It would be interesting to derive the above metric directly from the
Lagrangian of ref. [\bk] with a similar computation as the one done
in ref. [\Tseytlin].

We would like to make one final remark concerning the range of the
fields for $\epsilon=-1$.
We define $\phi ^*={\ap\o 2} \log (M k /2) $, that indicates the
value of $\phi$ at the event horizon. The range of the field $\phi $
consists of two parts. The region $\phi^* \leq \phi < \infty$
describes the Euclidean version of region~I (outside the horizon),
while $-\infty < \phi \leq \phi^*$ is the Euclidean version of region
III (between the singularity and the horizon).

\vfill
\endpage

\chapter{\chafive}

\no As a first step to see if there exist similarities between the
black hole CFT and standard non-critical string theory, we would like
to analyse the cohomologies of the two theories.

It is well known that the critical bosonic string in two dimensions
has in its spectrum, apart from the tachyon, an infinite number of
states at higher mass levels and discrete values of the momenta. They
are the so-called discrete states
\REF\disc{D.~J.~Gross, I.~R.~Klebanov, M.~J.~Newman, ``The Two-Point
Correlation Function of the One-Dimensional Matrix Model'', {\it
Nucl.~Phys.} {\bf B350} (1991) 621; E.~Witten, ``Ground Ring of
Two-Dimensional String Theory'', {\it Nucl.~Phys.} {\bf B373} (1992)
187; A.~M.~Polyakov, ``Selftuning Fields and Resonant Correlations in
2-d Gravity'', {\it Mod.~Phys.~Lett.} {\bf A6} (1991) 635.}
[\disc]
that play an important role in the black hole CFT, as we will later
see.

\section {\secfiveone}

\no Distler and Nelson [\dn] have used the {\sl} representation
theory and the standard coset construction to calculate all the
physical states that could (in principle) occur in the black hole
CFT. It is possible that the true spectrum is a subset of the
spectrum that is allowed by representation theory. A definitive
statement on the states that occur can be achieved through the
computation of the scattering amplitudes. This will be done in
chapter 6.

Since for the quantization of the black hole theory we have two BRST
charges to characterize the physical states of the coset theory, one
could in principle consider either the cohomology of $Q^{\rm total}$
or the iterated cohomology in which the states are annihilated by
each of the charges instead of by their sum.

\midinsert
\epsfysize=3.5in
\vskip -0.5cm
\centerline{\hskip 0.0cm \epsffile{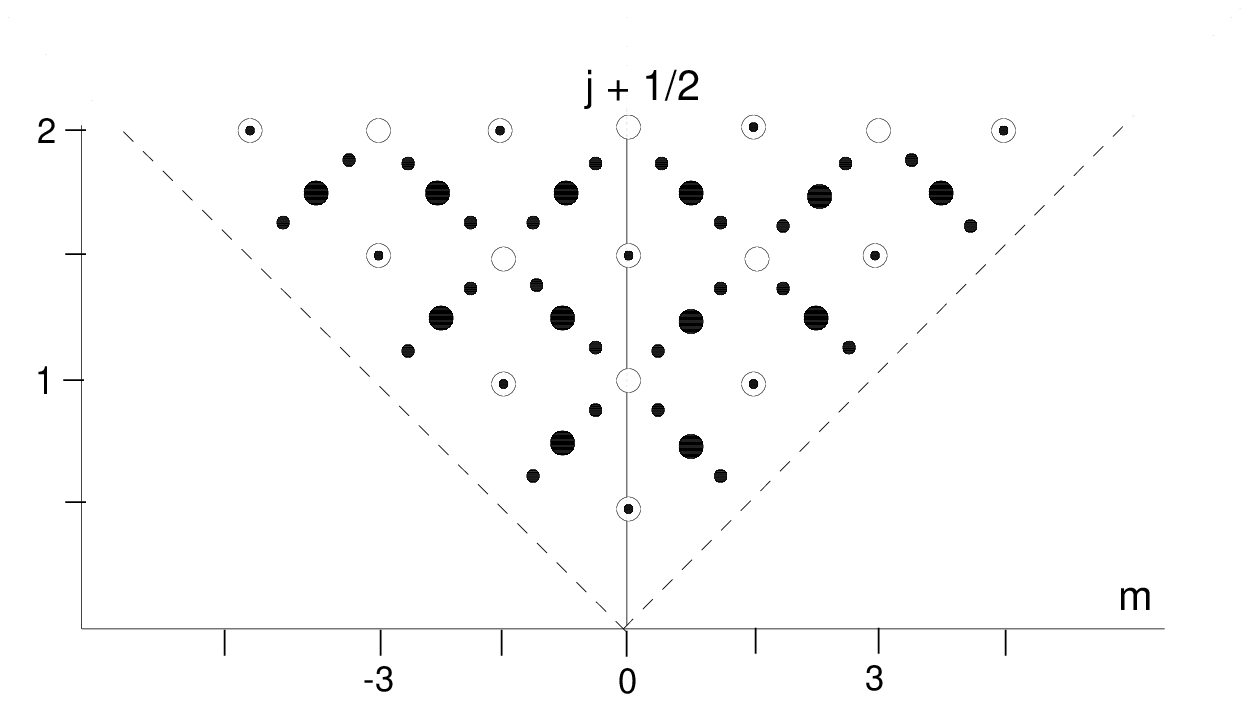}}
\vskip -1.3cm
\caption{{\bf Fig. 5}: Physical states of the Euclidean black hole:
open circle ${\bigcirc}$ ${\cal C}$, filled circle $\bullet$ ${\cal
D}$, dot ${\cdot}$ ${\widetilde {\cal D}}$, circled dot $\odot$
(double occupied). The special states satisfying $|m|=3(j+1/2)$, that
lay on the broken line are ``discrete tachyons'' that occur at zero
mass level.}
\endinsert

\no  In ref. [\dn] it is claimed that the two cohomologies agree.
In the spectrum of the Euclidean black hole there appear discrete
states besides the tachyon. This is familiar from the spectrum of
$c=1$ matter coupled to Liouville theory. These fields are no longer
Kac-Moody primaries. The (on-shell) tachyon vertex operators for
$c=1$ have the form

$$
\exp{(ip_X X)} \exp (p_\ph \ph ),
\eqn\siv
$$
\no where
$$
\pm p_X=p_\ph +\sqrt 2,
\eqn\ssv
$$
\no is fixed by the on-shell condition. Comparing this with the
on-shell condition of the coset model for $k=9/4$, we can identify
$$
p_\phi=2\sqrt{2}j  \qquad {\rm and} \qquad  p_X={2\o 3} \sqrt{2} m.
\eqn\ssvi
$$

We can compare the quantum numbers of the states that occur in $c=1$
coupled to Liouville theory with the states that may occur in the
Euclidean black hole.
The conclusion in ref. [\dn] is that the massless spectrum
corresponding to the tachyons is identical in $c=1$ and in the black
hole CFT. However, at higher mass levels there seem to appear
differences between the two theories. In addition to the $c=1$
discrete states, they obtained new discrete states which have no
counterpart in $c=1$. The discrete states classified in ref. [\dn]
are:

\Cx

\no where $r$ and $s$ are positive integers and $\cal N$ is the mass
level.

The states $\widetilde{\cal D}^{\pm}$ are the new
discrete states, while ${\cal D}^{\mp}$ appear in
$c=1$ matter coupled to Liouville theory, as well as ${\cal C}$, and
belong to the discrete and supplementary series of
{\sl}, respectively. We will later see how part of
this spectrum appears as poles in tachyon
correlation functions, as it is known to hold for $c=1$ coupled to
gravity [\dfk,\gl,\dtwo].

\midinsert
\epsfysize=3.5in
\centerline{\hskip 0.0cm \epsffile{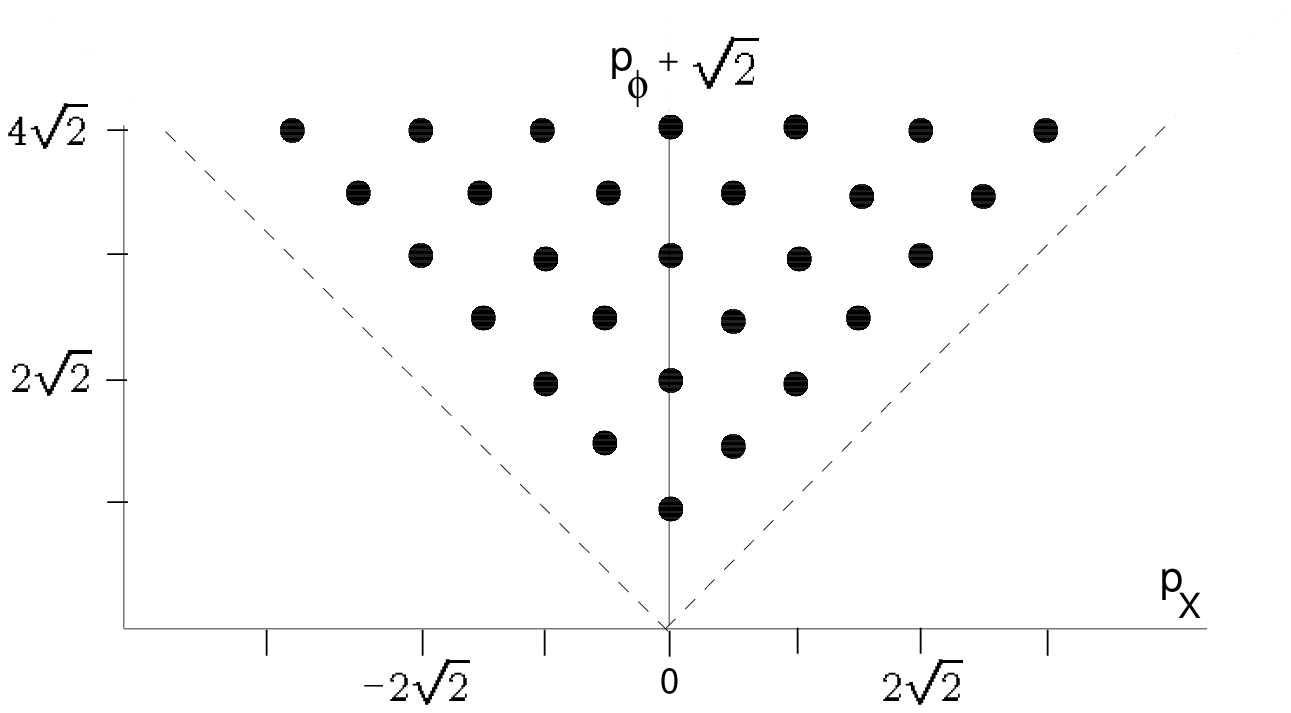}}
\vskip -1.5cm
\caption{\hskip 1.75cm {\bf Fig. 6}: Physical states of the $c=1$
non-critical string.}
\endinsert

\sec{\secfivetwo}
\no We have already seen how the tachyon operators, i.e. the
Kac-Moody primaries, look like in the free-field representation.
In this section we are going to see how the vertex operators of the
simplest discrete states that occur in the Euclidean black hole can
be represented in this approach. These states are created by acting
with negative modes of the currents on the Kac-Moody primaries and
are therefore no longer annihilated by the positive modes of them.
They satisfy the on-shell condition

$$
-{j(j+1)\over k-2}+{m^2\over k}+{\cal N}=1.
\eqn\sviii
$$

The form of the simplest discrete states has been computed in ref.
[\bk,\eky] and they have the form

$$
V_{\cal N}=\b^{\cal N} \exp\left( {2j\o \ap} \ph+ im \sqrt{2\o k}
X\right).
\eqn\svi
$$

\no The invariance under $Q^{\rm U(1)}$ implies $m={\cal N}+j$. If we
now consider states of dimension $(1,1)$, this implies that for
$k=9/4$ we get two series of solutions:

$$
\eqalign{
m&={3\o 2}{\cal N} -{3\o 8} \qquad \qquad j=-{3\o 8}+{{\cal N}\o
2},\cr
&\cr
m&={3\o 4}{\cal N}-{3\o 4} \qquad \qquad j=-{3\o 4}-{{\cal N}\o 4}
.\cr}
\eqn\svii
$$

\no The previous states are special cases of the general
classification of Distler and Nelson [\dn]. In particular we see that
the first series in eq. {\svii} realizes some of the new discrete
states. The second series describes discrete states that appear in
$c=1$ that are on the wrong branch, i.e. that satisfy $j<-1/2$.

We now consider the first examples to see explicitly if there is a
correspondence to discrete states of $c=1$.
The case ${\cal N}=1$ of the first series corresponds to the state
$(j,m)=(1/8,9/8)$. After simple computations one can show that this
state can be represented as

$$
\eqalign{
&V\left(j=1/ 8, m= 9/8\right)= \cr
&\cr
&\g^{-1}{\p \o  \p z} \exp\left({\ph\o 2\sqrt{2}} +i{3 \o 2\sqrt{2}}
X\right) +\left\{ Q^{\rm U(1)} ,b \g^{-1} \exp\left(- { \ph\o
2\sqrt{2}} +i {3 \o  2\sqrt{2} }X\right)  \right\}=\cr
&\cr
&\g^{-1}\left\{Q^{Diff},\left[b_{-1},\exp\left( {\ph\o
2\sqrt{2}}+i{3 \o 2\sqrt{2}}X\right) \right]\right\}+\left\{ Q^{\rm
U(1)} ,b \g^{-1} \exp\left(-{  \ph\o 2\sqrt{2} }+ i {3 \o 2\sqrt{2}}
X \right) \right\}.\cr}
\eqn\six
$$

\no Here $\{,\}$ and $[,]$ means the commutator and anti-commutator
respectively. We see that this new discrete state of Distler and
Nelson becomes BRST-trivial in the Wakimoto representation, since we
are left with a total derivative that decouples from the amplitudes.
It seems reasonable that the same holds for the rest of the new
discrete states. That this is indeed the case will be checked through
the computation of the scattering amplitudes. We will see that these
states do not appear as poles in the amplitudes. This is in agreement
with the result of the analysis of the free field cohomology carried
out by Eguchi et al. [\eky], which we will present in the next
section.

We now discuss the case ${\cal N}=1$ of the second series that
corresponds to $(j,m)=(-1,0)$ and represents the screening charge.
This operator was identified as the black hole mass operator in ref.
[\bk,\eky], where
the semiclassical interpretation of the background of the free-field
model has been found. In this computation one does not use the
``$L_0$-approach''. Instead of this, one identifies the action of the
free-field model {\Cvvvv} with the $\s$-model action {\zviii}. For
this purpose one has to transform the action formulated in terms of
Wakimoto coordinates to a $\s$-model form. This is equivalent to
eliminating the $\b$-$\g$ system of this description and to use only
the coordinates of $c=1$. Of course, if we are working directly at
the level of the action, i.e. in the full field theory description,
we have to be careful with the transformation of variables. However,
in the case of the representation in terms of Euler angles, Tseytlin
\REF\aat{A.~A.~Tseytlin, ``Effective Action of the Gauged WZW Model
and Exact String Solutions'', {\it Nucl.~Phys.} {\bf B399} (1993)
601.}
[\aat]
showed that a careful analysis of the quantum effective action leads
to the correct metric for finite $k$.  We will now restrict to the
semiclassical approach presented in ref. [\bk,\eky]. We start
eliminating the $\b$ contribution of the screening charge. This can
be done using the $Q^{\rm U(1)}$ charge. Using the bosonization
formula we have

$$
V=\b e^{-{2\o \ap} \ph}=-i\p v e^{iv-u-{2\o \ap }\ph}.
\eqn\ti
$$

\no After taking the antiholomorphic part into account, it is easy to
see that the interaction can be written as

$$
V\simeq e^{-u+iv}e^{-{\bar u}+i{\bar v}}\left( { \ap\o 2} \p
\ph+i\sqrt{k\o 2}\p X\right)\left( {\ap\o 2} {\bar \p} \ph+i\sqrt{k\o
2}{\bar \p} X\right) \exp \left(-{2\o \ap}\ph\right),
\eqn\tii
$$

\no where we have neglected a BRST-trivial part and we have chosen a
"-" sign  in eqn. {\mnm}. The exponential factors involve the $u,v$
dependence only through the combination $u-v$. In correlation
functions every contraction of $u$ will be cancelled by the
contraction of $v$. This means that the interaction term can be
written as

$$
V\simeq \left( { \ap\o 2} \p \ph+i\sqrt{k\o 2}\p X\right)\left(
{\ap\o 2} {\bar \p} \ph+i\sqrt{k\o 2}{\bar \p} X\right) \exp
\left(-{2\o \ap}\ph\right).
\eqn\tii
$$

\no We find that the interaction, for $k=9/4$, is the discrete state

$$
W_{1,0}^-\bar W_{1,0}^-=\p X {\bar \p} X e^{-2\sqrt{2} \ph}
\eqn\tttii
$$

\no of $c=1$, up to a total derivative. Therefore the action of the
{\slU} Euclidean black hole is identical to the 2D non-critical
string, with a different interaction. Instead of the cosmological
constant that is a tachyon operator, the interaction is a discrete
state

$$
{\cal S}={1\o 8\pi}\int \sqrt{\widehat g} {\widehat g}^{ab}(\p_a \ph
\p_b \ph+\p_a X \p _b X)-{1 \o 4\pi \ap}\int \sqrt{\widehat
g}{\widehat R}^{(2)} \ph +M \int V.
\eqn\tiii
$$

\no The parameter $M$ is free and it can be eliminated with a shift
in $\phi$. The metric that follows from the above $\s$-model is [\bk]

$$
\eqalign{
&G_{\ph \ph} =1+M {\ap^2 \o 4}\exp\left( -{2\o \ap} \ph \right)\cr
&\cr
&G_{\ph X}=i M {\ap \o 2}\sqrt{k \o 2} \exp\left( -{2\o \ap}
\ph\right)\cr
&\cr
&G_{XX}=1-M {k\o 2} \exp \left( -{2\o \ap} \ph \right). \cr}
\eqn\tttiii
$$

\no We observe that this metric is not diagonal. To diagonalize it we
introduce a new coordinate\foot{Here we have rescaled the fields
$\ph$ and $X$. The presence of the $i$ in the above metric can be
avoided with a change to Minkowski conventions $X \rightarrow iX$
[\eky].}:

$$
\vartheta=X+i f(\ph),\qquad {\rm with} \qquad f'(\ph)={1\o
e^{2\ph}-1}.
\eqn\tiv
$$

\no The parameter $M$ can be eliminated with a shift in $\phi$.
The resulting action takes the form:

$$
{\cal S}={k\o 4\pi} \int \sqrt{\widehat g}{\widehat g}^{ab}\left(
{\p_a \ph \p _b \ph\o 1-e^{-2\ph}} +\left(1-e^{-2\ph}\right)
\p_a\vartheta \p_b\vartheta\right) -{1\o 4\pi}\int \sqrt{\widehat g}
{\widehat R}^{(2)} \ph.
\eqn\tv
$$

\no After introducing the variable $r$

$$
\ph=\log \cosh r,
\eqn\tvi
$$

\no we are left with the {\slU} WZW action found by Witten {\wwwix}
in the semiclassical limit. We can do the similar considerations for
the dual $\s$-model in the semiclassical limit. We previously saw
that this dual model can be obtained from {\wwwix} with the shift $r
\rightarrow  r+i \pi /2$. One arrives at the conclusion [\bk] that it
corresponds to $M<0$, and it describes region V (behind the
singularity).

 The scalar curvature is given by [\eky]
$$
R=M\exp\left( -{\sqrt{2\o k}} \ph\right),
\eqn\tttvi
$$

\no so that the curvature singularity occurs for $\phi=-\infty$.

\sec{\secfivethree}
\no Eguchi et al. [\eky] have shown that there exists an isomorphism
between the algebraic structure of Liouville theory coupled to $c=1$
matter and the black hole CFT. To every state in $c=1$ we can
associate a state in the black hole CFT that has the same values of
$(p_{\phi},p_X)$. This isomorphism can be shown by realizing that
there exists a correspondence between the complete energy-momentum
tensors of both theories up to BRST-trivial terms. It gives a
systematic way to obtain the representation of the physical states of
the 2D black hole in terms of Wakimoto coordinates.

For $c=1$ matter coupled to Liouville, the energy-momentum tensor
(including the reparametrization ghosts) is

$$
T_{c=1}=-\h(\p\ph)^2-\sqrt{2}\p^2 \ph -\h (\p X)^2 -2b\p c-\p b c.
\eqn\si
$$

\no In analogy, the total energy-momentum tensor of the Euclidean
coset CFT can be written as

$$
T_{\rm SL(2,\R)/U(1)}=-\h(\p\ph')^2-\sqrt{2} \p^2\ph'-\h(\p
X')^2-2b'\p c-\p b' c+\left\{Q^{\rm total},-(\p \log \gamma)
B\right\}.
\eqn\sii
$$

\no Here we have used the total BRST charge $Q^{\rm total}=Q^{\rm
U(1)}+Q^{Diff}$ and we have defined the following transformation
between the variables

$$
\eqalign{
&\ph'=\ph+{1\o 2\sqrt{2}}\log \g\cr
&\cr
&X'=X+i{3\o 2\sqrt{2}} \log \g\cr
&\cr
&b'=b+B\p \log \g.\cr}
\eqn\siii
$$

\no The previous relations can be used to compute the physical states
of the black hole CFT out of the states of $c=1$ [\eky]. Of course
one has to ensure that these states are indeed BRST-invariant.  Using
the transformations {\siii} and the form of the tachyon operators of
$c=1$ {\siv}, this would imply that tachyon vertex operators in the
black hole CFT have the form

$$
\g^{-{1\o 2\sqrt{2}}(-3p_X+p_L)}\exp( ip_X X) \exp (p_L \ph),
\eqn\sv
$$

\no and this is precisely the form of the dressed Kac-Moody primaries
{\Ciii}.
A systematic analysis of the discrete states in the free-field
approach can be done [\eky] if the transformation of variables
{\siii} is used to obtain all the discrete states of the black hole
in terms of the discrete states of $c=1$.
In the $c=1$ model coupled to Liouville theory there appear operators
at certain discrete values of the momenta [\disc] that form a
$W_{\infty}$ algebra. These operators have the form

$$
W_{j,j}^+=\exp\left(i \sqrt 2 j X\right) \exp\left(\sqrt 2(j-1)
\phi\right) \qquad {\rm with} \qquad j=0,\h,1\dots,
\eqn\svzw
$$

$$
W_{j,m}^{+}=\left(\oint \exp\left( -i\sqrt{2} X(w)\right)
\right)^{j-m} W_{j,j}^{+}\qquad {\rm with} \qquad -j\leq m \leq j.
\eqn\ssvzw
$$

\no There also exist discrete states on the wrong branch

$$
W_{j,j}^-=\exp\left(i \sqrt 2 j X\right) \exp\left(-\sqrt
2\left(j+1\right) \phi\right)\qquad {\rm with} \qquad j=0,\h,1\dots,
\eqn\svzwz
$$

$$
W_{j,m}^{+}=\left(\oint \exp\left( -i\sqrt{2} X(w)\right)
\right)^{j-m} W_{j,j}^{+}\qquad {\rm with} \qquad -j\leq m \leq j.
\eqn\ssvzwz
$$

\no The corresponding operators in the black hole are obtained using
the previous substitutions. Some examples are collected in
ref.[\eky].

In this way it is possible to show the appearance of the $W_{\infty}$
algebra and the ground ring elements that have been discovered in
standard non-critical string theory
\REF\zwiebach{E.~Witten and B.~Zwiebach, ``Algebraic Structures and
Differential Geometry in 2D String Theory'', {\it Nucl.~Phys.} {\bf
B377} (1992) 55.} [\disc,\zwiebach] in the context of the black hole.
The states obtained are all BRST-invariant [\eky].
All the physical states computed in ref. [\eky] can be formulated
entirely in the $c=1$ language i.e. in terms of the coordinates
$(\phi, X)$. The $\beta$-$\g$ contribution can be dropped out taking
into account the BRST constraints and the bosonization formula for
$\g$.

\endpage

\chapter{\chasix}

\no In this chapter we are going to compute the scattering amplitudes
of tachyons in the black hole background and we will analyse the
connection to the correlation functions of tachyon operators in
standard non-critical string theory.

\sec{\secsixoneone}

We are interested in the computation of a scattering amplitude

$$
{\cal A}^{j_1.\,.\,.\,.\,j_N}_{m_1.\,.\,. m_N}
=\langle {\cal V}_{j_1\; m_1} \dots {\cal V}_{j_N\;
m_N}\rangle,
\eqn\svi
$$

\no where the average has to be performed with respect to the action
{\Cvvvv}. We are going to consider spinless fields, i.e. winding
modes so that we will  use a short-hand notation where the
antiholomorphic dependence with $m={\bar m}$ will be understood. The
tachyon vertex operators are described by the conformal invariant
expression

$$
{\cal V}_{j \; m}=\int \g^{j-m}{\bar\g}^{j-m} e^{{2\over \ap} j\phi}
e^{im\sqrt{{2\over k}}X} d^2z
\eqn\svii
$$

\no that satisfy the on-shell condition {\Cixa}. The $N$-point
correlation function of these vertex operators can be written as the
following path integral

$$
\int \prod_{i=1}^N d^2 z_i {1\o \rm Vol_{SL(2,{\tenpoint{\IC}})}}
\langle e^{im \sqrt{ 2\o k} X(z_1, {\bar z_1})} \dots\rangle_X
\langle e^{{2\o \ap } j \ph(z_1, {\bar z_1})} \dots\rangle_{\phi}
\langle \g^{j_1-m_1}(z_1) \dots \rangle
\langle {\bar \g}^{j_1-m_1}({\bar z_1})\dots \rangle_{\beta\gamma}.
\eqn\ssviii
$$

\no In order to render the amplitude finite we have to divide out the
volume of
the group ${\rm SL(2,\IC)}$

$$
{\rm Vol_{SL(2,{\tenpoint{\IC}})}}
=\int { d^2 z_1 d^2 z_2 d^2 z_3 \o |z_1-z_2|^2|z_1-z_3|^2
|z_2-z_3|^2}.
\eqn\six
$$

\no Unfortunately, the computation of the above path integral is not
an easy task since the involved field $\phi$ is not free and
perturbation theory in $M$ does not make sense.
However, there is a clever way to circumvent this problem. After
integrating out the zero modes of the fields we will see that we  are
left with the correlation functions of a free theory, and the
interaction plays the role of new insertions.

\sec{\secsixonetwo}

We will now carry out the integration of the zero modes of the fields
explicitly
\REF\gtw{A.~Gupta, S.~Trivedi and M.~Wise, ``Random Surfaces in
Conformal Gauge'', {\it Nucl. Phys.} {\bf B340} (1990) 475.}
[\gtw,\gl].
 We start with the zero mode of $\phi$. We introduce the notation
$\phi ={\widetilde \phi}+\phi_0$, where $\phi_0$ denotes the zero
mode of the field. Vertex operators depending only on ${\widetilde
\phi}$ will be denoted by a tilde. Using the identity

$$
\exp\left(-M \int \b {\bar \b} e^{ -{2 \o \ap} \ph }\right)=
\int_0^\infty d A \delta \left( \int \b {\bar \b} e^{ -{2 \o \ap} \ph
} -A \right) e^{-M A}
\eqn\sx
$$

\no we can write the $\phi$-path integral in the following way

$$\eqalign{
\Biggl\langle \prod_{i=1}^N {\cal V}_{{j_i}\;{m_i}}
\Biggr\rangle=
\Biggl\langle \prod_{i=1}^N {\widetilde {\cal V}}_{{j_i}\;{m_i}}
\Biggr\rangle
&\int_0^\infty d A \delta \left( \int \b {\bar \b} e^{ -{2 \o \ap}
\ph } -A \right) e^{-M A} \cr
&\cr
\times &\int d \ph_0 \exp\left({{\ph_0\o \ap} \left( 2\sum j_i +{1\o
\pi} \int R^{(2)}\right)}\right) .\cr}
\eqn\sxi
$$

\no We introduce the notation ${\tilde A}$ for the surface integral
of the screening operator depending only on $\tilde \phi$ and, using
with some standard formulas for the $\delta$-function, we find the
expression

$$
\delta\left(\int \b {\bar \b} e^{-{2\o \ap} \ph}-A\right)=-{\ap \o
2A} \delta\left( \ph_0-{\ap \o 2} \log \left({{\widetilde A}\o
A}\right) \right).
$$

\no To integrate the zero-mode we take into account the Gauss-Bonnet
theorem

$$
{1\o 2\pi} \int R^{(2)} d^2 z=(1-g),
\eqn\sxii
$$

\no where $g$ is the genus of the Riemann surface. Inserting these
conditions into the path integral we find the final result

\diii
where
\di

\noindent We have absorbed a factor $-\ap/2$
into the definition of the path integral and we have used the
identity

$$
M^s\Gamma(-s)=\int_0^\infty dA A^{-s-1}e^{-MA}.
\eqn\ddi
$$

\no From the last expression we see that the amplitude is divergent
if $s$ is a positive integer. This divergence comes from the
integration over the volume of the $\phi$ coordinate and can be
regularized using the formula [\dfk]

$$
\mu^s \G(-s)|_{s\rightarrow 0}\rightarrow \int_{\epsilon}^\infty { dA
\o A} e^{-\mu A} \rightarrow \log( \mu \epsilon).
\eqn\ddii
$$

\no This divergence has a nice space-time interpretation: the
incoming particles are in resonance with $s$ particles that form the
wall against which they scatter.

{}From the previous considerations it becomes clear that the
free-field approach is suitable to compute
the amplitudes that obey a special energy sum
rule where the number of screenings $s$ is a
positive integer. Any of the desired correlators can be determined
indirectly by an analytic continuation in $s$ as has been done with a
tachyon background in standard Liouville theory [\gl,\dfk,\dtwo].
{}From the zero mode of $X$ we obtain with similar methods the
conservation law

\dib

\noindent The number of zero modes of the $\b$-$\g$
system, determined by the Riemann-Roch theorem,
leads for spherical topologies to the condition:

\dii

\no This constraint is equivalent to {\di} and {\dib}
for states of the form ${\Ciii}$.

\sec{\secsixonethree}

We would like to study first the correlation functions that satisfy
the charge conservation and therefore do not need screening charges,
i.e. that satisfy $s=0$. These correlators have been analysed in ref.
[\bk,\dfk].
It is clear that since we have no interaction for $\beta$ and
$\gamma$, this system does not contribute to the amplitudes;
contractions coming from the $\g$'s of the vertex operators {\svii}
are equal to one. We will now see how the intermediate on-shell
states produce poles in the amplitudes that are located at the
positions where the discrete states of $c=1$ occur.
We introduce the notation

$$
p_i=\sqrt{ 2\o k}\; m_i\qquad {\rm and} \qquad \b_i={2\o \ap} j_i,
\eqn\ppii
$$

\no in terms of which the tachyon operators {\svii} take the form

$$
{\cal V}_p=\exp\left( ipX+\b(p) \phi\right).
\eqn\pppii
$$

\no The $\g$ part of the vertex operator can be dropped, since
contractions of $\g's$ do not give any contribution to the scattering
amplitudes if $s=0$. In the above notation the on-shell condition
{\sviii} takes the form \foot{The parameter $\cal N$ appearing in
this equation can be interpreted as the mass level, because the
energy has the form $E=\b+Q/2$ so that this equation is equivalent to
the equation $E^2=p^2+m^2$.}

$$
-{\b \o 2} \left( \b+Q \right) +{p^2 \o 2}+{\cal N} =1,
\eqn\pppiii
$$

\no where $Q=2/\ap$ and $Q=2\sqrt 2$ for $k=9/4$. The solution of the
above equation has two branches

$$
\b=-\sqrt{2}\pm \sqrt{p^2+2{\cal N}}.
\eqn\ppppiiii
$$

\no Operators with $\b>-\sqrt 2$ are called Seiberg states or
operators on the right branch, while operators with $\b<-\sqrt 2$ are
called anti-Seiberg states or operators on the wrong branch. When
computing correlation functions of tachyon operators, we are going to
consider operators on the right branch

$$
\b+\sqrt{2}=|p|,
\eqn\pppiv
$$

\no in order to compare our results with the correlation functions of
standard non-critical string theory.  However, as we saw in section
5.2. the anti-Seiberg states play an important role in the physics of
the 2D-black hole, since the black hole mass operator is a discrete
state of this type. Therefore,  in general, a correlation functions
will include also operators of anti-Seiberg type that come from the
screening charge.

For $N$ tachyon operators the $s=0$ amplitude is given by the
Shapiro-Virasoro integral

$$
{\cal A }_{s=0}(p_1,p_2, \dots, p_N)= \int \prod_{i=4}^N d^2 z_i |
z_i| ^{2( p_1 p_i -\b_1 \b_i)} |1-z_i| ^{2( p_3 p_i -\b_3 \b_i)}
\prod_{4\leq i < j\leq N} |z_i-z_j| ^{2( p_i p_j -\b_i \b_j)}.
\eqn\ppi
$$

\no In general, no closed expression for eq. {\ppi} is known. The
basic problem is the complicated pole structure of this integral;
there are many channels in which poles appear. To analyse them we
have to consider the region of the moduli integrals in eq. {\ppi}
where some of the $z_i$ approach each other
\REF\st{N.~Sakai and Y.~Tanii, ``Operator Product Expansion and
Topological States in $c=1$ Matter Coupled to 2D Gravity'', {\it
Prog.~Theor.~Phys.~Supp.} {\bf 110} (1992) 117.}
[\dfk,\st]. So for example, to analyse the limit $z_4, z_5,
\dots,z_{n+2} \rightarrow 0$ it is convenient to define the variables

$$
z_4 =\e \qquad \qquad z_5=\e y_5, \dots, z_{n+2}=\e y_{n+2},
\eqn\ppiii
$$

\no and to consider the contribution of the region $|\e|<<1$ to the
integral {\ppi}. There appear an infinite number of poles and the
residues of these poles are related to correlation functions of on
shell intermediate string states. In two space-time dimensions, it
turns out that most of the residues of the above poles vanish, so
that in this case one is able to obtain a closed expression for the
amplitude. This is different from the situation in higher dimensions.
This is why it is not possible to compute the above amplitude for $D
\neq 2$. Plugging eq. {\ppiii} into eq. {\ppi} we find explicit
expressions for these poles. For $D=2$ the first pole, for example,
takes the form

$$
{\cal A}(p_1, \dots p_N) \simeq
{\langle {\cal V}_{p_1} { \cal V}_{p_4} \dots {\cal V}_{p_{n+2}}
{\cal V}_{-{\widetilde p}}\rangle
 \langle {\cal V}_{{\widetilde p}} { \cal V}_{p_2}{ \cal
V}_{p_3}{\cal V}_{p_{n+3}} \dots  {\cal V}_{p_N} \rangle
\o
\left( \sqrt 2 +\sum_i \b_i \right)^2 -\left( \sum_{i} p_i\right)^2},
\eqn\ppiv
$$

\no where ${\widetilde p}=\sum_{i} p_i$ (where $i=1,4,5,6,\dots,
n+2$, because of momentum conservation). The pole in this amplitude
indicates the appearance of an on shell intermediate tachyon with
$\b=\sum_i \b_i$ and  $p=\sum_{i} p_i$. It is simple to obtain an
analog expression for the higher poles that are related to on shell
intermediate states at higher mass levels.

\midinsert
\epsfysize=1.7in
\vskip 0.5cm
\centerline{\hskip 0.0cm \epsffile{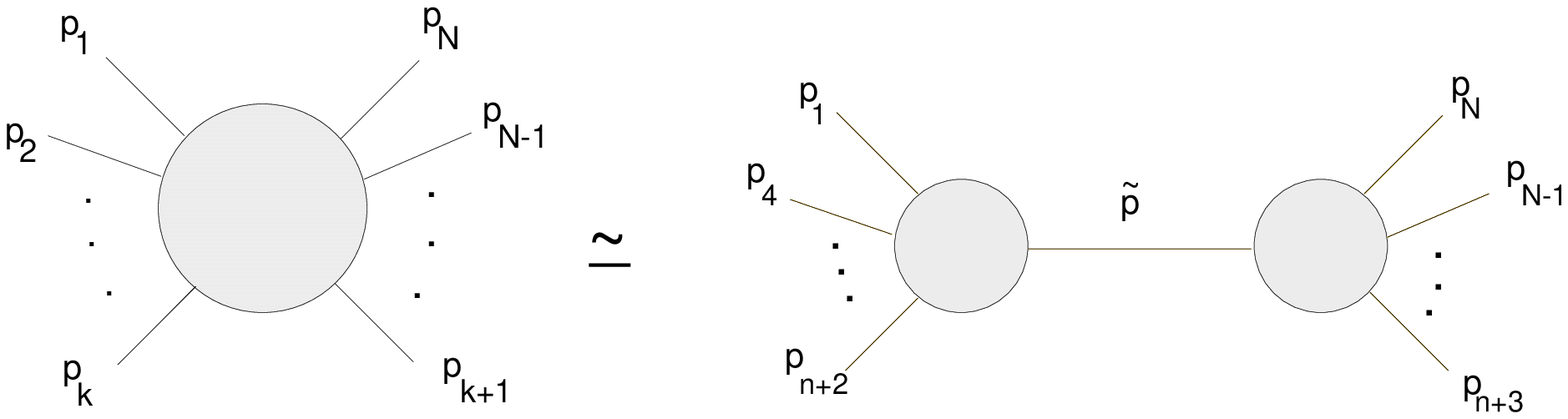}}
\caption{{\bf Fig. 7}: The factorization of the $N$-point tachyon
amplitude by the OPE of the operators $4,\dots, n+2$. The
intermediate resonance particle has momentum $\widetilde p$.}
\vskip 0.5cm
\endinsert

\no A closed expression for these amplitudes has been computed in
ref. [\dfk,\gl,\dtwo,\st]. Since the amplitudes are non-analytic in
the momenta {\pppiv} the result is different in the different
chirality configurations. If we consider a kinematical configuration
where all tachyons except one have the same chirality, e.g.
$(-,+,+,\dots,+)$, then the momentum conservation for this
kinematical configuration reads

$$
p_1+p_2+\dots+p_N=0
\eqn\ppvi
$$

\no and the energy conservation is

$$
-p_1+p_2+\dots+p_N=\sqrt2 (N-2)
\eqn\ppvii
$$

\no This fixes $(p_1,\b_1)$ as a function of $N$

$$
p_1={2-N\o \sqrt2} \qquad \qquad \b_1={N-4\o \sqrt 2}
\eqn\ppviii
$$

\no We are now going to see that while $p_1$ is fixed by the
kinematical constraints {\ppviii}, the amplitude exhibits
singularities as a function of the other momenta $p_2,p_3,\dots,p_N$

$$
p_i={\nu +1\o \sqrt2}\qquad {\rm with} \qquad i=2,\dots,N \qquad {\rm
and}\qquad \nu =0,1,\dots
\eqn\ppix
$$

\no but it has no singularities in combinations of the momenta.

To be more explicit, we consider the $s=0$ four-point amplitude of
tachyons

$$
{\cal A}_{s=0}(p_1, p_2 ,p_3,p_4)=\int d^2 \eta
|\eta|^{2{ \bf p_1} {\bf  p_4} }|1-\eta|^{2{ \bf p_3} {\bf p_4}},
\eqn\ppx
$$

\no where we have used the notations $\eta=z_{13} z_{24}/z_{12}
z_{34}$, ${\bf p}=(p,-i \b)$
and we have performed conformal transformations, to fix the positions
of three vertices. The volume of ${\rm SL(2,\IC)}$ has been dropped.
The result of this integral can be computed exactly. For on-shell
states it can be written in the form

$$
{\cal A}_{s=0}(p_1, p_2 ,p_3,p_4)=\pi \D( 1+ {\bf p_1} {\bf p_4})\D(
1+ {\bf p_2} {\bf  p_4})\D( 1+ {\bf p_3}{\bf  p_4}),
\eqn\ppxi
$$

\no where $\D(x)=\G(x)/\G(1-x)$. We see that this amplitude has an
infinite set of  poles where

$$
{\bf p_i} {\bf p_4} =-\h \left( \b_i +\b_4+{Q\o 2} \right)^2+\h
\left(p_i+p_4\right)^2-1=-{\cal N}-1,
\eqn\ppxii
$$

\no for $i=1$, $2$ or $3$. These are precisely the on-shell
conditions for discrete states at level {\cal N}. If we choose the
chirality configuration $(-,+,+,+)$ we obtain

$$
\eqalign{
&j_i=-\h+{2\o 3} m_i \qquad {\rm for } \qquad i=2,3,4 \cr
&\cr
&j_1=-\h-{2\o 3} m_1  \cr}
\eqn\qvii
$$

\no After using the conservation laws and the condition $s=0$, we
have

$$
(j_1,m_1)=(0,-3/2).
\eqn\qqqvii
$$

\no The amplitude takes the form

$$
{\widetilde {\cal A}}^{j_1\;\;\;j_2\;\;\;j_3\;\;\;
j_4}_{m_1\;m_2\;m_3\; m_4}
=\pi \D \left(-4j_2-1\right)\D \left(-4j_3-1\right)\D
\left(-4j_4-1\right).
\eqn\qviii
$$

\no As can be seen from the above formula the poles appear for
$j=(\nu -1)/4$. These are precisely the $c=1$ discrete states, in the
classification {\Cx}.

We can obtain these poles as singularities from the OPE of two vertex
operators that collide at one point.
The short-distance singularities corresponding to $\eta \rightarrow
0$, i.e. $z_4 \rightarrow z_1$, appear if we expand the integrand of
eq. {\ppx} around $\eta=0$. The result is

$$
{\cal A}_{s=0}(p_1, p_2 ,p_3,p_4)
\approx
\int_{|\eta|\leq \varepsilon} d^2 \eta |\eta|^{2{\bf p_1}{\bf
p_4}}\left|
\sum_{\nu=0}^\infty {\G (1+{\bf p_3}{\bf  p_4} ) \o \nu! \G(1+{\bf
p_3}{\bf  p_4}-\nu)}(-\eta)^\nu \right|^2.
\eqn\qiv
$$

\no This integral can be evaluated by transforming to polar
coordinates. The result is

$$
{\cal A}_{s=0}(p_1, p_2 ,p_3,p_4)
\approx
\sum_{\nu=0}^\infty {\pi \varepsilon^{2(\nu +1 +{\bf p_1}{\bf  p_4})}
\o \nu +1+{\bf p_1}{\bf  p_4}}
\left( {\G(1+{\bf p_3}{\bf  p_4}) \o \nu! \G(1+{\bf p_3}{\bf
p_4}-\nu)} \right)^2.
\eqn\qv
$$

\no If ${\cal V}_{p_3}$ is a generic tachyon, we see that the above
amplitude has poles as a function of ${\bf p_1}{\bf  p_4}$. These
poles come from the contribution of operators in the OPE of ${\cal
V}_{p_1}$ and ${\cal V}_{p_4}$, that satisfy eq. {\ppxii} for $i=1$.
The values of the external momenta ${\bf p_1}$ and ${\bf p_2}$ can be
obtained from the kinematic relations. Choosing the chiralities of
the four states as $(-,+,+,+)$ we obtain the relations

$$
\eqalign{
&{\bf p_1}=\left(-\sqrt{2},0\right)\qquad \qquad {\bf p_4}=\left({\nu
+1 \o \sqrt{2}},-i{\nu -1\o \sqrt{2}}\right) \cr
&\cr
&{\bf p_j}=\left(p_j,-i ( \sqrt{2} +p_j) \right) \qquad {\rm for}
\qquad i=2,3.\cr }
\eqn\qqvi
$$

\no We observe that the pole structure of the exact result {\ppxii}
is correctly reproduced by the poles that appear in the short
distance singularities of two vertex operators [\st]. The level of
the intermediate state is equal to $\nu$.

The above considerations can be generalized to $N$-point tachyon
correlators without screening charges. The amplitudes again factorize
in leg factors that have poles at all the discrete states belonging
to the BRST cohomology of $c=1$. Taking into account some symmetry
properties of the correlator and the high energy behavior, the result
for the $N$-point amplitude for the chirality configuration
$(-,+,\dots,+)$ is

$$
{\cal A}(p_1, \dots p_N)=\G(N-1)\prod_{i=1}^N \D\left({\b_i^2\o 2}
-{ p_i^2\o 2}\right)
\eqn\ppv
$$

\no The amplitudes that contain two or more states in each chirality
class vanish [\dfk,\st].
We conclude [\bk]: The scattering amplitudes in the black hole
background that satisfy $s=0$ have the same form as tachyon
scattering amplitudes that satisfy $s=0$ in standard non-critical
string theory {\ppv}.
In the next sections we are going to analyse the amplitudes with $s$
different from zero [\becker,\wir] to see whether we find a
correspondence with the amplitudes of standard non-critical string
theory. We will start by considering
the three-point function.

\sec{Three-Point Function with One
Highest-Weight State}

It turns out that the simplest way to address a
general three-point tachyon amplitude is to take
one state belonging to the discrete representation
of \sl.
We choose in this section one of the tachyons as a
highest-weight state, for example $j_1=m_1$. We
will see later that an arbitrary three-point
function, where no restriction is made to the
representation to which the tachyons belong, can be expressed as a
function of this one.
The reason for this is that the {\sl}
Clebsch-Gordan coefficients can be analytically
continued from one representation of {\sl} to the
others [\bih]. For the time being, the level of the
Kac-Moody algebra is still arbitrary and we will
need this restriction only when we want to make the
analytic continuation in the number of screenings
at the end.

Using SL(2,\IC) transformations on the integrand,
we can fix the three-tachyon vertex operators at
$(z_1,z_2,z_3)=(0,1,\infty)$\foot{We will drop the
zero mode in the next formulas.}:

\hskip 0.5cm

\div

\noindent This correlator has the following form
after bosonizing the $\b$-$\g$ system:

\dvi
\noindent where
\dv

\noindent The derivatives of the above expression
come from the bosonization of $\b$ {\Bxxiv}. We can
show that the following identity holds:

\dvii

\noindent The proof uses the definition of the
Vandermonde:

\dviii

\noindent where the sum goes over all permutations
of the indices. Inserting this expression for the
Vandermonde in the definition of ${\cal P}$ leads,
after a simple calculation, to eq. {\dvii}. We
introduce the notation $\rho=-2/\ap^2=-1/(k-2)$. The complete
expression
for the amplitude is:

\dix

\noindent We have used the identity:

\dxi

\noindent which holds for integer $s$. The
remaining integral:

\dx

\noindent can be solved using the Dotsenko-Fateev formula (B.9) in
ref. [\dfa].
A careful treatment of the regularisation for the
case $k=9/4$ is needed. We will discuss this
solution and the analytic continuation to arbitrary
$s$ later. From the above simple result, we can
already see that {\sl} fusion rules appear (see for
example the appendix of ref. [\cl]). If the second
tachyon is also in the highest-weight module, i.e.
$m_2=j_2-{\rm \IN}$, then the result vanishes
unless the conjugate of $j_3$ is also in the
discrete representation i.e. $M=J-{\rm \IN}$,
where $(J,M)=(-1-j_3,-m_3)$. We will now see how
this generalizes for an arbitrary three-point
correlator, where a proportionality to a
Clebsch-Gordan coefficient appears.

\sec{Three-Point Function of Arbitrary Tachyons}

After getting an expression for the three-point
function containing one highest-weight state and
two generic tachyons, we would like to see how we
can obtain the amplitude of three generic tachyons.
Acting with the lowering operator $J_0^-=\oint
J^-(z) dz$ we compute the amplitude $ {\bf {\cal
A}}^{j_1\;\;\;\;\;\;\;j_2\;\;j_3}_{j_1-k_1\;
m_2\;m_3}$, where the holomorphic $m_1=j_1-k_1$
dependence has been changed by an integer $k_1$.
We will make an analytic continuation in $k_1$ to
non-integer values, while for the time being $s$
will be taken as an integer. Our computation shows
that the general three-point function of (not
necessarily on-shell) tachyons has the form:

\dxvi

\noindent where ${\cal C}$ is essentially the {\sl}
Clebsch-Gordan coefficient, whose expression we now
calculate. The square takes the antiholomorphic
${\bar m}$ dependence into account. We will use the
Baker-Campbell-Hausdorff formula:

\dxiia

\noindent where we have defined

\dxiib

\noindent The lowering operator $J_0^-$ acts on the
holomorphic part of the vertex operators:

\dxiic

\noindent We are going to use the fact that $J_0^-$
commutes with the screening charge ${\cal Q}$,
which is actually only true up to a total
derivative. The
surface terms that appear are discussed in section~7.6. It is shown
that they can be neglected in a
particular region of $(j_i,m_i)$, and the other
regions can be obtained by analytic continuation
\REF\gs{\rgs}[\gs]. With the above formulas and the
fact that $J_0^-$ annihilates the vacuum, we get an
identity for the general amplitude as a function of
that with one highest-weight state, if we
identify in powers of $\a$ the r.h.s. and l.h.s.
of:

\dxiii

\vskip .5cm

\noindent Taking into account the antiholomorphic
${\bar m}$ dependence and formula {\dix} we get:

\dxvib

\vskip 0.5cm

\noindent Our aim is to give up the condition that
$k_1$ is an integer. First we notice that the above
sum can be extended to $\infty$ and written in
terms of the generalized hypergeometric function
$_3F_2$ (A.5), which has a definition in terms of
the Pochhammer double-loop contour integral that
possesses a unique analytic continuation to the
whole complex plane of all its indices [\bih].
Comparing eq. {\dxvi} and eq. {\dxvib}:

\dxvii

\vskip 0.5cm

\noindent The appearance of the generalized
hypergeometric function in our result is natural,
since this function is always present in the theory
of {\sl} Clebsch-Gordan coefficients [\bih]. It can
be expanded in $s$, which will be useful to check
the analytic continuation in $k_1$ in simple
examples, as we do in section~7.6. In general
$_3F_2$ has a complicated expansion as a sum of
$\G$ functions. Fortunately, for on-shell tachyons,
which is the case we are interested in, we have a
simple result.

We will consider three tachyons on the right branch
satisfying $j\geq -1/2$.
To fulfil the $m$ conservation law {\dib}, we take $m_2\leq 0$
without loss of generality. If we
choose $k=9/4$ the on-shell conditions are {\Cixb}:

\dxxi

\noindent This fixes $j_2$ as a function of the
screening,

\dxxia

\noindent With these kinematic relations it is easy
to check that the generalized hypergeometric
function is well-poised and we can apply Dixon's
theorem \REF\eder{\reder}[\eder] to simplify
$_3F_2$. With (A.6) it is straightforward to
obtain:

\dxxic

\hskip 0.40cm

\noindent We now have to evaluate the remaining
part of the amplitude {\dxvi}.
The integral ${\cal I}(j_i,k)$ can be solved using
the Dotsenko-Fateev (B.9) formula [\dfa]:

\dxix

\noindent where
\dxx

\vskip 0.5cm

\noindent This result holds for arbitrary level $k$
but integer screenings, so that it has to be
transformed in order to obtain an expression valid
for a non-integer $s$. This analytic continuation
will be done \`a la Di Francesco and Kutasov
[\dfk], using the on-shell condition and taking the
kinematics into account. From the above definition
we notice that ${\cal Y}_2$ has dangerous
singularities for $k=9/4$, i.e. $\rho=-4$, while
${\cal Y}_{13}$ is well defined since $j_1$ and
$j_3$ could be chosen arbitrarily.

We first consider ${\cal Y}_{13}$. Choosing
$j_1,j_3\notin \IZ /8$ and the kinematic relations
for $\rho=-4$, we obtain:

\dxxiv

\vskip 0.5cm

The product ${\cal Y}_2$ is more subtle because we
have to find a regularisation that preserves the
symmetries of the theory (see [\ms] for considerations related to
this subject). In standard non-critical string theory this is
achieved with the introduction of a background charge for the field
$X$ [\dfk]. However, correlation functions involving discrete states
are more delicate and the issues concerned with the regularisation
are much more tricky (see for example
\REF\mli{M. Li, ``Correlators of Special States in $c=1$ Liouville
Theory'', { \it Nucl.~Phys.} {\bf B382} (1992) 242.}
\REF\jlfbar{J.~L.~F.~Barbon, ``Perturbing the Ground Ring of 2D
String Theory'', {\it Int.~J.~Mod.~Phys.} {\bf A7} (1992) 7579.}
[\mli,\jlfbar]).

We will shift
the level $k$ away from the critical value and
introduce a small parameter $\e\rightarrow 0$, which can be set to
zero for ${\cal Y}_{13}$. This can be
done by setting $\rho=-4+16\e$, i.e. $k=9/4+\e$,
and taking the limit $\e\rightarrow 0$. With this
modification of the level, the total central charge
will be of order $\varepsilon$ and the Liouville
field $e^{\g \varphi_L}$ has to be taken into
account in order to get an anomaly-free theory. For
the on-shell condition we make the general ansatz:

\dxxii

\no which in the limit $\varepsilon
\rightarrow 0$ reduces our previous result.
Here $r(j,\g)$ could be, in principle, an arbitrary
function of the $j$'s and the Liouville dressing
$\g$. With the kinematics {\dxxii}, we get:
\dxxv

\no with ${\widetilde
r}=r(j_1,\g_1)-r(j_2,\g_2)+r(j_3,\g_3)$.
After simple transformations we obtain:

\vskip 0.2cm
\dxxvi
\vskip 0.5cm

\no In the above formula there appears a
multiplicative factor:

\dxxvia
\vskip 0.3cm

\no which comes from the regularisation. We have no
further constraint on the parameter $\widetilde{r}$
that appears in the above expression. The best we
can do is to fix it by physical arguments. Choosing
$\widetilde{r}=4$ will imply that the three-point
function factorizes in leg factors and the
four-point function will have obvious symmetry
properties. This imposes strong constraints on
$\widetilde{r}$. If we set
$\widetilde{r}=4$ the contribution of the
renormalization factor is one. We can gain more information about
this renormalization by considering other correlation functions, as
we will later do.

Our result for the on-shell tachyon three-point
amplitude is obtained from eqs. {\dxvi}, {\dxix},
{\dxxiv} and {\dxxvi}:
\dxxviia
\hskip 1cm

\no which can be transformed finally
to\foot{We absorb a factor of two in the definition
of the path integral.}:
\dxxviii

\no where $\widetilde{M}=-\pi
M\D(-\rho)\rho^{-2}$.

We can translate our notation into the one used in
$c=1$ where the $\beta$-$\gamma$ system is eliminated. The
vertex operators can be written as:

\dxxviiia
\vskip 0.3cm

\noindent Here $\pm$ denotes the tachyon vertex
operators on the right (wrong) branch, which
represent the incoming (outgoing) wave at infinity
\REF\eguchi{\re}[\eguchi]. The three-point function of on-shell
tachyons
in the black hole background then takes the form:

\dxxviiib

\hskip 0.1cm

\noindent This can be compared with the $c=1$
three-point function, with tachyonic background
[\dfk]:
\dxxviiic

\noindent There appear several remarkable features
in our result:
\item{\triangleright} In order to get finite correlation
functions, the parameter $M$ has been infinitely
renormalized, as is done for the cosmological
constant $\mu$ in Liouville theory. Here it is known to be
equivalent to the replacement
$e^{-\sqrt{2}\varphi}\rightarrow \varphi
e^{-\sqrt{2}\varphi}$, which may have interesting physical
consequences. Perhaps this will be the
case for the black hole model as well.

\item{\triangleright} From the zero mode integration of
the $N$-point function in both theories, we see
that the number of screenings for the black hole is
half of that in the $c=1$ model.

\item{\triangleright} The amplitude can be factorized in
leg poles, which (with this normalization) have
resonance poles where the $c=1$ discrete states are
placed. The new discrete states of Distler and
Nelson do not appear. The explanation of this is
that these extra states are BRST-trivial in the
Wakimoto representation. This has been shown by
Bershadsky and Kutasov for the first examples,
as we saw in section~5.2 [\bk].

\section{Illustrative Example}

\noindent As an illustrative example we will
consider in more detail the three-point function
with one screening.  In this case we have:
\hi
\noindent and the correlator is given by:

\hii
\vskip 0.5cm
\noindent To evaluate this integral directly,
without any restriction on the $m$ dependence of
the three vertex operators, we have to use partial
integration. We get:

\hiii
\vskip 0.5cm
\noindent where ${\cal B}(j_1,j_2)$ is the part
coming from the boundaries of the region of
integration. It is proportional to:
\hiiia
\vskip 0.3cm
\noindent This integral can be evaluated using (see
appendix A of ref. \REF\moore{\rmoore} [\moore])

\hiiib
\vskip 0.3cm

\noindent We obtain:

\hiiic

\noindent where the integral is around a small
circle of radius $\e$ around 1. The contribution of
the surface term is zero for $j_2>0$, finite for
$j_2=0$, and diverges for $j_2<0$. As argued by Green and Seiberg
[\gs], a finite contact
term must be added in the case where the boundary terms are
finite and an infinite term if they diverge in
order to render the amplitude analytic. These
contact terms can be avoided by calculating
amplitudes in an appropriate kinematic
configuration, where the contact terms are not
needed, and then analytically continuing to the
desired kinematics.
For $s>1$, our argumentation will be the same as
for $s=1$, and we will restrict the values of $j$
to the regions where the boundary terms vanish.
This also means that we will restrict to the
kinematic regions where $J_0^-$ commutes with the
screening charge. We can compare this result with
the one, following from eq. {\dxvi}, which is based on
the analytic continuation in $k_1=j_1-m_1$ to
non-integer values. We obtain ${\cal C}$ for
integer screenings by expanding eq. {\dxvii}:

\dxviii

\noindent which for $s=1$ gives:

\hva

\noindent With eq. {\dxvi} and eq. {\dx} the amplitude
becomes eq. {\xxiii}, where ${\cal B}(j_1,j_2)=0$. With
this explicit example, one can already see that the
analytic continuation in $k_1$ is correct. The
integral can be solved using (A.7) for $m=1$ and
the result is eq. {\dxxviii}.

In the next section we will calculate the two-point
function to see whether similar characteristics appear. If we
have $c\leq 1$ matter coupled to Liouville theory
the simplest way to construct a two-point function
of on-shell tachyons is to use the three-point
function. One of the operators is then set to be
the dressed identity and this is the derivative
with respect to the cosmological constant $\mu$ of
the two-point function [\dfk]. In this case the
situation is different because the interaction is
not a tachyon but a discrete state of $c=1$
[\eky,\bk]. Fortunately we can construct the two-point
function of (not necessarily on-shell) tachyons in
the black hole background. We will do this with two
independent methods, as a double check of our
computation.

\sec{The Two-Point Function}

We can perform a direct computation, fixing the
position of one of the screenings at $z_s=\infty$
and evaluating the remaining $(s-1)$ integrals:

\gviiia

\vskip 0.5cm

\no We can follow closely the steps of the
previous computation, although in this case it will
be much simpler. We use the representation {\Bxvi}
for the vertex operators and the bosonization
formulas for the $\b$-$\g$ system. The zero mode
integrations give:

\gviiib

\no Due to the kinematic relations satisfied
by the two-point amplitude, the part of the
integrand coming from the $\b$-$\g$ system can be
written as:

\gviii

\noindent where

\gix

\noindent This can easily be seen with the
substitution $z_i=1/y_i$.
After evaluating the $\phi$ contractions and taking
$z_s\rightarrow \infty$,
the complete ${\cal A}_{1\rightarrow 1}$ amplitude is reduced
to the evaluation of the following integral:

\gx

\vskip 0.3cm

\noindent The result is well defined for $\rho\neq
-4$ and can be obtained from (A.6):

\gxa

\vskip 0.5cm

\noindent In the next section we show that the
two-point function of (not necessary) on-shell
tachyons is only different from zero for
$j_1=j_2=j$, so that we set $s=2j+1$ and, from the
conservation law $m_1=-m_2=m\geq0$, we obtain for
an arbitrary level:

\gxi
\vskip 0.3cm

\noindent If $k\neq 9/4$ and the tachyons do not
belong to a discrete representation of {\sl}, the
above amplitude has one divergence, which appears
for $s$ integer and comes from the zero mode
integration. For $k=9/4$ we demand $j\notin \IZ/2$,
which implies that $s$ is non-integer. This can
always be done, since the above expression is well
defined in this case. The final expression for on-shell
tachyons is:

\gxii

\vskip 0.5cm
\no where $\widetilde{M}$ is the renormalized
black hole mass defined previously.

We can obtain the answer from the three-point
tachyon amplitude that contains one highest-weight
state ${\cal V}_{j_1\; j_1}$, taking the limit
$j_1=i\e \rightarrow 0$. In this way we are fixing
only two points of the SL(2,\IC)-invariant
${\cal A}_{1\rightarrow 1}$ amplitude. The result will contain
a divergence coming from the volume of the dilation
group [\sei,\gm].

First we can show that this amplitude is diagonal
by pushing $J_0^+$ and $J_0^-$ through the
correlator [\bih]. Here again we will use the fact that the
screening charge commutes with the currents.  We
obtain the relations

\gi

\vskip 0.5cm

\noindent From the $X$ zero-mode integration we get
$m_2=m$ and $m_3=-1-m$, so that this equation has
two solutions, one with $j_2=-1-j_3$, which
contains no screenings, and another one with
$j_2=j_3$, which is a two-point function with
$s=2j_2+1$ screenings. The first one is normalized
to one up to the divergence $\G(0)$ coming from the
zero mode integral. We now compute the two-point
function with screenings.

We first consider the case of arbitrary
$k$ and take the limit $k\rightarrow 9/4$ at the
end of the calculation. From eq. {\dix}, we obtain  for
$j_1=i\e$:

\gii

\noindent where

\giii

\no Here we have simplified the products of
well defined $\D$-functions. To evaluate the limit
we take (A.4) and the following
representation of the $\delta$-distribution:

\giv

\no into account. In total we obtain for the two-point
function of generic (in general off-shell)
tachyons:

\gvii

\no where $s=2j_2+1$ and the level is
arbitrary. This agrees with
eq. {\gxi} up to a factor $s$ and the volume of the
dilation group $\delta(j_2-j_3)$.

Using the $c=1$ language, we obtain for the
two-point function {\gxii} of two Seiberg on-shell
tachyons in the black hole background\foot{We have
absorbed a factor of 2 as before.}

\gxiii

\no The two-point function in standard non-critical string theory is:

\gxiv

\noindent Comparing with the two-point function in
the black hole background we find the same features
as for the three-point function. The pole structure
is the same as the one of the two-point function of
tachyons of $c=1$ at non-vanishing cosmological constant,
while the screenings differ by a factor of 2.

The two-point function of the deformed matrix model
is given by the expression [\jy]:

\gxv

\noindent Here only half of the states of $c=1$
(the supplementary series) appear as poles in the
leg factors. Both two-point functions can be
reconciled if we take into account that we can
renormalize the tachyon vertex operators in a
different way according to the {\sl}  representation
theory. If in the normalization {\Bxva} we use the
on shell condition and we take into account the
antiholomorphic piece, we renormalize our operators
as follows:

\gxvi

\no Here a regular function depending on $p$
has been dropped, which is not determined by
eq. {\Bxva}. The two-point function of these operators
in the black hole background agrees with eq. {\gxv}. It
is simple to see how the $N$-point function of
these differently normalized tachyons behaves. The
$N$-point function with chirality $(+,\dots,+,-)$,
will be divided by a factor $\D(-s-(N-1)/2)$,
coming from the state with the opposite chirality.
This will imply, that $(+,\dots,+,-)$ odd-point
functions of these operators vanish for positive
integer screenings, if they were previously finite.
For even point functions this factor is of course
irrelevant for integer $s$ so that they are finite
in this case. Which one is the correct physical
normalization is to be clarified. For related
problems see ref. [\gm].

\sec{\secsixsix}
\no In the previous sections we have seen that we are able to obtain
closed expressions for the two- and three-point tachyon correlation
functions in the Euclidean black hole background. Now we are going to
consider the four-point amplitude of tachyons\foot{We would like to
thank U.~Danielsson for discussions on this subject.}. As in the
computation of scattering amplitudes of minimal models coupled to
Liouville theory [\dfk], we are faced with the problem that the
integrals that have to be computed do not exist in the mathematical
literature. However, we are able to address this problem with similar
methods as those used in refs. [\dfk,\st] (see ref.
\REF\abdallabook{E.~Abdalla, M.~C.~B.~Abdalla, D.~Dalmazi and
A.~Zadra, ``2D-Gravity in Non-Critical Strings: Continuum and
Discrete Approaches'', Lecture Notes in Physics, Series M, Springer
Verlag (to appear).}
[\abdallabook]
for a review), although in this case the situation is more
complicated. The three-point function of on-shell tachyons previously
computed will be one of the basic ingredients. The results of this
section should be considered as preliminary and will be discussed in
more detail in ref. [\becker]. A general four-point function has the
form

$$
\eqalign{
{\widetilde {\cal A}}^{j_1\;\;\;j_2\;\;\;j_3\;\;\;
j_4}_{m_1\;m_2\;m_3\; m_4}=
& \int  d^2 \eta |\eta|^{2{\bf p_1}{\bf p_4}} |1-\eta|^{2{\bf
p_3}{\bf p_4} }\int \prod_{i=1}^s d^2 z_i \left( {\cal P}^{-1} {\p^s
{\cal P} \o \p z_1 \dots \p z_s}{\bar {\cal P}}^{-1}{\p^s {\bar {\cal
P}}\o \p{\bar z}_1\dots \p{\bar z}_s}\right)\cr
&\cr
&\times \prod_{i=1}^s  |z_i|^{-4\rho j_1} |1-z_i|^{-4\rho j_3}
|\eta-z_i|^{-4\rho j_4} \prod_{i< j} |z_i-z_j|^{4\rho}, \cr}
\eqn\jji
$$

\no where

$$
{\cal P}=\prod_{i=1}^s z_i^{m_1-j_1} (1-z_i)^{m_3-j_3}
(\eta-z_i)^{m_4-j_4} \prod_{i<j} (z_i-z_j).
\eqn\jjii
$$

\no We have fixed the positions of the four vertex operators at
$(z_1,z_2,z_3,z_4)=(0,\infty,1,\eta)$. The above integral is very
complicated and to get a closed expression for the solution we will
use

\item{\triangleright} the symmetries,

\item{\triangleright} the pole structure and

\item{\triangleright} the high energy behavior

\no of the amplitude. These properties will be enough to determine
the form of the solution using some powerful theorems for entire
functions. We will compute the amplitude for $j_i > 0$ and can obtain
more general amplitudes using analytic continuation in $j_i$.
We will begin with the determination of the symmetries.

\vskip 0.5cm
\no ${\rm \underline {Symmetries}}$
\vskip 0.5cm

\no The integral {\jji} exhibits several symmetries. Changing the
integration variables to $\eta \rightarrow 1-\eta$ and $w_i
\rightarrow 1-w_i$ we obtain

$$
{\widetilde {\cal A}}^{j_1\;\;\;j_2\;\;\;j_3\;\;\;
j_4}_{m_1\;m_2\;m_3\; m_4}=
{\widetilde {\cal A}}^{j_3\;\;\;j_2\;\;\;j_1\;\;\;
j_4}_{m_3\;m_2\;m_1\; m_4}.
\eqn\jjjxiii
$$

\no With a change of variables of the form $\eta \rightarrow {1/
\eta}$ and
$w_i \rightarrow {1/w_i }$ we obtain the symmetry

$$
{\widetilde {\cal A}}^{j_1\;\;\;j_2\;\;\;j_3\;\;\;
j_4}_{m_1\;m_2\;m_3\; m_4}={\widetilde {\cal A}}^{-j_1-j_3-j_4-1+s
\;\;j_2\;\;\;j_3\;\;\; j_4}_{-m_1-m_3-m_4 \;\;\;\;\;\;m_2\;m_3\;
m_4}.
\eqn\jjjxiv
$$

\vskip 0.5cm
\no ${\rm \underline {Pole \; \; Structure}}$
\vskip 0.5cm

\no The amplitude {\jji} has a complicated pole structure.
These poles can be computed by evaluating the short-distance
singularities that appear if some vertex operators (and some
screening charges) collide at one point. In these arguments the
screening charge has to be treated carefully, because it is not a
tachyon operator but a discrete state of $c=1$. In the case where the
perturbation is a cosmological constant we can compute the $N$-point
function with $s$ screenings out of the $(N+s)$-point function
without screenings, by sending $s$ of the momenta to zero. In the
black hole we have to distinguish between tachyon operators as
external insertions and the screening charge that is a discrete
state. We will begin by analysing the poles that appear if the vertex
operator ${\cal V}_{j_4 \; m_4}(\eta)$ approaches ${\cal V}_{j_1 \;
m_1}(0)$, while leaving the screenings far away from zero. The other
possibilities, where $\eta$ approaches $1$ or $\infty$, are
equivalent. The OPE of two vertex operators takes the form:

$$
:e^{i {\bf p_4} \cdot {\bf X}(\eta)}::e^{i {\bf p_1}\cdot {\bf
X}(0)}: =\sum_{n=0}^\infty \left({1\o n!}\right)^2 |\eta|^{2 {\bf
p_1}\cdot {\bf p_4}+2n} V_n(0)
\eqn\jjjii
$$

\no where ${\bf  X}=(X,\phi)$ and we have taken the same for the
holomorphic and the antiholomorphic part since these are the only
contributions when we integrate over $\eta$. The operators on the
r.h.s. are given by

$$
V_n=:e^{i{\bf p_4}\cdot {\bf  X}} \p^n {\bar \p}^n e^{i {\bf
p_1}\cdot {\bf  X}}:=:\left( -{\bf  p_1}\cdot \p^n {\bf  X} {\bf
p_1} \cdot {\bar \p}^n {\bf X}+\dots \right) e^{i {\bf  p} \cdot {\bf
 X}}:,
\eqn\jjjiii
$$

\no where ${\bf p}={\bf p_1}+{\bf p_4}$.
In the black hole CFT, the OPE between two vertex operators of the
form {\Ciii} is

$$
{\cal V}_{j_4 \; m_4}(\eta,{\bar \eta}){\cal V}_{j_1 \; m_1}(0,0)
=|\eta|^{2{\bf p_1}\cdot {\bf p_4}} {\cal V}_{j_1+j_4 \;
m_1+m_4}(0,0)+\sum_{n=1}^\infty|\eta|^{2 {\bf p_1}\cdot {\bf p_4}+2n}
{\cal V}_n(0,0).
\eqn\jjiii
$$

\no All the operators from the r.h.s. satisfy $(J,M)=(j_1+j_4,
m_1+m_4)$.
The first operator is a discrete tachyon that appears at zero mass
level, while ${\cal V}_n$ denote the discrete states that appear at
higher mass levels.

To analyze the pole structure we insert the OPE {\jjiii} into the
four-point function and near the poles that come from
$\eta\rightarrow 0$ we obtain

$$
\eqalign{
{\widetilde {\cal A}}^{j_1\;\;\;j_2\;\;\;j_3\;\;\;
j_4}_{m_1\;m_2\;m_3\; m_4}\approx & \int_{|\eta|\leq \ve} d^2 \eta
|\eta|^{2{\bf p_1}{\bf p_4}}\langle {\cal V}_{j_1+j_4\;
m_1+m_4}(0){\cal V}_{j_2\; m_2} (\infty) {\cal V}_{j_3\;
m_3}(1)\rangle + \cr
&\cr
&\sum_{n=1}^\infty \int_{|\eta|\leq \ve} d^2 \eta |\eta|^{2{\bf
p_1}{\bf p_4}+2n }
\langle {\cal V}_{n}(0){\cal V}_{j_2\; m_2} (\infty) {\cal V}_{j_3\;
m_3}(1)\rangle.\cr}
\eqn\jjv
$$

\no The residues of the poles of the above integral are described by
three-point functions that, in general, contain one discrete state
and two tachyon operators as external states. The poles appear if the
integral over $\eta$ diverges for $\ve \approx 0$. To evaluate the
integral we use polar coordinates and obtain

$$
\int_{|\eta|\leq \ve} d^2 \eta |\eta|^{2{\bf p_1}{\bf p_4}+2n}\approx
\pi {\ve^{2{\bf p_1} {\bf p_4}+2n+2} \o {\bf p_1} {\bf p_4}+n+1}.
\eqn\jjvi
$$

\no Therefore, the poles appear at

$$
{2 \o k} m_1 m_4 -{4 \o \a_+^2} j_1 j_4 +n+1=0.
\eqn\jjvii
$$

\no The above equation is precisely the condition that the
intermediate state is on-shell:

$$
-{(j_1+j_4)(j_1+j_4+1)\o k-2}+{(m_1+m_4)^2 \o k}+n=1,
\eqn\jjiv
$$

\no where the mass-level of the intermediate particle is ${\cal
N}=n$. We consider first the chirality configuration $(+,+,+,-)$. We
are interested in the case $k=9/4$; the kinematical relations are
then

$$
\eqalign{
&2j_i+1={2\o 3} m_i \qquad {\rm for } \qquad i=1,2,3\cr
 &\cr
&2j_4+1=-{2\o 3} m_4.\cr}
\eqn\jjviii
$$

\no Using the above kinematics we obtain the relations

$$
j_4={s\o 2} \qquad {\rm and } \qquad     j_2=-j_1-j_3-1+{s\o 2}.
\eqn\jjix
$$

\no This means that the independent kinematic variables that describe
this process are $j_1$, $j_3$ and $s$. The poles appear in the
amplitude if

$$
j_1={{\cal N} \o 4(2s+1)} -{1\o 4}.
\eqn\jjx
$$

\no Since the intermediate state satisfies $(J,M)=(j_1+j_4,m_1+m_4)$
we obtain from eqs. {\jjix} and {\jjx}:

$$
J=j_1+j_4 ={1\o 4} \left( {{\cal N} \o 2s+1}-1+2s\right).
\eqn\jjxi
$$

\no We will first analyse the case where the intermediate state is a
discrete tachyon:

$$
J=j_1+j_4={2s-1 \o 4}.
\eqn\jjxiii
$$

\no The corresponding four-point function is represented in Fig. 8.
While the first three-point function contains $s$ screenings, the
second three-point function satisfies $s=0$ and is therefore equal to
one.

\midinsert
\epsfysize=1.5in
\vskip 0.5cm
\centerline{\hskip 0.0cm \epsffile{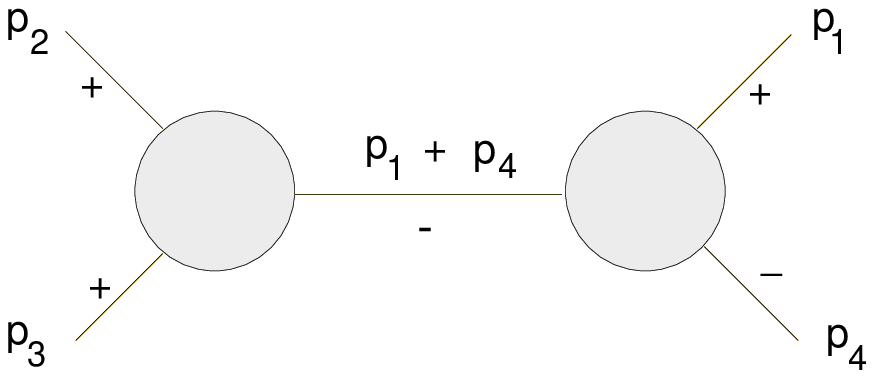}}
\caption{{\bf Fig. 8}: Near the pole, the four-point amplitude with a
chirality configuration $(+,+,+,-)$ can be factorized into two
three-point functions. The intermediate state is on-shell, has
negative chirality and is on the right branch.}
\vskip 0.5cm
\endinsert

Using the kinematical relations {\jjviii} and {\jjxiii} we obtain
that one of the external states is fixed at $(j_1,m_1)=(-1/4,3/4)$,
while $(j_4,m_4)=(s/2,-3(s+1)/2)$. Since the intermediate state has
negative chirality the residuum of the pole is described by the
three-point function:
$$
\langle {\cal V}^{-}_{j_1+j_4\; m_1+m_4} (0) {\cal V}^{+}_{j_2\; m_2}
(\infty){\cal V}^{+}_{j_3\; m_3} (1)\rangle,
\eqn\jjxiv
$$

\no where the upper index denotes the chirality. Using the expression
for the on-shell three-point function {\dxxviii} it is easy to see
that this residuum is different from zero. We conclude that for the
value of $J$, given in equation {\jjxiii}, there appears a pole in
the four-point function that corresponds to an intermediate discrete
tachyon.

We now consider the case where the intermediate particle is a state
at higher mass level and we would like to determine those values of
$J$ from eq. {\jjxi} for which there appear poles in the amplitude.
First we observe from eq. {\jjxi} that the new discrete states of
Distler and Nelson [\dn] do not appear.
To obtain the poles, we can use the arguments of ref. [\st] for $c=1$
based on the decoupling of null-states. Only if ${\cal N}=\alpha
(2s+1)$, where $\a$ is a positive integer does the intermediate state
describe an on-shell physical string state of the classification
{\Cx}

$$
J={\a -1 \o 4} +{s\o 2}\qquad { \rm with} \qquad \a\in \IN.
\eqn\jjxii
$$

\no In this case we have $j_1=(\a-1)/4$.

To make clear the above condition we consider a simple example [\st].
Using the $c=1$ language [\eky,\bk] the on-shell vertex operator of
the black hole at level ${\cal N}=1$ can be written in the form

$$
{\cal V}_{1}={\bf p}\cdot \p {\bf X} {\bf  p} \cdot {\bar \p} {\bf
X} e^{ i {\bf  p} \cdot {\bf  X}}=-\p {\bar \p}e^{ i {\bf  p} \cdot
{\bf  X}} .
\eqn\jjxiii
$$

\no This state is clearly a null state for generic values of ${\bf
p}$, so that it decouples from correlation functions and the residuum
vanishes. The situation is different for ${\bf p}=0$. Then this state
has $(J,M)=(0,0)$ and corresponds to a discrete state. It appears as
a pole in amplitudes that satisfy $s=0$, where $\a =1$. Similarly we
can argue for the other residues. They are described by three-point
functions involving one state at higher mass-level and vanish if this
state is not an on-shell physical string state described by eq.
{\jjxii}.
This decoupling of null states is a property of tachyon correlation
functions of $c=1$ matter coupled to Liouville theory and it is
plausible that the same property is satisfied by the black hole CFT
as well.

We can carry out a similar analysis of the pole structure by scaling
a number $r$ of screening charges to zero as $\eta \rightarrow 0$. It
turns out that the poles occur for $2j_1$ integer or half-integer

$$
j_1={n \o 4(2(s-r)+1)}+{r\o 2} -{1\o 4}={\a-1 \o 4} +{r\o 2} \qquad
{\rm with} \qquad s\geq r.
\eqn\hhxiii
$$

\no The residues of these poles are described by products of three
point functions of on-shell states. These are the only poles that
occur for $j_i>0$.

Summarizing, the pole structure for $j_i>0$ and the symmetries of the
integral are fully captured, if we make the following ansatz for the
amplitude

$$
{\widetilde {\cal A}}(j_1,j_2,j_3,j_4) = f(j_1,j_{3},s)
\D(-4j_1-1)\D(-4j_3-1)\D\left(4\left(j_1+j_3-{s\o
2}+1\right)-1\right).
\eqn\jjjjxiii
$$

\no To determine the four-point amplitude we have to find the
function $f(j_1,j_3,s)$.

\vskip 0.5cm
\no ${\rm \underline {High \;\;Energy \;\;Behavior}}$
\vskip 0.5cm

\no Next, we would like to understand the asymptotic behavior of the
amplitudes for high energies. We consider, for example, the $j_1$
dependence. Introducing  the notation $\a=-2 \rho j_1$ we analyse the
behavior of the integral in the limit $\a \rightarrow \infty$. Making
the transformation of variables:

$$
\eta=\exp\left( {x\o \a}\right) \qquad {\rm and} \qquad
z_i=\exp\left( {y_i \o \a}\right)
\eqn\jjjxv
$$

\no we are able to keep the relevant contributions to the integral
{\jji}.  We obtain the asymptotic formula

$$
{\widetilde {\cal A}}^{j_1\;\;\;j_2\;\;\;j_3\;\;\;
j_4}_{m_1\;m_2\;m_3\; m_4}
\approx \a^{8 j_3-4s+2}\qquad {\rm for} \qquad \a \rightarrow \infty.
\eqn\jjjxvi
$$

\no We can compute now the asymptotic behavior as $\a \rightarrow
\infty$ of the r.h.s. of equation {\jjjjxiii} using Stirling's
formula. We obtain a power law growth as a function of $\a$ with the
same exponent as in eq. {\jjjxvi}. We conclude that in the asymptotic
region $f(j_1,j_3,s)$ is independent of $j_1$ and, by the symmetry
{\jjjxiii}, of $j_3$ as well.

The results {\jjjjxiii}, {\jjjxvi} and the symmetries {\jjjxiii} and
are actually enough to evaluate the whole amplitude in the case that
the perturbation is a cosmological constant [\dfk], because it is
possible to show that the function $f(j_1,j_3,s)$ is an analytic
function of $j_1$ and of $j_3$ as well. With the asymptotic behavior
{\jjjxvi} and using powerful theorems for entire functions
\REF\boas{R.~Boas, ``Entire Functions'', Academic Press, NY 1954.}
[\boas]
 one can conclude that $f$ is independent of $j_1$ and $j_3$, so that
$f=f(s)$. Then it was possible fix the unknown function $f(s)$ by
choosing some convenient values of the momenta; by sending $N-3$ of
the momenta to zero the amplitude is identified with derivatives of
the three point function with respect to the cosmological constant
and this fixes $f$ uniquely.
In the case of the black hole we will assume that $f$ is an entire
function as well. Since it depends only on $s$ in the asymptotic
region, we know that $f$ does not depend on $j_1$ and by symmetry it
does not depend on $j_3$ in the whole $j_i$ plane. Summarizing, the
four point function of tachyon operators has the form

$$
{\widetilde {\cal A}}(j_1,j_2,j_3,j_4) = f(s)
\D(-4j_1-1)\D(-4j_3-1)\D\left(4\left(j_1+j_3-{s\o
2}+1\right)-1\right).
\eqn\jjjjy
$$

The computation of $f(s)$ is more complicated in this case because
the interaction is a discrete state and not a tachyon. However, we
can calculate a four-point function where some of the $j_i's$ are
fixed at the most convenient values, while keeping $s$ arbitrary.
Taking

$$
j_1=-{1\o 4}-{\ve \o 4},
\eqn\jjjxvii
$$

\no where $\ve$ is a small parameter, we obtain from the ansatz
{\jjjjxiii} that the order $1/ \ve$ is given by the expression

$$
{\widetilde {\cal A}}(j_1,j_2,j_3,j_4)\approx {1\o \ve}
\D(-4j_2-1)\D(-4j_3-1)f(s).
\eqn\jjjxviii
$$

\no On the other hand, we can directly compute the order $1/ \e$ of
this  amplitude using the factorization formula {\jjv}. In this case,
we observe that the residuum is described by a three-point function
{\jjxiv} of on-shell tachyons, that we already know how to calculate.
Therefore, we have to consider

$$
{\widetilde {\cal A}}^{j_1\;\;\;j_2\;\;\;j_3\;\;\;
j_4}_{m_1\;m_2\;m_3\; m_4}
\approx {- \pi \o (4j_1+1)(4j_4+1)}\langle {\cal V}_{j_1+j_4\;
m_1+m_4 }^{-}(0)  {\cal V}_{j_2\; m_2}^{+}(\infty) {\cal V}_{j_3\;
m_3}^{+}(1)\rangle,
\eqn\jjjxix
$$

\no taking into account the kinematical constraints. After using our
expression for the three-point function {\dxxviii}  we obtain the
result

$$
f(s)=-\pi (2s+1){ \D(-4j_4-1) \o \G(-s)}.
\eqn\jjjxx
$$

\no In this computation we have assumed that we have no other
contributions to the residuum, which may come from the scaling of the
screenings, while leaving $\eta$ fixed. If this contribution is
different from zero, the function $f(s)$ will be different. This
point has to be carefully understood.
Our result for the four-point function of tachyons in the black hole
background is

$$
{\cal A}(j_1,j_2,j_3,j_4)={\widetilde M}^s (2s+1)\prod_{i=1}^4 (-\pi)
\D(-4j_i-1).
\eqn\jjjxxi
$$

\no We have introduced a factor $(-\pi)^3$ into the definition of the
path integral. This is in agreement with the standard notation of
$c=1$.
{}From the above result we see that we have agreement with the
four-point function of tachyon operators in $c=1$ coupled to
Liouville theory perturbed by the cosmological constant, if we take
into account that the number of screenings between both theories
differ by a factor of two and that the cosmological constant is
square of the black hole mass. To do this computation we have used
the three-point function of on-shell tachyons as a basic ingredient.
We can conclude, that the regularisation procedure previously used
for the three-point function is fully consistent with the four-point
function.

\endpage
\chapter{CONCLUDING REMARKS}
\no Quantum black hole physics involves many problems, which are hard
to solve in four dimensions. One of them is the determination of the
endpoint of black hole evaporation. The first difficulty to be faced
is that at the late stages in the evaporation process quantum gravity
effects become relevant. Therefore string theory could play an
important role.
A model to describe the propagation of strings in a black hole
background was proposed by Witten [\Wit].  The exact CFT describing a
black hole in two dimensions has a Lagrangian formulation in terms of
a gauged WZW model based on the non-compact group {\sl}.
This model is exact in the sense that it solves the $\beta$-function
equations of the string to all orders in the string coupling
constant. This is important because, near the singularities,
higher-order effects in $\a'$ are expected to be relevant.
Since this black hole is described in terms of a coset model we can
hope to address problems in black hole physics using well-known
methods of CFT.

One of these well-known methods is the free-field description of a
CFT. This formulation of Witten's black hole has been proposed by
Bershadsky and Kutasov [\bk]. The problem of considering scattering
processes becomes much simpler in this formulation. Using the
Wakimoto representation of the {\sl} current algebra, they obtained
the action formulated in terms of Wakimoto coordinates. With this
approach the {\sl} symmetry of the theory is manifest, while a
natural derivation of the action coming directly from the Lagrangian
of the WZW model can be achieved using the Gauss decomposition [\vv].
In the semiclassical limit, the space-time structure of this model
has been analysed in ref. [\bk]. It was found that it reproduces all
the regions of an ordinary Schwarzschild black hole of classical
general relativity.

However, Dijkgraaf, E.~Verlinde and H.~Verlinde [\dvv] have shown
that Witten's solution is only correct in the semiclassical limit of
$k \rightarrow \infty$. The exact expressions for the dilaton and the
metric receive corrections of order $1/k$ that can be computed with a
mini-superspace description of the conformal field theory.

Carrying out the quantum mechanical analysis of the black hole in
terms of Wakimoto coordinates, we have been able to find the same
space-time interpretation for finite $k$ [\becker]. This is important
for two reasons. First it clarifies the relation between the
free-field model of ref. [\bk] and Witten's black hole for finite $k$
and shows the equivalence between the two models. Secondly it
provides us with a space-time interpretation for $k=9/4$ that is the
interesting case in order to perform the computation of the
scattering amplitudes of tachyons in the black hole background. It
would be nice to derive the obtained expressions for the metric and
the dilaton for finite $k$ directly from the Lagrangian of ref. [\bk]
with a careful treatment of the quantum effective action [\Tseytlin].

Having a suitable formulation of the exact 2D black hole solution of
string theory we have performed the computation of the amplitudes
describing the interaction of tachyons in the Euclidean black hole
background. Specially interesting is the question whether there
exists a connection between these scattering amplitudes and the $\cal
S$-matrix describing the interaction of tachyons in standard
non-critical string theory and the deformed matrix model of Jevicki
and Yoneya [\jy]. These are systems for which we have a powerful
nonperturbative description in terms of the matrix model formalism so
that we can study the issues of singularities in string theory.

Using the Wakimoto free-field representation of the
{\slU} Euclidean black hole, we have found that
tachyon two-, three- and four-point correlation functions
share a remarkable analogy with the tachyon amplitudes of
$c=1$ coupled to Liouville at non-vanishing
cosmological constant. This observation was made by
Bershadsky and Kutasov for those amplitudes where no screening charge
is
needed to satisfy the total charge balance.

In order to have non-vanishing correlation
functions, we have infinitely renormalized the
black hole mass. This is a well-known phenomenon
for $c=1$, and in this case it has interesting
physical consequences. Perhaps this is the case for
the black hole mass as well; further investigation
is desirable.
The amplitudes factorize in leg factors, which have
poles at all the discrete states of $c=1$. The new
discrete states of Distler and Nelson [\dn] do not
appear, because they are BRST-trivial in the
Wakimoto representation, as checked in ref. [\bk] for
the first examples. This is in agreement with the analysis of the
BRST cohomology carried out by Eguchi et al. [\eky]. The
scaling of the correlators is different from $c=1$,
but can be reproduced with the substitution
$\mu^2=M$. With our renormalization of the
operators, we do not reproduce the pole structure
of the correlators of the deformed matrix model
[\jy]. Whether this discrepancy could be merely a
normalization of the operators was discussed.

There are many interesting
questions, suggested by these observations. An
important one is to see whether this relation
to $c=1$ persists for the $N$-point functions. More results in this
direction  will appear in ref. [\becker].
It would be nice to see if the correlators can be computed using the
Ward identities of $c=1$
\REF\klepa{I.~R.~Klebanov, ``Ward Identities in Two-Dimensional
String Theory'', {\it Mod.~Phys.~Lett.} {\bf A7} (1992) 723; I.~R.
Klebanov and A.~Pasquinucci, ``Correlation Functions From
Two-Dimensional String Ward Identities'', {\it Nucl.~Phys.} {\bf
B393} (1993) 261.}
[\klepa].

Previously there appeared several papers in the literature, where a
relation between the 2D black hole and standard non-critical string
theory is found [\das,\wadia,\ms]. It will be interesting to explore
whether it is possible to see a direct connection to these
approaches.

Recently Vafa and Mukhi
\REF\vm{\rvm}
[\vm]
have proposed a topological field theory with which it is possible to
compute tachyon correlation functions in non-critical string theory
on higher-genus Riemann surfaces in the continuum approach
\REF\gmm{D.~Ghoshal and S.~Mukhi, ``Topological Landau-Ginzburg Model
of Two-Dimensional String Theory'', preprint MRI-PHY-13-93,
hep-th/9312189; S.~Mukhi, ``The Two-Dimensional String as a
Topological Field Theory, preprint TIFR-TH-93-61, hep-th/9312190.}
\REF\hop{A.~Hanany, Y.~Oz and R.~Plesser, ``Topological Landau
Ginzburg Formulation and Integrable Structure of 2D String Theory'',
preprint IASSNS-HEP-94/1, hep-th/9401030}
[\gmm,\hop].
The generalization of their methods to correlation functions of
discrete states could be used to determine the $\cal S$-matrix of
Witten's black hole for non-spherical topologies in the continuum
formulation.

An important question is the connection between the considered 2D
black hole solution and more realistic black holes in four dimensions
\REF\gps{S.~B.~Giddings, J.~Polchinski and A.~Strominger, ``Four
Dimensional Black Holes in String Theory'', {\it Phys.~Rev.} {\bf
D48} (1993) 5784.}
\REF\ghpss{S.~B.~Giddings, J.~A.~Harvey, J.~G.~Polchinski,
S.~H.~Shenker and A.~Strominger, ``Hairy Black Holes in String
Theory'', preprint NSF-ITP-93-58 hep-th/9309152.}
\REF\hw{G.~T.~Horowitz and D.~L.~Welch, ``Exact Three Dimensional
Black Holes in String Theory'', {\it Phys.~Rev.~Lett.} {\bf 71}
(1993) 328.}
[\gps,\ghpss,\hw].
In higher dimensions there exists an infinite number of propagating
massive string modes and it would be interesting to analyse their
role in black holes physics. Perhaps some new stringy effects will be
discovered which will lead us to the conclusion that string theory is
really the correct quantum theory of gravitation.
\vfill
\endpage
\noindent {\bf Acknowledgements}

\no I would like to thank my advisor W.~Nahm for giving me the
opportunity to carry out the research work of my thesis at the Theory
Division of CERN. I am grateful to my advisor L.~Alvarez-Gaum\'e for
his guidance and important discussions during these years. Very
specially I would like to thank M.~Becker for a fruitful
collaboration. It is a pleasure to acknowledge discussions with
L.~Alvarez-Gaum\'e, J.~L.~Barb\'on, M.~Bershadsky, U.~Danielsson,
J.~Distler, V.~Dotsenko, M.~Douglas, E.~Kiritsis, D.~Kutasov,
W.~Lerche, G.~Moore, W.~Nahm, R.~Plesser, N.~Seiberg, S.~Shenker,
A.~Tseytlin, E.~Verlinde, H.~Verlinde, C.~Vafa and E.~Witten. I would
like to thank W.~Nahm for a careful reading of the manuscript.
Finally, I thank the Theory Division of CERN for hospitality and the
Deutsche Forschungsgemeinschaft for financial support.
\vfill
\endpage
\centerline{\bf APPENDIX  }
\noindent For convenience we will collect the
identities of $\G$ functions that we have used in
our computations:
\ki

\kia

\kiii

\noindent To regularize the result of the singular
integrals we use:

\kii

\noindent The definition of the hypergeometric
function, which we need in section~3 to compute the
three-point function of generic tachyons, is:

\dxvia

\noindent The following identity, known as
``Dixon's theorem'', is useful to evaluate the
three-point function of on-shell tachyons:

\kv

\vskip 0.5cm

\noindent To evaluate the integrals for the three-point function, we
have used the Dotsenko-Fateev
(B.9) formula :

\kvi

\vfill
\endpage
\refout
\end